\documentclass[aps,pra,showpacs,preprintnumbers,amsmath,amssymb,reprint,superscriptaddress]{revtex4-2}
\usepackage{graphicx,dcolumn,bm}
\usepackage[utf8]{inputenc}
\usepackage[T1]{fontenc}
\usepackage{mathptmx}
\usepackage{bm}
\usepackage{siunitx}
\usepackage{booktabs}
\usepackage{longtable}
\usepackage{dcolumn}
\newcolumntype{d}[1]{D{.}{.}{#1}}
\begin{document}

\preprint{}

	\title[Spectroscopic characterization of aluminum monofluoride with relevance to laser cooling and trapping]{Spectroscopic characterization of aluminum monofluoride with relevance to laser cooling and trapping}
	
	\author{S. Truppe}
	\email{truppe@fhi-berlin.mpg.de}
	\affiliation{Fritz-Haber-Institut der Max-Planck-Gesellschaft, Faradayweg 4-6, 14195 Berlin, Germany}
	\author{S. Marx}
	\affiliation{Fritz-Haber-Institut der Max-Planck-Gesellschaft, Faradayweg 4-6, 14195 Berlin, Germany}
	\author{S. Kray}
	\affiliation{Fritz-Haber-Institut der Max-Planck-Gesellschaft, Faradayweg 4-6, 14195 Berlin, Germany}	
	\author{M. Doppelbauer}
	\affiliation{Fritz-Haber-Institut der Max-Planck-Gesellschaft, Faradayweg 4-6, 14195 Berlin, Germany}
	\author{S. Hofs\"{a}ss}
	\affiliation{Fritz-Haber-Institut der Max-Planck-Gesellschaft, Faradayweg 4-6, 14195 Berlin, Germany}
	\author{H. C. Schewe}
	\affiliation{Fritz-Haber-Institut der Max-Planck-Gesellschaft, Faradayweg 4-6, 14195 Berlin, Germany}
	\author{N. Walter}
	\affiliation{Fritz-Haber-Institut der Max-Planck-Gesellschaft, Faradayweg 4-6, 14195 Berlin, Germany}
	\author{J. Pérez-Ríos}
	\affiliation{Fritz-Haber-Institut der Max-Planck-Gesellschaft, Faradayweg 4-6, 14195 Berlin, Germany}
	\author{B. G. Sartakov}
	\affiliation{General Physics Institute, Russian Academy of Sciences, Vavilovstreet 38, 119991 Moscow, Russia}
	\author{G. Meijer}
	\email{meijer@fhi-berlin.mpg.de} 
	\affiliation{Fritz-Haber-Institut der Max-Planck-Gesellschaft, Faradayweg 4-6, 14195 Berlin, Germany}
	
\date{\today}
\begin{abstract}
Here we report on spectroscopic measurements of the aluminum monofluoride molecule (AlF) that are relevant to laser cooling and trapping experiments. We measure the detailed energy level structure of AlF in the X$^1\Sigma^+$ electronic ground state, in the A$^1\Pi$ state, and in the metastable a$^3\Pi$ state. We determine the rotational, vibrational and electronic branching ratios from the A$^1\Pi$ state. We also study how the rotational levels split and shift in external electric and magnetic fields. We find that AlF is an excellent candidate for laser cooling on any Q-line of the A$^1\Pi$ - X$^1\Sigma^+$ transition and for trapping at high densities.  
        
The energy levels in the X$^1\Sigma^+, v=0$ state and within each $\Omega$-manifold in the a$^3\Pi, v=0$ state are determined with a relative accuracy of a few kHz, using laser-radio-frequency multiple resonance and ionization detection schemes in a jet-cooled, pulsed molecular beam. To determine the hyperfine and $\Lambda$-doubling parameters we measure transitions throughout the 0.1 MHz -- 66 GHz range, between rotational levels in the X$^1\Sigma^+, v=0$ state and between rotational and $\Lambda$-doublet levels in all three spin-orbit manifolds of the a$^3\Pi, v=0$ state. We measure the hyperfine splitting in the A$^1\Pi$ state using continuous wave (CW) laser-induced fluorescence spectroscopy of the A$^1\Pi, v=0 \leftarrow \textrm{X}^1\Sigma^+, v''=0$ band. The resolution is limited by the short radiative lifetime of the A$^1\Pi, v=0$ state, which we experimentally determine to be $1.90\pm 0.03$~ns. The hyperfine mixing of the lowest rotational levels in the A$^1\Pi$ state causes a small loss from the the main laser cooling transition of 10$^{-5}$. The off-diagonal vibrational branching from the A$^1\Pi, v=0$ state is measured to be $(5.60\pm0.02)\times10^{-3}$ in good agreement with theoretical predictions. The strength of the spin-forbidden A$^1\Pi, v=0 \rightarrow \textrm{a}^3\Pi, v'=0$ transition is measured to be seven orders of magnitude lower than the strength of the A$^1\Pi, v=0 \rightarrow \textrm{X}^1\Sigma^+, v''=0$ transition. We determine the electric dipole moments $\mu(\textrm{X})=1.515\pm 0.004$ Debye, $\mu(\textrm{a})=1.780\pm0.003$ Debye and $\mu(\textrm{A})=1.45\pm 0.02$ Debye in X$^1\Sigma^+$, $v=0$, a$^3\Pi, v=0$ and A$^1\Pi, v=0$, respectively, by recording CW laser excitation spectra in electric fields up to 150~kV/cm.  
\end{abstract}
	
\maketitle

%

\section{\label{sec:introduction} Introduction and motivation}

Over the last two decades, there has been great progress in cooling and trapping  neutral molecules in the gas phase \cite{Carr2009, Bohn2017}. Ultracold molecules can be used for studying collisions and chemistry at low temperature \cite{Krems2008, Balakrishnan2016}, for precision measurements to test fundamental symmetries \cite{DeMille2017, Safronova2018} and for quantum information and simulation \cite{DeMille2002, Moses2016, Blackmore2018, Ni2018}. The molecules can be associated at ultralow temperatures from pre-cooled atoms \cite{Sage2005, Ni2008, Liu2018} or they can be produced in a molecular beam and subsequently cooled and trapped. Experimental techniques involve buffer gas cooling \cite{Weinstein1998}, Stark deceleration \cite{Bethlem1999} and Zeeman deceleration \cite{Vanhaecke2007a, Narevicius2008}, Sisyphus cooling \cite{Zeppenfeld2012} and laser cooling. Laser cooling has so far been demonstrated for four diatomic species SrF \cite{Shuman2010}, YO \cite{Hummon2013}, CaF \cite{Zhelyazkova2014}, and YbF \cite{Lim2018} and one polyatomic species SrOH \cite{Kozyryev2017}. Magneto-optical traps (MOTs) for SrF \cite{Barry2014}, CaF \cite{Williams2017, Anderegg2017} and YO \cite{Collopy2018} have been demonstrated and sub-Doppler temperatures have been reached \cite{Truppe2017}. The laser cooling of BaF \cite{Chen2017, TheNL-eEDMcollaboration2018, Albrecht2019}, BaH \cite{Iwata2017}, TlF \cite{Norrgard2017}, MgF \cite{Xu2016}, CaOH \cite{Kozyryev2019} and YbOH \cite{Nakhate2019} is also being pursued. Magnetic \cite{Williams2018, McCarron2018} and optical \cite{Anderegg2018, Anderegg2019} trapping of laser cooled molecules has been demonstrated and sympathetic \cite{Lim2015, Son2019} and evaporative cooling \cite{Stuhl2012} is being explored. Ultracold molecular samples produced this way typically have a density many orders of magnitude lower compared to the association methods \cite{DeMarco2019}. New methods are steadily being developed to deliver more molecules at low speeds to the trapping region \cite{Fitch2016, Jayich2016, Wu2017, Galica2018, Petzold2018}.

For all the diatomic molecules that have been laser-cooled so far, a $^2\Pi_{1/2}$ $\leftarrow$ $^2\Sigma^+$ transition is used \cite{McCarron2018a}. In this case, rotational branching is suppressed on the P(1) line by angular momentum selection rules \cite{Stuhl2008}. The molecule must have an excited state that decays at a high rate to just one or a few vibrational levels in the ground state, and there should be no accessible intermediate electronic state. Preferably, the hyperfine structure of the molecule should be simple. Molecules with a $^1\Sigma$ ground state and a $^1\Pi$ excited state are very attractive candidates for laser cooling \cite{DiRosa2004, Hendricks2014, Tarbutt2018}. However, none have been laser-cooled so far. All Q-lines of a $^1\Pi$ $\leftarrow$ $^1\Sigma$ transition are rotationally closed, the hyperfine splitting in the $^1\Sigma$ state is typically within the natural linewidth of the optical transition, and molecules in $^1\Sigma$ states are intrinsically more stable than radicals.

\begin{figure}[htb!]
	\centering
	\includegraphics[width=\linewidth]{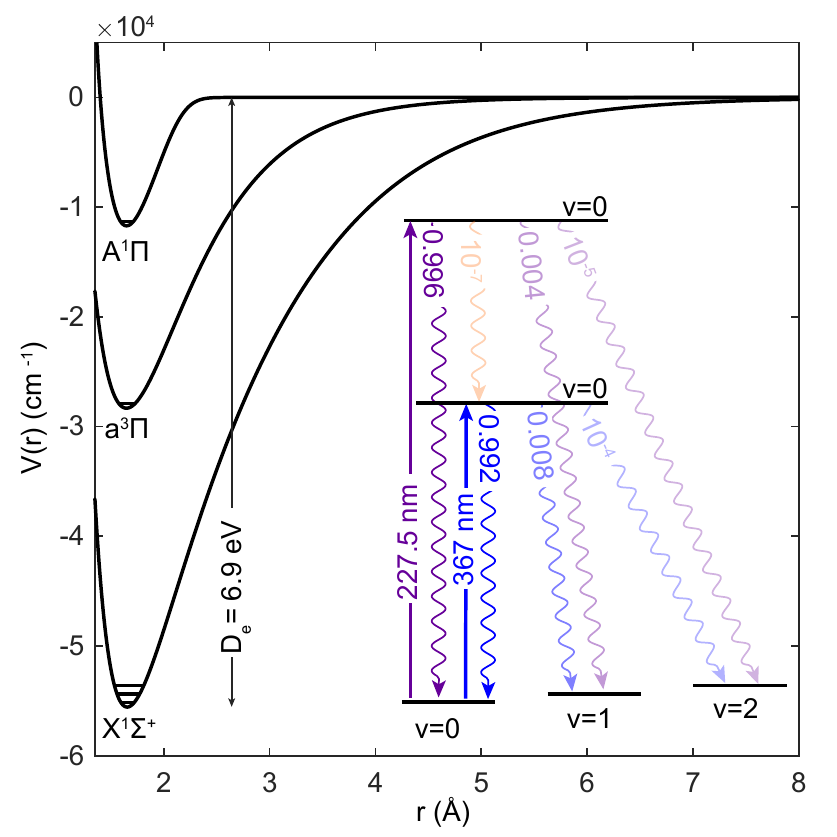}
	\caption{Potential energy curves of the three electronic states of AlF relevant for laser cooling. The inset on the right shows the transition wavelengths and calculated Franck-Condon factors presented in Sec. \ref{sec:fc}.}
	\label{fig:potentials}
\end{figure}

The feasibility of laser cooling of aluminum monofluoride (AlF) has been investigated using $\it{ab~initio}$ quantum chemistry \cite{Wells2011}. The study concludes that the A$^1\Pi, v=0 - \textrm{X}^1\Sigma^+, v''=0$ band of AlF around 227.5 nm has a calculated Franck-Condon factor of 0.99992 and is an excellent candidate for cooling with just a single laser. The spontaneous decay rate of the excited state is very high with a calculated value of $\Gamma=1/\tau=2\pi \times 84$ MHz, where $\tau=1.89$ ns is the radiative lifetime of the A$^1\Pi, v=0$ state \cite{Langhoff1988}. The spin-forbidden a$^3\Pi, v'=0 - \textrm{X}^1\Sigma^+, v''=0$ band is highly diagonal as well; it has a calculated Franck-Condon factor of 0.9967. Furthermore, the P$_1(1)$ line and all Q-lines of the a$^3\Pi - \textrm{X}^1\Sigma^+$ transition are rotationally closed. These transitions can be used for laser cooling to final temperatures in the low \si{\micro\kelvin} range, after pre-cooling on the much stronger A$^1\Pi - \textrm{X}^1\Sigma^+$ band. Figure \ref{fig:potentials} shows the potential energy curves for the three electronic states relevant to laser cooling. The main transition wavelength and the calculated Franck-Condon factors, based on experimental spectroscopic data, are shown. 

AlF is a promising candidate to reach high densities of ultracold molecular samples. It has a binding energy of almost 7 eV and forms as a stable constituent of aluminum-fluorine systems at high temperatures via the reaction
\begin{equation}
2 \textrm{Al(l)} + \textrm{AlF}_3\textrm{(g)} \rightarrow 3 \textrm{AlF}\textrm{(g)}.
\end{equation}
Vapour pressures of tens of mbar can be reached at temperatures around 1200-1350 K \cite{Ko1965}.
Therefore, a bright beam of AlF can be produced, either pulsed or CW. Because of the high spontaneous decay rate on the A$^1\Pi - \textrm{X}^1\Sigma^+$ band, the distance needed for laser slowing a molecular beam to rest is only several centimeters and the capture velocity of a MOT will be exceptionally large. The bimolecular rearrangement channel in which two AlF molecules react to form a fluorine molecule and an aluminum dimer is strongly endothermic. This reaction channel will therefore not limit the lifetime of AlF molecules in dense, cold samples. 

The spectroscopy of AlF is intrinsically interesting, as it is similar to carbon monoxide, one of the most-studied diatomic molecules. AlF is heavier and has eight electrons more than CO, shifting its electronic transitions to the more accessible region of the spectrum. The spin-forbidden a$^3\Pi \leftarrow \textrm{X}^1\Sigma^+$ band, for example, is centered at 206 nm for CO (Cameron band), whereas for AlF it appears around 367 nm. The most abundant isotopomer of CO has no nuclear spin. Aluminum and fluorine have a single stable isotope, with a nuclear spin of I$_\textrm{Al}=5/2$ for $^{27}$Al and I$_\textrm{F}=1/2$ for $^{19}$F. There are not many diatomics for which the most abundant isotopomer has this nuclear spin combination and for which the hyperfine structure has been completely resolved. The spectroscopic data on AlF has until now been restricted to absorption and emission measurements in samples at high ($\geq 900$ K) temperatures with at best Doppler-limited resolution. This data is suitable to get accurate information on the electronic potential curves and on the ro-vibrational energy level structure from which Franck-Condon factors can be calculated precisely. The fine and hyperfine structure, however, has barely been resolved \cite{Barrow1974,Brown1978,Lide1963,Lide1965,Wyse1970,Hoeft1970,Honerjager1974}.

Cooling and trapping experiments rely on measurements of the fine and hyperfine structure of the lowest rotational levels in X$^1\Sigma^+$, A$^1\Pi$ and a$^3\Pi$. First, the working principle of a MOT is determined by the hyperfine structure and the magnetic g-factors. Second, the degree of mixing of rotational levels due to hyperfine interactions determines how far branching to other rotational levels, i.e. loss from the optical cycling transition, occurs \cite{Norrgard2017}. The vibrational branching ratios can be calculated, but typically not to the required accuracy. Therefore, it is essential to compare these calculations to the vibrational branching ratios determined by experiments. Molecules that decay on the spin-forbidden A$^1\Pi \rightarrow \textrm{a}^3\Pi$ transition are also lost from the optical cooling cycle. We measure the strength of this transition to quantify this loss channel. The predicted spontaneous decay rate on the A$^1\Pi, v=0 - \textrm{X}^1\Sigma^+, v=0$ band requires experimental verification. Besides, it is useful to measure how the hyperfine levels split and shift in external electric and magnetic fields. We use a pulsed beam of jet-cooled AlF in combination with radio-frequency, microwave, and optical fields to experimentally determine this data.  

This paper is organized in eleven sections. The early spectroscopic characterization of AlF is discussed in Sec. \ref{sec:history}, with a particular focus on the electronic states relevant for this study. Then the experimental setup is described, together with the various multiple resonance excitation and detection schemes that are used. In Sec. \ref{sec:theory}, the Hamiltonian we use to describe the energy level structure of AlF is given. In Sec. \ref{sec:aToX}, overview excitation spectra of the a$^3\Pi, v=0 \leftarrow \textrm{X}^1\Sigma^+, v=0$ band are presented. Radio-frequency and microwave measurements in the a$^3\Pi, v=0$ state together with the spectroscopic parameters we determine from a fit to the Hamiltonian are presented in Sec. \ref{sec:hyp-a}. We first analyze the hyperfine structure in the a$^3\Pi, v=0$ state, because the long radiative lifetime of this state allows for high spectral resolution. Moreover, the hyperfine structure in the a$^3\Pi_1$ state is expected to resemble the one in the A$^1\Pi$ state. As shown below, the detailed knowledge of the hyperfine structure in the a$^3\Pi, v=0$ state is essential to resolve the hyperfine structure in the X$^1\Sigma^+, v=0$ state. This measurement is presented in Sec. \ref{sec:hyp-X}. Section \ref{sec:hyp-A} presents UV excitation spectra from which we infer the hyperfine structure and radiative lifetime of the A$^1\Pi$ state. In Sec. \ref{sec:losses} we discuss the potential loss channels from the main optical cycling transition. We measure the intensities of the A$^1\Pi,v=0 \rightarrow \textrm{X}^1 \Sigma^+,v''=1$ and the spin-forbidden A$^1\Pi,v=0 \rightarrow \textrm{a}^3\Pi,v'$ transitions relative to the A$^1\Pi,v=0 \rightarrow \textrm{X}^1\Sigma^+,v''=0$ transition to quantify these loss channels. In Sec. \ref{sec:dipole}, the measurements of the electric dipole moments in the X$^1\Sigma^+$, the a$^3\Pi$ and the A$^1\Pi$ state are presented. The paper closes with Sec. \ref{sec:summary} which summarizes the most important results of this study and discusses the prospects for laser cooling and electric field manipulation of AlF. Throughout the manuscript experimental data is shown as solid black curves or black open circles, simulated spectra are shown as solid blue curves and a fit to the experimental data as a solid red curve.

\section{\label{sec:history}Early characterization of AlF}

Over 80 years ago, George Rochester observed the first band-spectrum of AlF \cite{Rochester1939}. He observed a system of five strong absorption bands in the 220-235 nm region by heating aluminum fluoride (AlF$_3$) in a graphite furnace to temperatures up to 1973 K. Rochester did not resolve any rotational structure but concluded from the appearance of the bands that these must be vibrational sequences between two electronic states with nearly equal vibrational and rotational constants. He attributed these spectra to the A$^1\Pi \leftarrow \textrm{X}^1\Sigma ^+$ band of AlF, based on a comparison to previously measured spectra of AlCl and AlBr. 

The A$^1\Pi \rightarrow \textrm{X}^1\Sigma^+$ band was first observed in emission by Rowlinson and Barrow in 1953. They used a mixture of Al and AlF$_3$ in a hollow-cathode discharge \cite{Rowlinson1953}. From a detailed analysis of the separations of the band-heads, they obtained a first estimate of the rotational constant in the X$^1\Sigma^+, v=0$ state of $B= 0.54$~cm$^{-1}$. The relative magnitude of rotational constants could be determined more accurately, and its value in the A$^1\Pi, v=0$ state was concluded to be only about 1$\%$ larger than in the electronic ground state. In the mid 1950's, high resolution emission spectra of AlF were recorded by Naud\'{e} and Hugo throughout the 220-870 nm range \cite{Naude1953a,Naude1953b,Naude1954}. This work improved the vibrational constants, but could not reveal the rotational structure of the A$^1 \Pi \rightarrow \textrm{X}^1\Sigma^+$ band. They reported on several new band systems of AlF, spanning from the blue to the near infrared. These bands had the A$^1\Pi$ state as the common lower state. One band, centered around 725 nm, appeared particularly strong and could be rotationally resolved. It was assigned to the C$^1\Sigma^+ \rightarrow \textrm{A}^1\Pi$ band, and a first value for the rotational constant in the A$^1\Pi, v=0$ state \cite{Naude1953b} could be determined.

Rowlinson and Barrow also reported the first observation of, what they then called, the b$^3\Sigma ^+ \rightarrow \textrm{a}^3\Pi$ band of AlF in the 345-372 nm spectral region \cite{Rowlinson1953}. In 1974 a detailed study of the electronic spectrum of gaseous AlF revised the labelling of this excited $^3\Sigma^+$ state \cite{Barrow1974}. From then on it is referred to as the c$^3\Sigma^+ \rightarrow \textrm{a}^3\Pi$ band. A first estimate of the energy of the triplet states relative to the singlet states was inferred from spectral perturbations that were attributed to the interactions of vibrational levels in the singlet and triplet manifolds \cite{Barrow1974}. This estimate was confirmed when the a$^3\Pi \rightarrow \textrm{X}^1\Sigma^+$ intercombination emission was observed by Rosenwaks et al. in 1976 \cite{Rosenwaks1976}. They recorded low resolution spectra from flames produced by reactions of Al with SF$_6$, NF$_3$ and F$_2$. Shortly after this, the first rotationally resolved absorption spectrum of the spin-forbidden a$^3\Pi \leftarrow \textrm{X}^1\Sigma^+$ band in AlF was recorded \cite{Kopp1976}. 

The thermodynamic properties of gaseous aluminum monofluoride were also studied from early on. The heat of formation of AlF was derived \cite{Gross1948} and a first value for its dissociation limit of about 6.87 eV was deduced \cite{Gross1954}. The dissociation limit was also derived from spectroscopic data on the A$^1\Pi$ state \cite{Barrow1956} to 7.24 eV, significantly higher than the thermodynamic value. More than four decades later, the latter value was revised by Bernath's group to 6.9 eV \cite{Zhang1997}, using improved spectroscopic data on the X$^1\Sigma^+$ state.

\begin{figure}
	\centering
	\includegraphics[width=\linewidth]{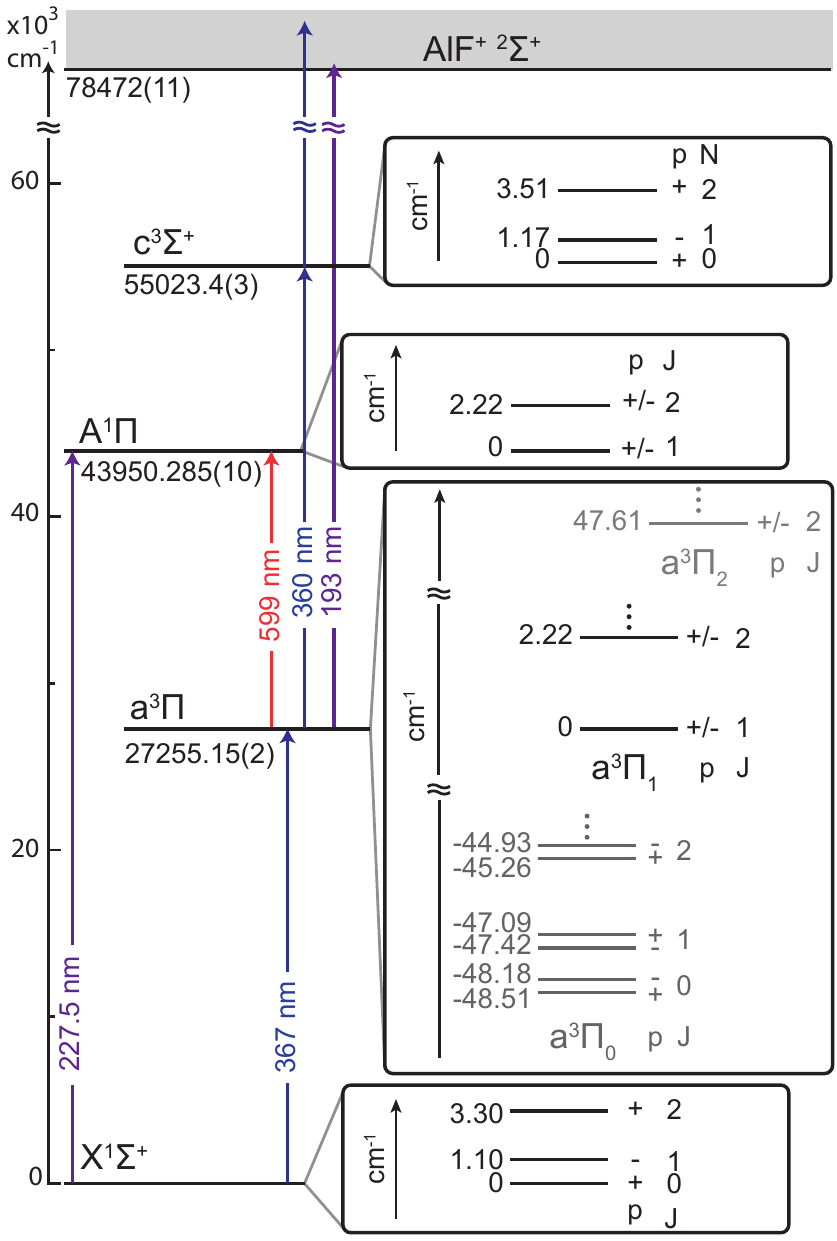}
	\caption{Energy level diagram of the relevant electronic states of AlF using accurate spectroscopic data. The optical transitions used for excitation and ionization are indicated. The lowest rotational levels in the electronic states are shown on an expanded scale. The $\Lambda$-doubling can only be seen in the a$^3\Pi_0$ manifold.}
	\label{fig:Electronic-Energy-Level-Scheme}
\end{figure}

Lide reported the first microwave spectrum of AlF in 1963 \cite{Lide1963}. He heated a mixture of powdered Al and AlF$_3$ up to 900-975 K inside a quartz tube that contained a microwave waveguide and recorded the spectrum in absorption via Stark modulation spectroscopy. He observed a triplet structure for the $J=0 \rightarrow 1$ transition in the $v=0$ and $v=1$ levels of the X$^1\Sigma^+$ electronic ground-state, characteristic of a molecule containing a single quadrupolar nucleus of spin 5/2. He found a quadrupole coupling constant of $eq_0Q = -37.6 \pm 1.0$ MHz and determined the electric dipole moment in the $v=0$ state to be $1.53 \pm 0.10$ Debye  \cite{Lide1963,Lide1965}. A more extensive set of millimeter and sub-millimeter measurements, including rotational levels up to $J=14$ and vibrational levels up to $v=4$, confirmed and further constrained the value for $eq_0Q$ and allowed for a more precise determination of the vibrational and rotational constants \cite{Wyse1970,Hoeft1970}. In 1974, the Zeeman effect of the $J= 0 \rightarrow 1$ transition in the X$^1\Sigma^+, v=0$ state was measured in fields of up to 4.4 T, from which values for the $g$-factors were determined \cite{Honerjager1974}. The Doppler-limited linewidth in these measurements was about 200~kHz and no splitting due to the $^{19}$F nuclear spin could be detected.

Barrow et al. and Brown et al. obtained limited information on the hyperfine structure in the triplet states by analyzing the line shapes of twenty unblended rotational lines in the b$^3\Sigma^+ \rightarrow \textrm{a}^3\Pi$ and c$^3\Sigma^+ \rightarrow \textrm{a}^3\Pi$ bands \cite{Barrow1974,Brown1978}. These lines showed partly resolved sub-structure, i.e. they appeared as doublets or triplets. Assuming that the magnetic interaction of the nuclear spin of Al with the electrons dominates the hyperfine structure, they simulated the experimental spectra and determined values for the Fermi contact parameter $b_F\textrm{(Al)}$ for each of the three electronic states. We briefly revisit this work in Sec. \ref{sec:hyp-a}, which presents the hyperfine structure in the a$^3\Pi$ state in more detail.

From the available spectroscopic data, the energies of the low-lying rotational levels in the various electronic states are known to within 0.01 cm$^{-1}$. Figure~\ref{fig:Electronic-Energy-Level-Scheme} shows the electronic states of AlF that are relevant to the experiments presented here, together with the optical transitions between them.

\section{\label{sec:experiment}Experimental setup }

\begin{figure}[htb!]
	\centering
	\includegraphics[width=\linewidth]{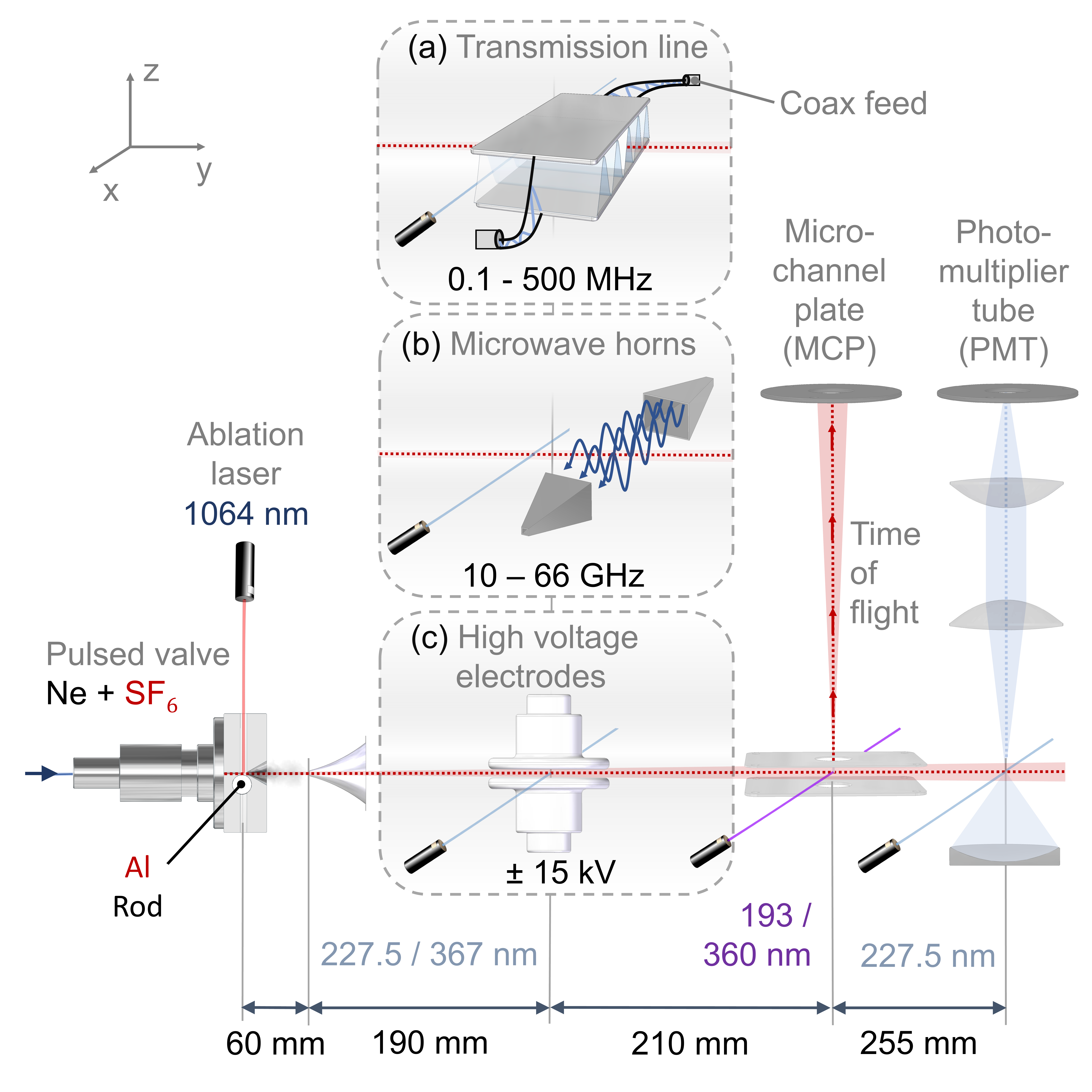}
	\caption{Schematic of the experimental setup. A supersonic molecular beam passes through a skimmer into the preparation chamber, where the internal quantum state of the AlF molecules is prepared via optical pumping. The optical pumping can be done under field-free conditions or in a high electric field (c). Alternatively, radio-frequency or microwave transitions can be driven using a transmission line (a) or microwave horn antennae (b). In the first detection chamber, resonant excitation and ionization, followed by mass-selective detection of the parent ions can be performed. In the second detection chamber, laser-induced fluorescence of the molecules is imaged onto a photomultiplier tube.}
	\label{fig:Experimental-Setup}
\end{figure}

The experimental setup is shown schematically in Fig.~\ref{fig:Experimental-Setup}. It consists of a molecular beam machine with a source chamber, a differentially pumped preparation chamber and two differentially pumped detection chambers. The AlF molecules are produced by laser ablation and cooled in a supersonic expansion. A solenoid valve (General Valve, Series 9) emits short pulses of carrier gas (Ne or Ar; 3 bar backing pressure) mixed with SF$_6$ (2\%) into a short reaction channel. A diode-pumped, pulsed Nd:YAG laser (1064 nm, 3 mJ/pulse, 10 ns pulse duration) with a near-Gaussian beam profile is focused to a spot size of approximately 0.5 mm to ablate an aluminum rod that rotates back and forth inside the reaction channel. The reaction of aluminum atoms with SF$_6$ produces vibrationally excited AlF molecules in the a$^3\Pi$ state \cite{Rosenwaks1976}. These highly excited molecules are efficiently quenched in the high pressure region of the reaction channel. The AlF molecules are translationally and internally cooled through collisions with the carrier gas, while expanding from the reaction channel into the source chamber through a conical nozzle. The AlF molecules thermalize to a rotational temperature of about 10 K and have a mean velocity of 780~m/s, when Ne is used as carrier gas). The experiment is operated at a repetition rate of 10 Hz.  About 60~mm downstream from the laser ablation point, i.e. at $y = 60$ mm, the molecular beam passes through a 5~mm diameter skimmer that separates the source chamber from the preparation chamber. 

In the preparation chamber, centered 190 mm behind the skimmer, the state of the molecules can be prepared by optical pumping or resonant excitation, driving transitions in either the A$^1\Pi \leftarrow \textrm{X}^1\Sigma^+$ or the a$^3\Pi \leftarrow \textrm{X}^1\Sigma^+$ band. For optical pumping on the A$^1\Pi \leftarrow \textrm{X}^1\Sigma^+$ band near 227.5 nm we use a CW Ti:Sa laser that is frequency-doubled twice using commercial, resonant bow-tie cavities, to produce up to 200 mW of UV radiation (MSquared). For resonant excitation to selected rotational levels in the metastable a$^3\Pi$ state, i.e. to drive transitions within the a$^3\Pi \leftarrow \textrm{X}^1\Sigma^+$ band near 367 nm, we either use a frequency-doubled pulsed dye laser ($\approx$ 10 mJ, 5 ns pulse duration, about 0.1 cm$^{-1}$ bandwidth), a frequency-doubled pulsed dye amplifier (PDA) seeded by a CW Ti:Sa laser ($\approx$ 10 mJ, 5 ns pulse duration, about 250 MHz linewidth) or a frequency-doubled CW Ti:Sa laser (up to \SI{1}{\watt}, sub-MHz linewidth). Both, the PDL and PDA are pumped by an injection seeded Nd:YAG laser (350 mJ, 5 ns). The molecules can be excited under field-free conditions or in static electric fields up to 150 kV/cm (Fig.~\ref{fig:Experimental-Setup} (c)); during operation, the pressure is maintained below 10$^{-6}$ mbar in this region.

Shortly after optical pumping or resonant excitation with (near) UV radiation, the molecules can be exposed to radio-frequency radiation to drive electric dipole allowed transitions between $\Lambda$-doublet or rotational levels. The rf and microwave radiation is generated by a synthesizer (Rohde \& Schwarz; SMR20) that is phase-locked to a stable GPS frequency reference (Quartzlock). The output of the synthesizer is connected to a fast, high-isolation switch and coupled to free space. For frequencies in the range of 0.1 to 500 MHz a TEM (transverse electromagnetic) transmission line is used (Fig.~\ref{fig:Experimental-Setup} (a)). The transmission line consists of two square copper plates with a side length of 57 mm that are separated along $z$ by 7.5 mm. The TEM mode propagates along $x$, orthogonal to the direction of the molecular beam. To produce radiation near 10 GHz the output of the synthesizer is coupled to free space through a horn antenna (Fig.~\ref{fig:Experimental-Setup} (b)). To produce radiation near 66 GHz, the output of the synthesizer is connected to a frequency quadrupler (Militech, AMC-15-RFH00), a precision attenuator and coupled to free space through a standard pyramidal gain horn (Fig.~\ref{fig:Experimental-Setup} (b)). The quadrupler delivers microwave powers of up to 10 dBm. The interaction time of the molecules with the radiation is set by the switch to values ranging from 40 to 80 \si{\micro\second}. In the preparation chamber, the ambient magnetic field is compensated by three pairs of coils that are close to Helmholtz configuration. They reduce the magnetic field in the interaction region to \SI{3}{\micro\tesla}. 

For state preparation in well-defined electric fields, two high voltage electrodes are installed in the preparation chamber to form a parallel plate capacitor. The electrodes are polished, stainless steel disks with rounded edges, a diameter of 55~mm and a thickness of 5 mm. The distance between the accurately spaced plates can be set to $2.020 \pm 0.010$~mm, $3.020 \pm 0.010$~mm or $4.005 \pm 0.010$~mm. The disks are mounted such that a high voltage of opposite polarity can be applied to each of them, thus creating a well-defined static electric field along the $z$-axis. Two power supplies (Spellman SL1200) are used to produce voltages of up to $\pm 15.5$~kV. The voltage is measured with a calibrated high voltage probe (CPS 250-M-01) to an accuracy of better than $5 \times 10^{-4}$. The molecular beam and the excitation laser beam cross between the circular high voltage electrodes, within 3~mm from their center. A finite element simulation of the plate geometry (COMSOL) shows that electric field variations due to finite size effects are suppressed to below 10$^{-5}$. The electric field can, therefore, be determined with a fractional accuracy that is mainly determined by the uncertainty in the distance between the plates.

In the first detection chamber, centered 210 mm downstream from the center of the preparation chamber, molecules in the metastable a$^3\Pi$ state can be single-photon ionized using 193 nm light from an ArF excimer laser ($\approx$ 2 mJ, 10 ns pulse duration). The molecules in the a$^3\Pi$ state can also be state-selectively ionized via resonant excitation on the c$^3\Sigma^+ \leftarrow \textrm{a}^3\Pi$ transition, followed by single-photon ionization from the c-state, i.e. via a (1+1)-resonance enhanced multi-photon ionization (REMPI) process, using a frequency-doubled pulsed dye laser near 360 nm. A linear time-of-flight (ToF) setup extracts the ions perpendicular to the molecular beam and accelerates them to a microchannel plate (MCP) detector. The signal from the MCP detector is amplified and read into a computer on a fast (5~ns/channel) digitizer card. The ToF electrodes are grounded during ionization and switched to high voltage to extract the ions. This ensures field-free excitation and ionization in the detection region. The mass-resolution $m/\Delta m$ of the ToF mass spectrometer is about 100, sufficient to unambiguously identify the AlF$^+$ parent cations at mass 46~amu. A grounded metal plate with a vertical slit is attached to the entrance of the ToF setup. The width of the slit can be adjusted between 5 and 15 mm. This shields the ionization region from stray electric fields and can reduce the residual Doppler broadening of the transitions in the a$^3\Pi \leftarrow \textrm{X}^1\Sigma^+$ band to about 30~MHz. The field-free REMPI ionization is parity selective and permits detecting the transitions between $\Lambda$-doublet levels or rotational levels in the a$^3\Pi$ state against zero background. This optical-radio-frequency/microwave-optical triple resonance ionization detection scheme has large similarities to schemes that have been used in the past to characterize the metastable a$^3\Pi$ state of CO, for instance \cite{Freund1965}. 

In the second detection chamber, at $y = 715$ mm from the source, the AlF molecules can be detected via laser-induced fluorescence (LIF) on the A$^1\Pi, v=0 - \textrm{X}^1\Sigma^+, v''=0$ band. The UV fluorescence is imaged onto a photomultiplier tube, and the photo-electron pulses are either counted (in case of low fluorescence intensity) or the photo-current is amplified and measured. 

\section{\label{sec:theory}The Molecular Hamiltonian}

The Hamiltonian to describe the energy level structure of a diatomic molecule has been considered in many articles \cite{Frosch1952,Zare1973,Brown1979a,Brown1979b,Brazier1986} and text books \cite{Gordy1984,Carrington2003}. It can be written in a general form as
\begin{equation}
{H}=H_{\textrm{ev}}+H_{\textrm{rot}}+H_{\textrm{fs}}+H_{\textrm{hfs}},
\end{equation}
where $H_{\textrm{ev}}$ contains the terms describing the electronic and vibrational part, $H_{\textrm{rot}}$ the rotational, $H_{\textrm{fs}}$ the fine-structure and $H_{\textrm{hfs}}$ the hyperfine structure. The electronic and vibrational term of the Hamiltonian determines the vibrational energy levels $E_\textrm{v}$ for a given electronic state and can be approximated by 
\begin{equation}
E_{\textrm{v}}=T_e+\hbar\omega_e(v+1/2)-\hbar\omega_e x_e(v+1/2)^2+\ldots, 
\end{equation}
where $T_e$ is the electronic term energy, i.e. the minimum of the potential energy, $\omega_e$ the vibrational energy, with its first order correction term $\omega_e x_e$. The rotational term can be written as
\begin{equation}
H_{\textrm{rot}}=A_v\left(\mathbf{L}\cdot\mathbf{S}\right)+B_v\,(\mathbf{J}-\mathbf{L}-\mathbf{S})^2-D_v\,(\mathbf{J}-\mathbf{L}-\mathbf{S})^4,
\end{equation}
where $A_v$ is the electron spin-orbit coupling constant, $B_v=B_e+\alpha_e(v+1/2)$ the rotational constant with $B_e$ the rotational constant at equilibrium, $\alpha_e$ the anharmonicity correction and $D_v$ the centrifugal distortion constant, $\mathbf{J}$ the angular momentum of the molecule, $\bf{L}$ the total orbital angular momentum of electron motion and $\mathbf{S}$ the total spin of the electrons. The spin-orbit interaction splits electronic states with non-zero values of $L$ and $S$ into $\Omega$ manifolds, where $\Omega=\Lambda+\Sigma$, with $\Lambda$ the projection of $\mathbf{L}$ and $\Sigma$ the projection of $\mathbf{S}$ along the internuclear axis. The fine-structure part of the Hamiltonian can be expressed as \cite{Zare1973,Brown1979a}
\begin{equation}
H_{\textrm{fs}}=\gamma\left(\mathbf{N}\cdot\mathbf{S}\right)+2\lambda\left(S_\textrm{Z}^{\,2} -\,\frac{1}{3} \mathbf{S}^{\,2}\right)+H_{\Lambda},
\end{equation}
where the first term describes the interaction of the electron spin $\bf{S}$ with the angular momentum $\bf{N}=\bf{J}-\bf{S}$, i.e. with the angular momentum of pure spatial rotation of the diatomic molecule. Note that for the fine and hyperfine structure we omit the subscript $v$ for the interaction parameters. The second term describes the electron spin-spin interaction, resulting from the spin-orbit interaction between a given and other electronic states. It appears as a second order perturbation to the spin-orbit interaction. Both of these terms do affect the $\Omega$ sub-levels but they do not affect the $\Lambda$-splitting of the energy levels. The term $\,H_{\Lambda}$ can be written as 
\begin{equation}
\begin{split}
H_{\Lambda}=&\frac{1}{2}\{-o (\Lambda_+^2 S_-^2+\Lambda_-^2 S_+^2 )+p(\Lambda_+^2 S_- N_-+\Lambda_-^2 S_+ N_+ )\\
&-q(\Lambda_+^2 N_-^2 + \Lambda_-^2 N_+^2 )\},
\end{split}
\end{equation}
and contains the terms describing the second order perturbation with respect to the spin-orbit interaction that do cause the $\Lambda$-splitting \cite{Brown1979b}. The $+$ and $-$ subscripts designate the appropriate axial components of the corresponding vector.

The hyperfine structure part of the Hamiltonian, $H_{\textrm{hfs}}$, describes the interaction of the nuclear spins with the electronic and rotational degrees of freedom. The strongest hyperfine interactions are magnetic interactions between nuclear spins and the electrons' degrees of freedom, as considered in detail by Frosch and Foley \cite{Frosch1952}. For each nucleus, the magnetic interaction with the electrons can be described using four parameters $a$, $b$, $c$ and $d$. For the hyperfine part of the Hamiltonian we follow the notation used by Brown and co-workers, in which the explicit expressions for each of the terms is given \cite{Brown1978}. We also use the Fermi contact parameter $b_F = b + (1/3) c$ and $c$, instead of $b$ and $c$. The interaction strength of the electric quadrupole moment $Q$ of the Al-nucleus with the electric field gradient at the nucleus is determined by two parameters $eq_0Q$ and $eq_2Q$ \cite{Brown1978}, where $q_0$ is the electric field gradient in the direction of the internuclear axis and $q_2$ the field gradient in the perpendicular direction. 

Both, the interaction between the nuclear magnetic moments and the interaction between the pure rotational angular momentum of the molecule with the nuclear magnetic moments are a factor $m_p/m_e$ smaller than the electronic contribution to the hyperfine structure, where $m_{p}$ and $m_{e}$ are the mass of the proton and electron, respectively. The magnitude of the hyperfine interaction involving the electron magneton is in the 200 -- 2000 MHz range whereas the nuclear interactions are  in the 10 -- 100 kHz range. The Hamiltonian for the nuclear magnetic hyperfine interactions is similar to the one for the electron hyperfine interactions, but with the electrons' degrees of freedom replaced by $\bf{N}$. It is convention to use the parameters $C_I$ and $C'_I$ for each of the nuclei to describe the nuclear spin-rotation interaction between $\Lambda'=\Lambda$ and $\Lambda'=-\Lambda$ basis wave functions, respectively, and use the parameter $D_1$ to describe the spin-spin interaction between the nuclei \cite{Townes1975, Klaus1997}.

In view of the complexity of the set of basis wave functions the numerical calculations are based on the tensor representation of the terms in the Hamiltonian. This facilitates the spectroscopic analysis and numerical calculations for many polyatomic molecules \cite{Klaus1997,Champion1992,Veldhoven2004} because it is not necessary to derive the explicit formulae for matrix elements. It also simplifies the calculation of the spectra in external electric fields. 

\section{\label{sec:aToX}The $a^3\Pi, v'=0 \leftarrow X^1\Sigma^+, v''=0$ band}

\begin{figure}[b]
	\centering
	\includegraphics[width=\linewidth]{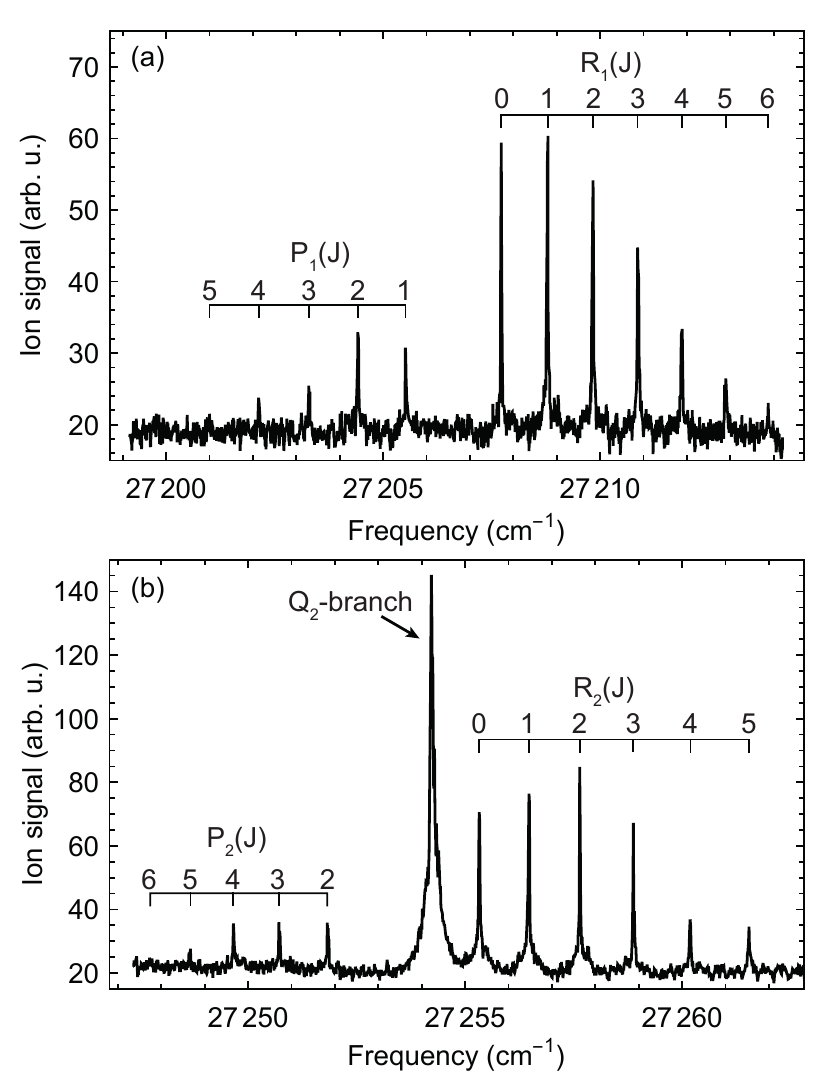}
	\caption{Excitation spectrum of the a$^3\Pi_0, v'=0 \leftarrow \textrm{X}^1\Sigma^+, v''=0$ (a) and a$^3\Pi_1, v'=0 \leftarrow \textrm{X}^1\Sigma^+, v''=0$ band (b) of AlF. The molecules are excited with a pulsed dye laser, followed by ionization with an ArF excimer laser. The signal of the parent ions is shown as a function of the excitation laser frequency. The pulse energy of the excitation laser used to record spectrum (a) is ten times higher than the energy used to record spectrum (b).}
	\label{fig:Overview-Spectra-a-X}
\end{figure}

The overall rotational structure of the spin-forbidden a$^3\Pi, v'=0 \leftarrow \textrm{X}^1\Sigma^+, v''=0$ band of AlF is investigated using resonant excitation with a pulsed dye laser around 367~nm. The beam of the dye laser is spatially overlapped in the ionization detection region with the beam from an ArF excimer laser, whose pulse is delayed relative to the dye laser by about 50~ns. Both laser beams have a diameter of approximately 3 mm. The current best value for the ionization potential (IP) of AlF is $78,472 \pm 11$ cm$^{-1}$ ($9.729 \pm 0.001$ eV), obtained by extrapolating the energies of a series of Rydberg states \cite{Dearden1991}. Single-photon ionization from the a$^3\Pi, v=0$ state with a 6.40 - 6.42~eV photon from the ArF laser creates AlF$^+$ cations with an internal energy of maximally 0.05 - 0.07~eV. The cations are therefore produced only in the $v=0$ level of their $^2\Sigma^+$ electronic ground state \cite{Dearden1991}. Fig.~\ref{fig:Overview-Spectra-a-X} shows the low-resolution excitation spectra obtained by detecting the signal of the parent ions in the ToF setup. 

Figure~\ref{fig:Overview-Spectra-a-X} (b) shows the spectrum of the spin-forbidden a$^3\Pi_1, v'=0 \leftarrow  \textrm{X}^1\Sigma^+, v''=0$ band. The spin-orbit coupling admixes a small amount of A$^1\Pi$ character to the wave function of the a$^3\Pi_1$ state, and the transition becomes weakly allowed, similar to the Cameron band of CO. The spin-orbit coupling mixes levels with the same $\Omega$ quantum number, and the resulting spectrum is equivalent to that of a $^1\Pi$ $\leftarrow$ $^1\Sigma^+$ transition. The observed relative intensities of the rotational lines are consistent with a rotational temperature of 10~K. 
Figure~\ref{fig:Overview-Spectra-a-X} (a) shows the spectrum of the spin-forbidden a$^3\Pi_0, v'=0 \leftarrow \textrm{X}^1\Sigma^+, v''=0$ band. This band is about ten times weaker than the a$^3\Pi_1, v'=0 \leftarrow \textrm{X}^1\Sigma^+, v''=0$ band shown in Fig.~\ref{fig:Overview-Spectra-a-X} (b), and is shifted to lower frequency by about 50~cm$^{-1}$. It is missing a Q-branch and thus resembles a $^1\Sigma^+$ $\leftarrow$  $^1\Sigma^+$ band. This means that the transition becomes weakly allowed due to the mixing of the a$^3\Pi$ state with $^1\Sigma^+$ states. In contrast, for metastable CO, the a$^3\Pi_0 \leftarrow \textrm{X}^1\Sigma^+$ band mainly gets its intensity via the $\Omega=1$ character in the wave functions of the rotational levels in the $\Omega=0$ manifold, due to deviations from a pure Hund's case (a) description. This effect is much less pronounced in AlF, because the $A/B$ ratio in the a$^3\Pi$ state is about four times larger for AlF than for CO.

Using the excitation and detection scheme described above, we could not find the a$^3\Pi_2, v'=0 \leftarrow \textrm{X}^1\Sigma^+, v''=0$ band because it is expected to be almost three orders of magnitude weaker than the a$^3\Pi_1, v'=0 \leftarrow \textrm{X}^1\Sigma^+, v''=0$ band. The detection of this weak band is further complicated by the occurrence of the a$^3\Pi_1, v'=2 \leftarrow \textrm{X}^1\Sigma^+, v''=2$ hot band in the same spectral region. 

When the molecules are ionized with the ArF excimer laser, all rotational levels in the a$^3\Pi$ state are detected with the same efficiency. Using two independent ArF excimer lasers, we determine the absolute ionization cross-section from the a$^3\Pi, v=0$ state at 193~nm in a pump-probe experiment to be ($3.6 \pm 0.5)\times 10^{-17}$ cm$^2$, or $36 \pm 5$ Mbarn which is almost identical to the photoionization cross-section of the Al atom in this wavelength region \cite{Kohl1973}. The excitation spectra show a constant background signal due to the ionization of metastable AlF molecules. These are produced in the source and are still present when the molecular beam reaches the detector. These triplet molecules mainly reside in the a$^3\Pi_0$ manifold due to cooling in the expansion and due to the longer radiative lifetime of the rotational levels in this manifold. The $J=1$ level in the a$^3\Pi_1, v=0$ state has the shortest lifetime. Nevertheless, molecules in this level live long enough to travel the 210~mm from the preparation chamber to the first detection chamber without significant loss. This indicates a radiative lifetime of at least 200 \si{\micro\second}. The lifetime of the corresponding rotational level in CO has been accurately determined in Stark deceleration and electrostatic trapping experiments as $2.63 \pm 0.03$~ms \cite{Gilijamse2007}. Early calculations on the radiative lifetime of this level in CO, using perturbation theory, give a value of 2.93 ms \cite{James1971}. We apply the same model to AlF, assuming diagonal Franck-Condon factors, and find a value for the radiative lifetime of about 1 ms for the $J=1$ level in the a$^3\Pi_1, v=0$ state.

\begin{figure}
	\centering
	\includegraphics[width=\linewidth]{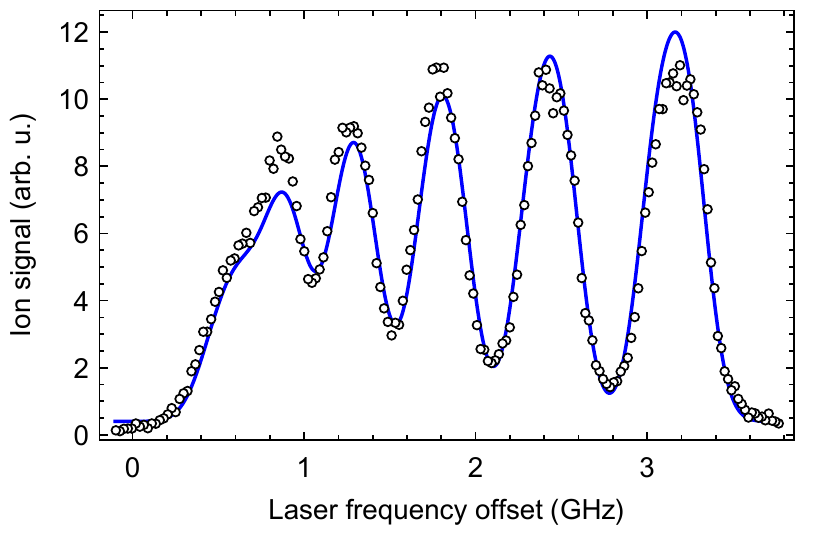}
	\caption{Excitation spectrum of the R$_3$(3) transition of the a$^3\Pi_2, v'=0 \leftarrow \textrm{X}^1\Sigma^+, v''=0$ band near 27304 cm$^{-1}$, using excitation with the PDA and (1+1)-REMPI detection via the c$^3\Sigma^+, v=0 \leftarrow \textrm{a}^3\Pi_1, v=0$ band around 360~nm. The simulated spectrum (solid curve) is generated using the parameters presented in Sec. \ref{sec:hyp-a} and \ref{sec:hyp-X}}.
	\label{fig:PDA-R3}
\end{figure}

To excite the AlF molecules on weak rotational lines of the a$^3\Pi, v'=0 \leftarrow \textrm{X}^1\Sigma^+, v''=0$ band, we use the frequency doubled PDA described in Sec. \ref{sec:experiment}. For state-selective ionization and to reduce the background ion signal we use the frequency doubled pulsed dye laser to ionize the AlF molecules from the a$^3\Pi$ state via the c$^3\Sigma^+, v=0$ state. With this excitation and ionization detection scheme, we can observe even the weak low-$J$ transitions to the $\Omega=2$ manifold. In Fig.~\ref{fig:PDA-R3} the spectrum of the R$_3$(3) transition, reaching the $J=4$, $e$-level in the a$^3\Pi_2$ manifold, is shown. The partly resolved hyperfine structure in this spectrum spans over almost 3 GHz and is well reproduced by the simulations
({\it vide infra}).

\begin{figure}
	\centering
	\includegraphics[width=\linewidth]{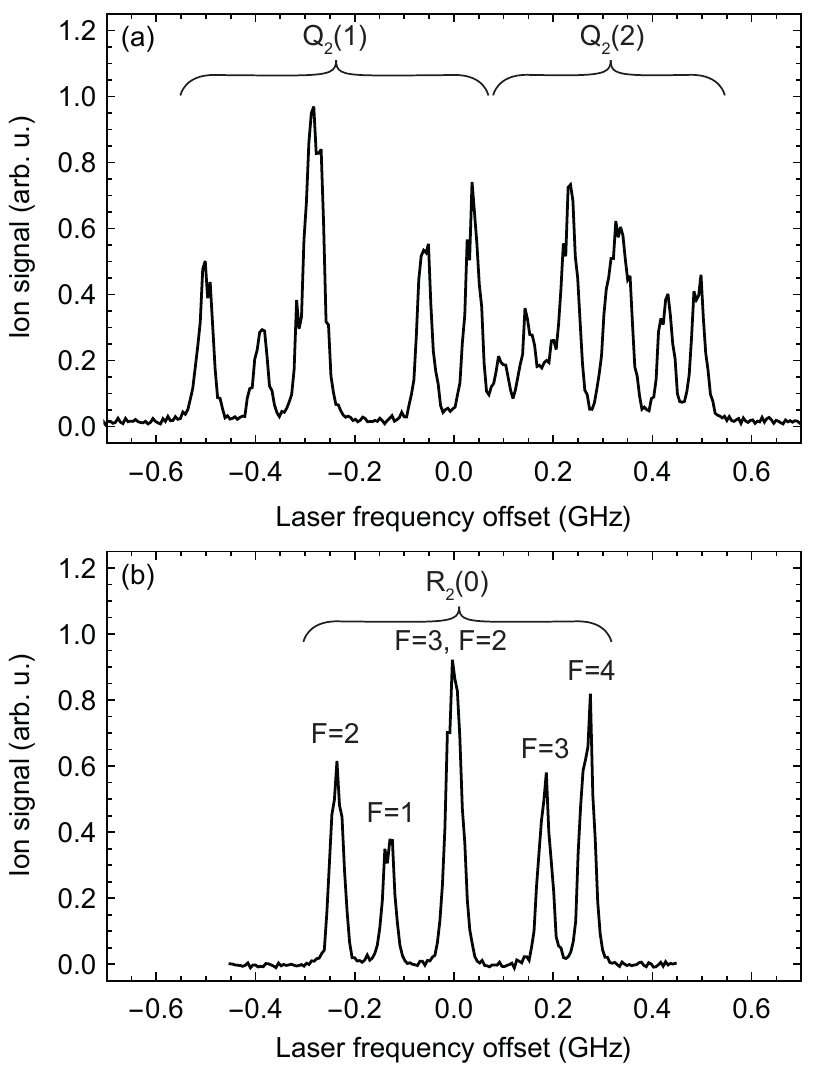}
	\caption{(a) High-resolution excitation spectrum of the Q$_2$(1) and Q$_2$(2) transitions of the a$^3\Pi_1, v'=0 \leftarrow \textrm{X}^1 \Sigma^+, v''=0$ band, using CW laser excitation and (1+1)-REMPI detection via the Q-branch of the c$^3\Sigma^+, v=0 \leftarrow \textrm{a}^3\Pi_1, v=0$ band around 360~nm. (b) High-resolution excitation spectrum of the R$_2$(0) line of the a$^3\Pi_1, v'=0 \leftarrow \textrm{X}^1\Sigma^+, v''=0$ band. The assignment of the $F$ quantum numbers in the $J=1$ level of the metastable state is based on the measurements presented in Sec. \ref{sec:hyp-a}.}
	\label{fig:High-Resolution-Q2-Branch}
\end{figure}

The AlF molecules can also be excited on the spin-forbidden transition using a frequency-doubled CW Ti:Sa laser. This laser is used to record the hyperfine resolved excitation spectra of the two lowest rotational transitions in the Q$_2$-branch as well as on the R$_2$(0) line, shown in Fig.~\ref{fig:High-Resolution-Q2-Branch}. We can saturate the rotational transitions of the a$^3\Pi_1, v'=0 \leftarrow \textrm{X}^1\Sigma^+, v''=0$ band with about 300~mW of CW radiation at 367~nm and a laser beam diameter of 2.0 mm. With about \SI{1}{\watt} of CW radiation we can also detect molecules that have been excited on the weak P$_1$(1) line.

\section{\label{sec:hyp-a}Hyperfine measurements in the $a^3\Pi, v=0$ state}

To determine the hyperfine structure in the metastable state with higher accuracy, we measure rf transitions between $\Lambda$-doublet components. In the preparation chamber, we prepare the AlF molecules in the $J=1$ level of the a$^3\Pi_1, v=0$ state via excitation with the PDA on the R$_2$(0) line. The molecules are now distributed across the negative parity components of the $J=1$ level, i.e. in the $J=1$, $e$-levels \cite{Brown1975}. We then pulse the rf radiation on for a short period of 40 - 80~\si{\micro\second} at the time when the molecular pulse is inside the rf transmission line. If the rf is tuned to a resonance frequency, the population is transferred to a level with positive parity. Provided the electric fields inside the molecular beam machine are kept sufficiently low, parity remains well-defined, and parity-selective detection can be used to probe the transferred population. In the ionization detection chamber, the AlF molecules are resonantly excited from the metastable a$^3\Pi$ state to the c$^3\Sigma^+, v=0$ state and are subsequently ionized by the same laser. By choosing a rotational level in the c$^3\Sigma^+$ state with an odd rotational quantum number $N$, i.e. with a negative parity, only molecules in the level with positive parity in the a$^3\Pi$ state are detected. In this particular case, we ionize the molecules via the $N=3$ level to record the radio-frequency spectra background-free.

\begin{figure}
	\centering
	\includegraphics[width=\linewidth]{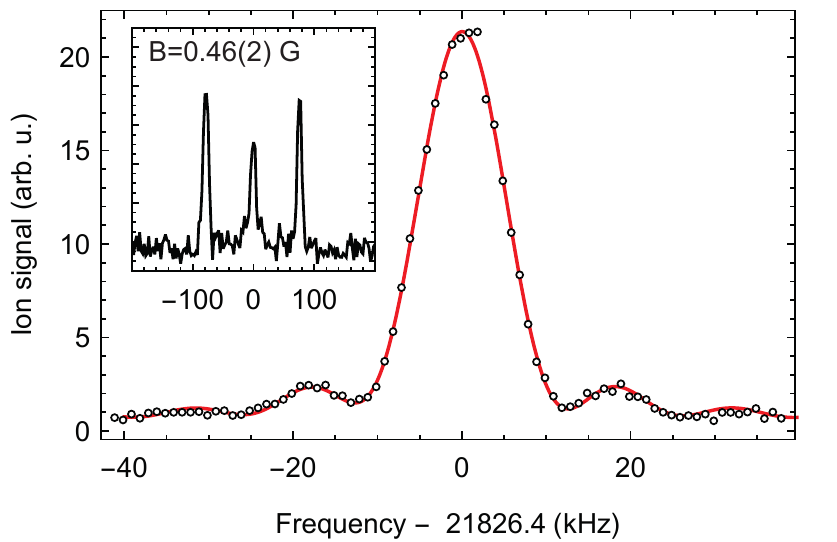}
	\caption{Observed $F=4,+ \leftarrow F=4,-$ transition in the $J=1$ level of the a$^3\Pi_1, v'=0$ state. The line shows a fit to the data using the Rabi line shape, with the background, amplitude and central frequency as free parameters. The inset shows the Zeeman splitting of the line when the compensation coils along $x$ and $z$ are turned off.}
	\label{fig:Sinc-Lineshape}
\end{figure}

In the $J=1$ level, there are twelve hyperfine components in total with 26 possible, electric-dipole allowed transitions between them. We determine the transition frequencies by recording the ion signal originating from the positive parity components as a function of the rf frequency. The strengths of these transitions vary by over four orders of magnitude, and the rf power coupled into the transmission line is adjusted accordingly. Transitions between levels with small differential $g_F$-factors have the expected Rabi line shape, and a fit to the data determines the line center with an accuracy of well below 1 kHz. This is shown in Fig.~\ref{fig:Sinc-Lineshape} for the transition around 21.8264~MHz. The residual magnetic field in the interaction region broadens the line, which can cause a systematic frequency shift. We model this effect, place an upper bound to the frequency shift and add a magnetic field related uncertainty to the error budget. Transitions with a large differential $g_F$-factor appear considerably broader, and with a Gaussian lineshape. In this case, a systematic uncertainty associated with the unknown lineshape is added to the statistical uncertainty obtained from the fit. For some of the weaker transitions, i.e. when high rf powers are needed, ac Stark shifts of several kHz are observed. To find the transition frequency of these lines, we determine the line center at different rf powers and extrapolate to zero power. The transition frequencies with their statistical and systematic uncertainties are given in the Appendix. 

\begin{figure}
	\centering
	\includegraphics[width=\linewidth]{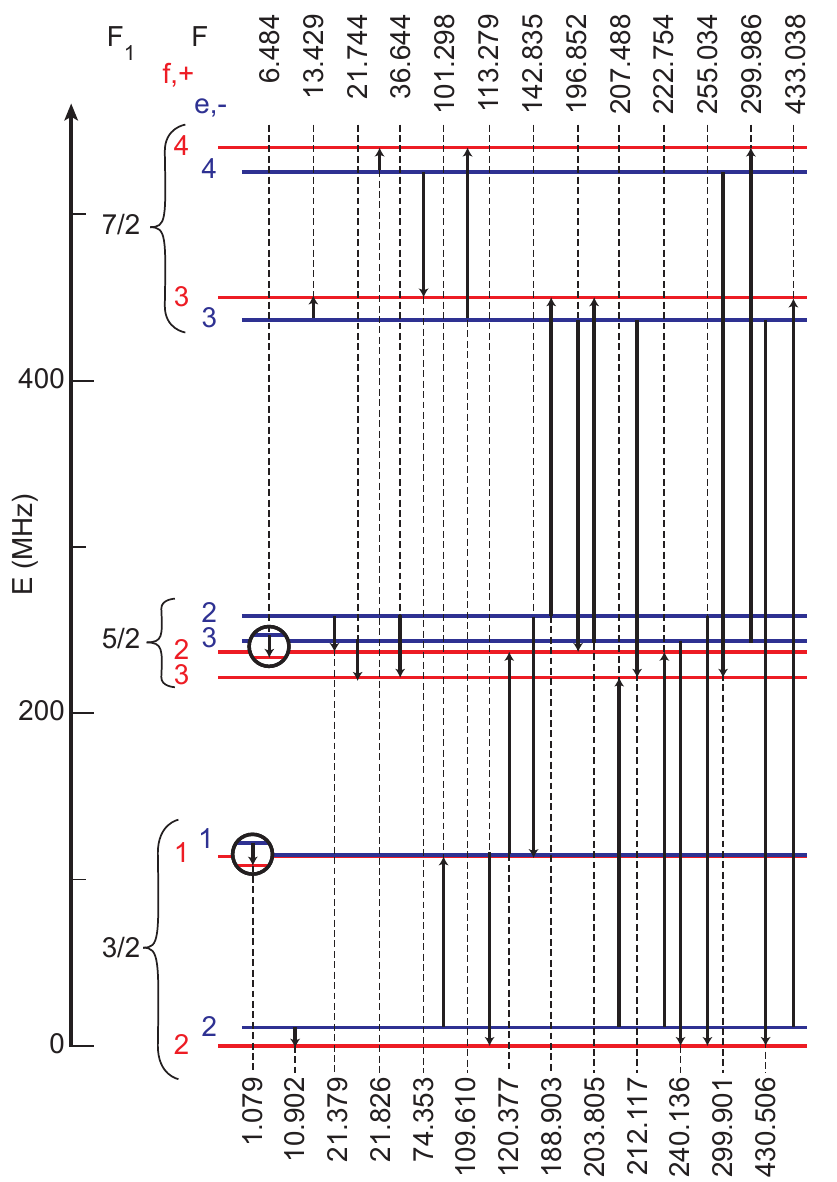}
	\caption{Hyperfine energy level diagram of the $J=1$ level in the a$^3\Pi_1, v=0$ state. All 26 allowed transitions are shown with their transition frequencies in MHz. To aid the identification of the levels, the intermediate quantum number $F_1$ is given as well.}
	\label{fig:J=1-Level-Scheme}
\end{figure}

All 26 transitions are observed and we uniquely assign the $F$ quantum numbers using angular momentum selection rules. We verify this assignment by applying a static magnetic field in the preparation chamber along $x$ and/or $y$ to measure the Zeeman splitting of some of these transitions. The inset of Fig.~\ref{fig:Sinc-Lineshape} shows the Zeeman splitting of the $F=4,+\leftarrow F=4,-$ transition when the compensation coils along $x$ and $z$ are turned off. The splitting in three relatively narrow $\Delta M_F=0$, $\pm$1 components indicates that this transition connects two $F$ levels with almost identical $g_F$-factors. The ambient magnetic field has a strength of approximately 46 $\pm$ 2 \si{\micro\tesla} with an inclination of 67 degrees between the lines of magnetic flux and the ($x$, $y$) plane. The $\Delta M_F=\pm 1$ lines are split by +/- 77 kHz which gives a magnetic $g_F$-factor of 0.115 $\pm$ 0.005, consistent with the theoretically predicted value of 0.113 (see Appendix). The experimentally determined hyperfine energy level diagram of the $J=1$ level, together with all the observed transitions, is shown in Fig.~\ref{fig:J=1-Level-Scheme}. 

The level diagram shows six pairs of opposite parity $F$-levels. To identify the hyperfine components we introduce the intermediate quantum number $F_1$ that labels the vectorial coupling of the rotational angular momentum $\mathbf{J}$ with the spin of the aluminum nucleus $\mathbf{I}_\textrm{Al}$, i.e. $\mathbf{F}_1 = \mathbf{I}_\textrm{Al} + \bf{J}$. The spin of the fluorine nucleus, $\mathbf{I}_\textrm{F}$, is then vectorially added to $\mathbf{F}_1$ to get the final vector $\mathbf{F}$, i.e. $\mathbf{F} = \mathbf{F}_1 + \mathbf{I}_\textrm{F}$. The three sets of four $F$-levels labeled by different $F_1$ quantum numbers in Fig.~\ref{fig:J=1-Level-Scheme} reflect the expected triplet structure arising from the nuclear spin of $^{27}\textrm{Al}$. The hyperfine splitting due to the nuclear spin of $^{19}\textrm{F}$ is of the same order of magnitude. 

\begin{figure}
	\centering
	\includegraphics[width=\linewidth]{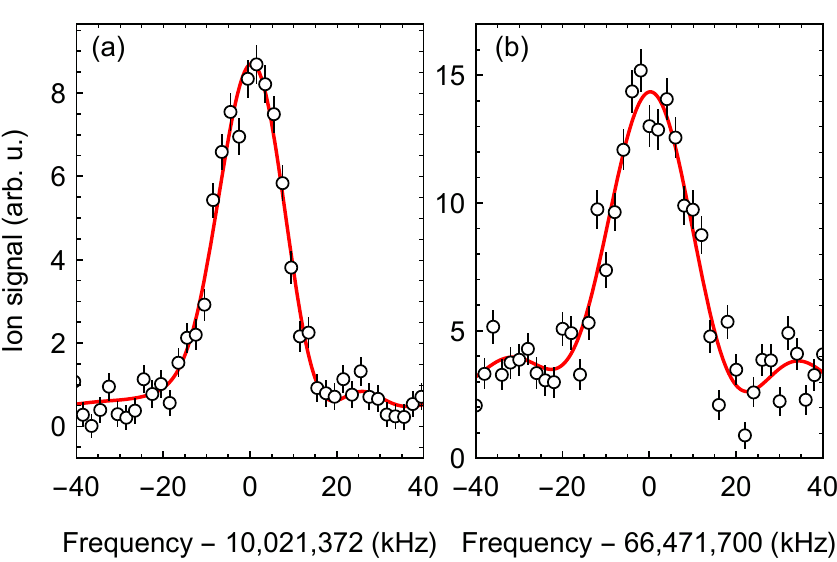}
	\caption{Observed line shapes for the $F=3, + \leftarrow F=2, -$ $\Lambda$-doublet transition in the a$^3\Pi_0, v=0, J=1$ level (a) and for the $F=3,+ \leftarrow F=3, -$ component of the $J=2\leftarrow J=1$ rotational transition in the a$^3\Pi_1, v=0$ state (b).}
	\label{fig:MW-Transitions-a}
\end{figure}

We fit the eigenvalues of the Hamiltonian to the measured hyperfine energy levels of the $J=1$ level to determine preliminary values for the spectroscopic parameters. We then use these parameters to predict the hyperfine levels, transition frequencies and transition dipole moments in the $J=2$ and $J=3$ level of the $\Omega=1$ manifold. The experimentally determined transition frequencies are typically within several MHz of the predicted frequencies. We then use the measured values to improve the parameters in the fit or if necessary we include higher order terms. This procedure is followed for the $J=7$ level in the $\Omega=1$ manifold, as well as for the $J=4$ and $J=7$ levels in the $\Omega=2$ manifold. All relevant hyperfine parameters can be determined this way. To accurately determine the $\Lambda$-doubling parameters, we measure the hyperfine resolved $\Lambda$-doubling transitions in the $J=0$, 1 and 2 levels of the $\Omega=0$ manifold around 10~GHz. To accurately determine the rotational splitting in the a$^3\Pi, v=0$ state, we measure several hyperfine resolved lines of the $J=2 \leftarrow 1$ transitions in both the $\Omega=1$ and $\Omega=0$ manifolds around 66~GHz. Selected hyperfine resolved transitions for the measurements around 10~GHz and 66~GHz are shown in Fig.~\ref{fig:MW-Transitions-a}. Compared to the rf transmission line, the electric field distribution of the microwave radiation in free space is not very well defined which affects the observed line shape. All the measured frequencies of the rf and microwave transitions, together with their experimental uncertainties, as well as the calculated frequencies with their assignments and calculated intensities are given in the Appendix.

\begin{figure}
	\centering
	\includegraphics[width=\linewidth]{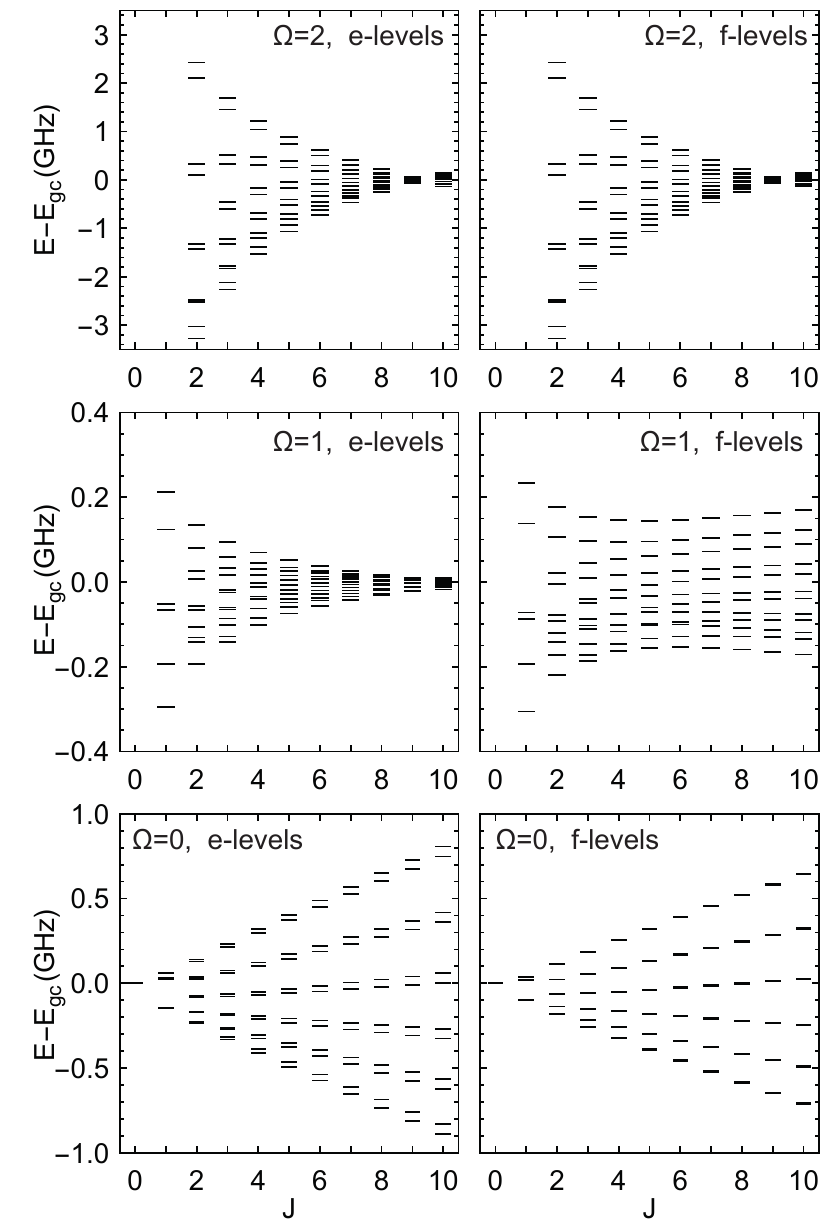}
	\caption{Hyperfine structure for rotational levels up to $J=10$ in the a$^3\Pi, v=0$ state, calculated using the parameters given in Table \ref{tab:a3Param}. The energies of the levels, $E$, are given relative to the energy of the gravity center, $E_{gc}$. Note that each $\Omega$-manifold has its own energy scale, which is the same for the $e$- and the $f$-levels.}
	\label{fig:Hyperfine-Overview}
\end{figure}

The parameter $A$ and its second order correction $\lambda$ can not be determined accurately from the rf and microwave measurements. We therefore measure the center frequencies of selected rotational transitions from the X$^1\Sigma^+, v''=0$ state to each of the $\Omega$-manifolds in the a$^3\Pi, v'=0$ state with a wavemeter (HighFinesse WS6-600). P and R lines reach $e$-levels and Q-lines reach $f$-levels (see Fig.~\ref{fig:Hyperfine-Overview}). In particular, we choose the P$_1(1)$, the R$_2(8)$ and the Q$_3(9)$ lines because they reach levels in the metastable state for which the total span of the hyperfine structure is smaller than 40~MHz. Therefore, they appear as single lines and the resolution is mainly limited by the bandwidth of the narrow-band PDA. The P$_1(1)$ transition appears 1,762,797 $\pm$ 100~MHz lower and the Q$_3(9)$ transition appears 1,098,984 $\pm$ 120~MHz higher in frequency than the R$_2(8)$ transition. The uncertainty in the relative frequencies is determined by tracking the wavelength of a temperature stabilized helium neon laser coupled to the same wavemeter. Using the known rotational energies in the electronic ground state, this gives a value of $A= 47.395 \pm 0.007$ cm$^{-1}$ and $\lambda= 0.0887\pm 0.0011$ cm$^{-1}$. We also fit the rotationally resolved absorption spectra for the a$^3\Pi, v'=0 \leftarrow \textrm{X}^1\Sigma^+ ,v''=0$ band measured by Kopp, Lindgren and Malmberg \cite{Kopp1976}, to the eigenvalues of our Hamiltonian for the a$^3\Pi$ state. However, we keep the values for the rotational constant and the $\Lambda$-doubling parameters fixed to the values determined from our fit to the microwave data. This way we obtain a value of $A=47.393 \pm 0.031$ cm$^{-1}$ and $\lambda = 0.086 \pm 0.020$ cm$^{-1}$, in agreement with the values derived from our own measurements. The term-value extracted from these Doppler-limited spectra is $E_0= 27253.04 \pm 0.01$ cm$^{-1}$.

Using the second order gradient technique, we determine the hyperfine structure parameters, the parameters describing the $\Lambda$-doubling, the spin-rotation interaction parameter $\gamma$ and the rotational constant with high accuracy, and the results are summarized in Table \ref{tab:a3Param}. For each of the parameters, the standard deviation (SD) as well as the product of the standard deviation with the square-root of the quality-factor Q is given; the latter is the better measure for the accuracy with which each parameter is determined by a fit whose parameters are correlated \cite{Watson1977}. The parameter for the spin-spin interaction between the nuclei is calculated from the known equilibrium distance between the nuclei and their magnetic moments to be $D_1$ = 0.0066~MHz. The centrifugal distortion parameter $D_0$ = 0.0314 MHz in the $v=0$ level is obtained by refitting the data presented by Kopp et al. \cite{Kopp1976} using the Hamiltonian presented above. For convenience, the vibronic energy $E_0$ is set to zero. In the final fit of the radio-frequency and microwave data, both $A$ and $\lambda$ are fixed at $A_0 = 47.395$ cm$^{-1}$ and $\lambda$ = 0.0887 cm$^{-1}$. The  centrifugal distortion term to the spin-orbit splitting is calculated from known spectroscopic constants to $A_D$ = 0.1 MHz \cite{Brazier1986,Veseth1970}.
\begin{table}
\caption{\label{tab:a3Param} Spectroscopic parameters of the a$^3\Pi, v=0$ state. Rotational constant $B_0$, fine-structure constant $\gamma$, $\Lambda$-doubling parameters and hyperfine structure parameters for the a$^3\Pi, v=0$ state obtained from the best fit to the experimental data together with their standard deviation (SD) and the product of SD and $\sqrt{Q}$ (all values in MHz). For the $\Lambda$-doubling parameter $o$ and for four hyperfine structure parameters, the centrifugal distortion term has been included. These terms are labelled with a superscript $(R)$ and are given directly underneath the parameter. The term named $eq_0Q_{LS}(\textrm{Al})$ describes the spin-orbit correction to $eq_0Q\textrm{(Al)}$. In the fit, the parameters $A_0 = 47.395$ cm$^{-1}$, $\lambda = 0.0887$ cm$^{-1}$ and $D_0 = 0.0314$ MHz are fixed to the best known experimental values. The centrifugal distortion term of the spin-orbit splitting, $A_D$, as well as the parameter for the  spin-spin interaction between the nuclei, $D_1$, are fixed to the calculated values of $A_D$ = 0.1 MHz and $D_1$ = 0.0066 MHz.}
\begin{ruledtabular}
\begin{tabular}{ld{2.4}d{2.4}d{2.4}}
	Parameter         & \multicolumn{1}{c}{Value (MHz)}   & \multicolumn{1}{c}{SD}   & \multicolumn{1}{c}{$\textrm{SD}\cdot\sqrt{\textrm{Q}}$} \\
    \hline	
	$B_0$                       &      16634.7458   &          0.0010   &         0.0024    \\
	$\gamma$                    &         -7.6089   &          0.0526   &         0.1369    \\
	$o$                         &       4968.3175   &          0.0510   &         2.9243    \\
	$o^{(R)}$                   &         -0.0061   &          0.0012   &         0.0097    \\
	$p$                         &        -24.3462   &          0.0510   &         4.7224    \\
	$q$                         &         -1.8176   &          0.0023   &         0.1798    \\
	$a\textrm{(Al)}$            &        199.1620   &          0.0353   &         0.4525    \\
	$b_F\textrm{(Al)}$          &       1247.9697   &          0.2783   &         9.2715    \\
	$b_F^{(R)}\textrm{(Al)}$      &          0.0222   &          0.0045   &         0.1638    \\
	$c\textrm{(Al)}$            &        -21.0093   &          0.4124   &        13.7410    \\
	$c^{(R)}\textrm{(Al)}$          &         -0.0568   &          0.0114   &         0.5826    \\
	$d\textrm{(Al)}$            &        121.9077   &          0.0155   &         0.0935    \\
	$eq_0Q\textrm{(Al)}$        &        -12.9921   &          0.0159   &         0.0286    \\
	$eq_0Q_{LS}\textrm{(Al)}$   &         -0.0392   &          0.0042   &         0.0097    \\
	$eq_2Q\textrm{(Al)}$        &         51.1137   &          0.0054   &         0.0062    \\
	$C_I\textrm{(Al)}$          &         -0.0569   &          0.0169   &         1.3288    \\
	$C'_I\textrm{(Al)})$        &         -0.0115   &          0.0007   &         0.0043    \\
	$a\textrm{(F)}$             &        207.1350   &          0.0109   &         0.0156    \\
	$a^{(R)}\textrm{(F)}$           &         -0.0688   &          0.0038   &         0.0338    \\
	$b_F\textrm{(F)}$             &        169.6289   &          0.1605   &         0.5853    \\
	$b_F^{(R)}\textrm{(F)}$           &         -0.0346   &          0.0019   &         0.0152    \\
	$c\textrm{(F)}$             &        122.8043   &          0.2596   &         1.3787    \\
	$d\textrm{(F)}$             &        119.2769   &          0.0868   &         0.4241    \\
	$C'_I\textrm{(F)}$          &          0.0346   &          0.0041   &         0.0202    \\
\end{tabular}
\end{ruledtabular}
\end{table}

To reach a standard deviation of the fit comparable with the accuracy of the radio-frequency and microwave measurements of a few kHz, we first include centrifugal distortion terms to each $\Lambda$-doubling and hyperfine parameter. We then analyze the correlations between the parameters and exclude them one by one, testing each time whether the standard deviation of the fit increases significantly. The five remaining centrifugal distortion terms, labelled with a superscript $(R)$, are listed in Table \ref{tab:a3Param}, directly underneath the corresponding parameters; if any of these is set to zero, the standard deviation of the fit increases significantly. We find that it is relatively straightforward to get a good fit to all the measured  hyperfine levels within a certain $\Omega$-manifold, but that some hyperfine parameters need a correction depending on the product of $\mathbf{L} \cdot \mathbf{S}$, i.e. depending on the $\Omega$-manifold that they are in. We first include spin-orbit correction terms to several hyperfine parameters, but find that only the term $(L_0\cdot S)\cdot D^{(1)}\cdot I^2\textrm{(Al)}$ is significant. It describes the spin-orbit correction to $eq_0Q\textrm{(Al)}$. We name this parameter $eq_0Q_{LS}\textrm{(Al)}$ and list it underneath $eq_0Q\textrm{(Al)}$ in Table \ref{tab:a3Param}. This term has, to the best of our knowledge, not been described in the literature before. It means that the quadrupole component of the electrostatic potential ($q_0$) is slightly different in the different $\Omega$-manifolds due to different admixing of other electronic wave functions via the $\mathbf{L} \cdot \mathbf{S}$ interaction.

Figure~\ref{fig:Hyperfine-Overview} summarizes the results of this section. It shows the calculated hyperfine structure for each $\Omega$-level and for rotational levels up to $J=10$ using the parameters given in Table \ref{tab:a3Param}. The energies of the hyperfine levels for a given $J$ are shown relative to the energy of the gravity center, i.e. relative to the energy of the $J$-level in absence of any hyperfine interaction. The accuracy of the predicted energy levels is better than 10~kHz, much less than the thickness of the lines that indicate the energy levels in the figure. The size of the hyperfine splitting varies significantly across the different $\Omega$-manifolds. In particular, the different behaviour of the $e$- and the $f$-levels in the $\Omega=1$ manifold is remarkable.

Brown and co-workers observed a sub-structure in Doppler-limited rotational lines of optical spectra that involved the a$^3\Pi, v=0$ state \cite{Brown1978}. They neglected the interaction of the fluorine nucleus with the electron, and determined a value for the Fermi contact parameter of the Al nucleus in the a$^3\Pi, v=0$ state of $b_F(\textrm{Al}) = 0.046 \pm 0.005$ cm$^{-1}$. This value agrees with the Fermi contact parameter of $b_F(\textrm{Al}) = 0.041674\pm 0.000004$ cm$^{-1}$ found here. As can be seen in Fig.~\ref{fig:Hyperfine-Overview}, the energy levels in the $\Omega=0$ manifold follow a regular pattern with relatively large hyperfine splittings, which are mainly determined by $b_F(\textrm{Al})$. The "unblended lines" that Brown and co-workers analyzed originated exclusively from high-$J$ ($\geq 15$) rotational levels in either the $\Omega=0$ or the $\Omega=2$ manifold. This allowed them to extract a first value for $b_F(\textrm{Al})$ from the optical spectra, despite their limited resolution \cite{Brown1978}.

\section{\label{sec:hyp-X}Hyperfine measurements in the $X^1\Sigma^+$ state}

\begin{figure}
	\centering
	\includegraphics[width=\linewidth]{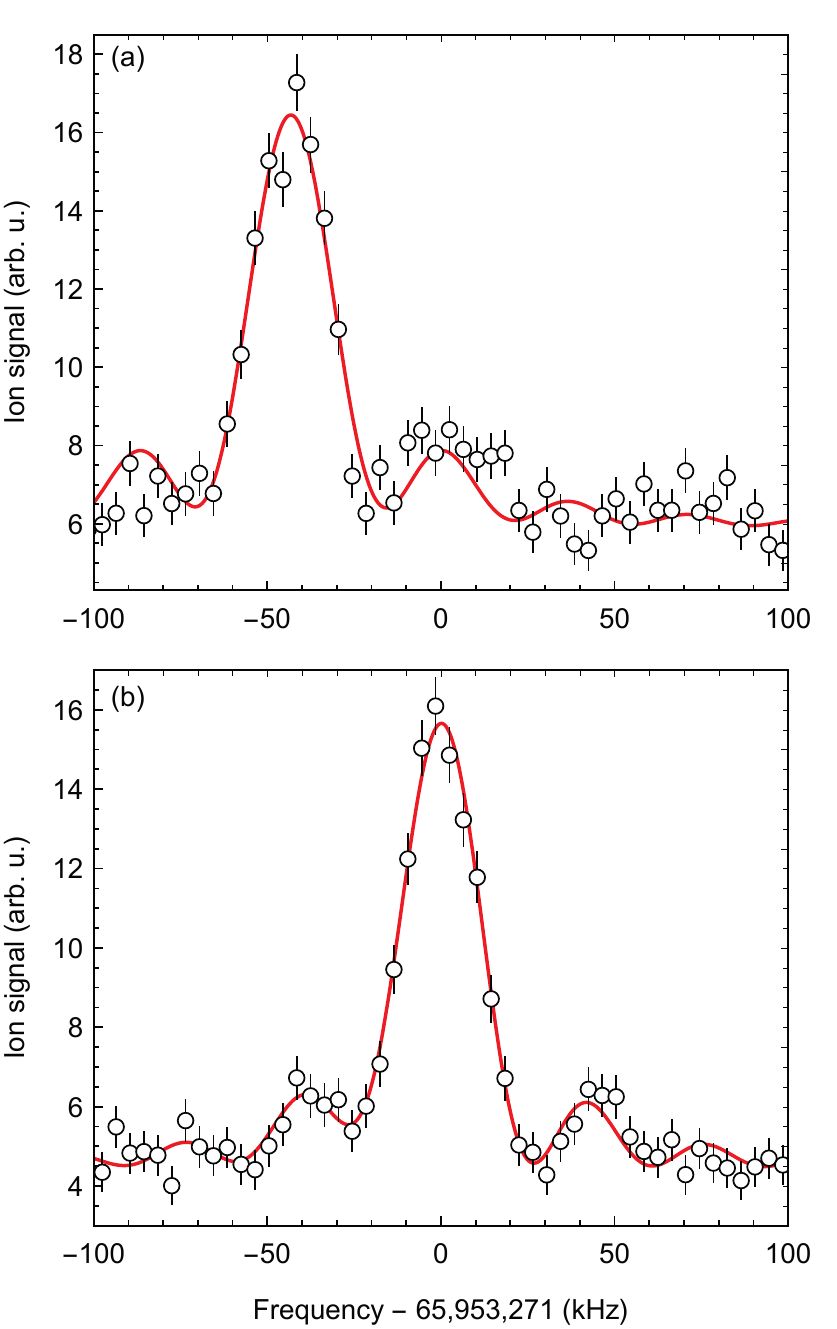}
	\caption{Observed hyperfine resolved components of the $J''=\rightarrow 1$ rotational transition in the X$^1\Sigma^+, v''=0$ state. Shown are the $F''=4, F''_1=9/2 \rightarrow F''=3, F''_1=7/2$ component, detected via the $F'=3, F'_1 = 7/2$ hyperfine component of the a$^3\Pi_1, v'=0, J'=1$ level (a) and the $F''=5, F''_1=9/2 \rightarrow F''=4, F''_1=7/2$ component, detected via the $F'=4, F'_1 = 7/2$ hyperfine component of the a$^3\Pi_1, v'=0, J'=1$ level (b).}
	\label{fig:MW-Transitions-X}
\end{figure}

To measure the hyperfine structure in the X$^1\Sigma^+, v''=0$ state we drive and detect the $J'' = 2 \rightarrow 1$ transition near 66~GHz using the setup described in Sec. \ref{sec:experiment} and illustrated in Fig.~\ref{fig:Experimental-Setup} (b). To maximize the signal-to-noise ratio, we first deplete the population in the $J''=1$ level by optical pumping on the R(1) line of the A$^1\Pi, v=0\leftarrow \textrm{X}^1\Sigma^+, v''=0$ band. We intersect the molecular beam with a laser beam (30~mW, 3~mm diameter, CW) in the preparation chamber to empty the population in the $J''=1$ level. Subsequently, the molecules interact with microwave radiation for a duration of 60~\si{\micro\second}, to transfer population from the $J''=2$ to the $J''=1$ level. The molecules in the $J''=1$ level are detected in the first detection chamber by driving a specific hyperfine component of the Q$_2(1)$ line to the a$^3\Pi, v'=0$ state using a CW laser near 367 nm. From the a$^3\Pi, v'=0$ state the molecules are ionized via the c$^3\Sigma^+, v=0$ state and mass-selectively detected. This detection scheme enables the selective recording of the hyperfine components of the $J''= 2 \rightarrow 1$ transition even when these are spectrally overlapping. This is demonstrated by the measurements shown in Fig.~\ref{fig:MW-Transitions-X}. 

\begin{table}
\caption{\label{tab:xFreq} Observed frequencies f$_\textrm{o}$ of the hyperfine resolved components of the $J''=2 \rightarrow 1$ transition in the X$^1\Sigma^+, v''=0$ state together with the observed-minus-calculated values f$_\textrm{o}$-f$_\textrm{c}$, their calculated intensity (in Debye$^2$) and the $F$, $F_1$ assignment of the levels involved.}
\begin{ruledtabular}
\begin{tabular}{crd{2.4}ccccc}
    f$_\textrm{o}$ (MHz) & f$_\textrm{o}$-f$_\textrm{c}$   & \multicolumn{1}{c}{Int. (D$^2$)} & $F$  &  $F_1$ & $\rightarrow$ & \text{$F$}  & \text{$F_1$} \\
 \hline   
	$65,944.464 \pm 0.002$    & -0.002    &   1.30    &   2   &   5/2 & $\rightarrow$   & 1 &   3/2\\
	$65,944.510 \pm 0.002$    & -0.002    &   2.04    &   3   &   5/2 & $\rightarrow$   & 2 &   3/2\\
	$65,945.968 \pm 0.004$    &  0.003    &   2.24    &   3   &   7/2 & $\rightarrow$   & 3 &   7/2\\
	$65,945.972 \pm 0.002$    &  0.002    &   2.93    &   4   &   7/2 & $\rightarrow$   & 4 &   7/2\\
	$65,949.835 \pm 0.004$    & -0.002    &   2.94    &   2   &   3/2 & $\rightarrow$   & 2 &   3/2\\
	$65,953.229 \pm 0.001$    & -0.001    &  8.20     &   4   &   9/2 & $\rightarrow$   & 3 &   7/2\\
	$65,953.271 \pm 0.001$    & -0.001    & 10.30     &   5   &   9/2 & $\rightarrow$   & 4 &   7/2\\
	$65,953.854 \pm 0.002$    &  0.002    & 3.98      &   3   &   7/2 & $\rightarrow$   & 2 &   5/2\\
	$65,953.912 \pm 0.002$    &  0.002    & 5.41      &   4   &   7/2 & $\rightarrow$   & 3 &   5/2\\
	$65,961.061 \pm 0.004$    &  -0.001   & 1.30      &   2   &   3/2 & $\rightarrow$   & 3 &   5/2\\
\end{tabular}
\end{ruledtabular}
\end{table}

\begin{table}
\caption{\label{tab:xParam} Spectroscopic parameters determined from the best fit to the data, together with their standard deviation (SD) and the product of SD with $\sqrt{Q}$ (all values in MHz). The rotational centrifugal distortion parameter and the nuclear-nuclear spin-spin interaction term are fixed at $D_0=0.0312$ MHz and $D_1=0.0066$ MHz, respectively.}
\begin{ruledtabular}
\begin{tabular}{ld{5.4}d{2.4}d{2.4}}
	Parameter               &    \multicolumn{1}{c}{Value (MHz)}        & \multicolumn{1}{c}{SD} &\multicolumn{1}{c}{$\textrm{SD}\cdot\sqrt{Q}$}\\
	\hline
	$B_0                $   &      16488.3548       &   0.0003    &      0.0003  \\
	$eq_0Q\textrm{(Al)} $   &        -37.5260       &   0.0069    &      0.0074  \\
	$C_I(\textrm{Al})   $   &          0.0104       &   0.0004    &      0.0004  \\
	$C_I(\textrm{F})    $   &          0.0360       &   0.0016    &      0.0016  \\
	$D_1                $   &          0.0066       &             &              \\
\end{tabular}
\end{ruledtabular}
\end{table}

Only the terms $eq_0Q$, $C_I$(Al), $C_I$(F) and the nuclear-nuclear spin-spin interaction term $D_1$ contribute to the hyperfine structure in the X$^1\Sigma^+, v''=0$ state. The quadrupole interaction splits the $J''=1$ or $J''=2$ level into three or five $F''_1$ components, respectively. The parameter $C_I$(Al) describes a small perturbation to the main splitting. The $C_I$(F) term gives rise to a splitting of the $F''_1$ levels into the final $F''$ levels. Fig.~\ref{fig:MW-Transitions-X} (a) shows a resonance shifted by 42 kHz from the resonance shown in Fig.~\ref{fig:MW-Transitions-X} (b). This shift is the frequency difference of the splittings due to the interaction of the fluorine nuclear spin in the $J''=1$, $F''_1=7/2$ and $J''=2$, $F''_1=9/2$ levels. Table \ref{tab:xFreq} lists the ten observed hyperfine resolved transitions together with the difference between the observed and calculated transition frequencies, f$_\textrm{o}$-f$_\textrm{c}$, their calculated intensity (in Debye$^2$) and the $F$, $F_1$ assignment of the levels involved. We fit the data to the eigenvalues of the Hamiltonian to determine the spectroscopic parameters. In the fit we fix the rotational centrifugal distortion parameter at its best known value of $D_0$ = 0.0312~MHz \cite{Kopp1976}, and calculate the nuclear-nuclear spin-spin interaction parameter $D_1$ from the known distance between the nuclei and their magnetic moments as $D_1$ = 0.0066~MHz. The results of the fit are presented in Table \ref{tab:xParam}. The values obtained for $eq_0Q(\textrm{Al})$ and $C_I(\textrm{Al})$ agree with the values reported in the literature \cite{Honerjager1974}. There has been no previous value for $C_I(\textrm{F})$, which is found to be considerably larger than $C_I(\textrm{Al})$.

\begin{figure}
	\centering
	\includegraphics[width=\linewidth]{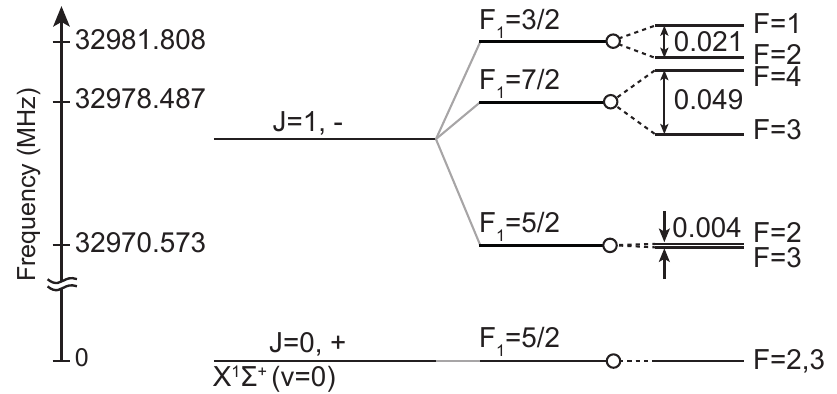}
	\caption{Hyperfine energy level structure for the $J=0$ and $J=1$ rotational levels in the X$^1\Sigma^+, v=0$ state.}
	\label{fig:Energy-Levels-X}
\end{figure}

The hyperfine energy level structure for the lowest two rotational levels in the X$^1\Sigma^+, v=0$ state is shown in Fig.~\ref{fig:Energy-Levels-X}. The $F=2$ and $F=3$ hyperfine components of the rotational ground-state level are nearly degenerate. There is a small splitting of less than 1 kHz due to second order interactions between levels with different $J$ but equal $F$ quantum numbers.

\section{\label{sec:hyp-A}Hyperfine measurements in the $A^1\Pi$ state}

\begin{figure}
	\centering
	\includegraphics[width=\linewidth]{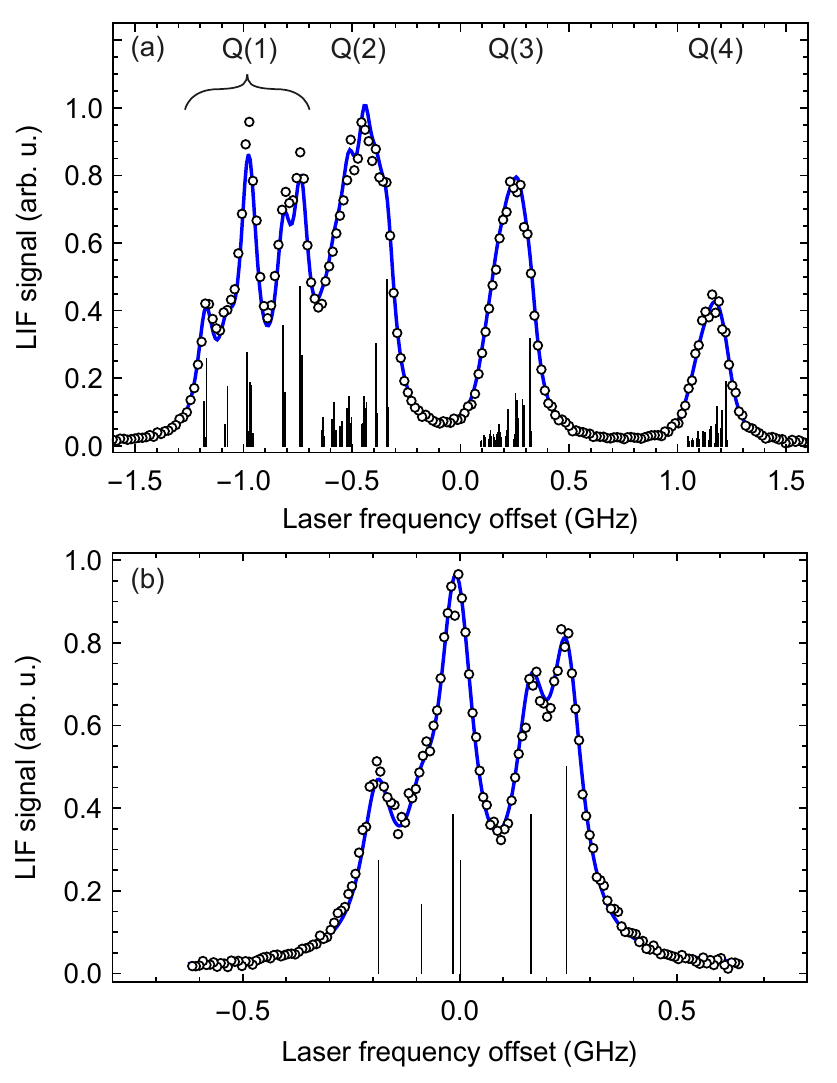}
	\caption{Observed laser-induced fluorescence excitation spectra of the Q(1), Q(2), Q(3) and Q(4) lines (a) and of the R(0) line (b) of the A$^1\Pi, v=0 \leftarrow \textrm{X}^1\Sigma^+, v''=0$ band of AlF, together with the simulated spectra (solid curves). Note that the frequency scale in panel (b) is a factor two smaller than in panel (a). The sticks underneath the spectrum show the position of the individual hyperfine components of the rotational lines and their intensities.}
	\label{fig:LIF-A-X}
\end{figure}

The accuracy with which we can measure the hyperfine structure in the A$^1\Pi$ state is intrinsically limited by its short radiative lifetime. The hyperfine structure can only be partly resolved for the lowest rotational levels in the A$^1\Pi$ state. This can be seen in the Q-branch spectrum shown in Fig.~\ref{fig:LIF-A-X}. The  Q(1) transition shows substructure, whereas all other Q-lines appear broadened. 

To record the LIF spectra with negligible Doppler broadening, we decrease the transverse velocity spread in the molecular beam. For this we reduce the speed of the molecular beam to about 600 m/s by using Ar as a carrier gas, we reduce the skimmer opening to 2 mm diameter, and we mount a 3 mm wide vertical slit in the molecular beam directly in front of the LIF detection zone. We measure the residual Doppler broadening in this setup by recording several hyperfine resolved rotational lines of the A$^2\Sigma^+, v'=0 \leftarrow \textrm{X}^2\Pi_{1/2}, v''=0$ band of NO. We choose NO because its optical transition occurs at 226 nm, very close to that of AlF, whereas its excited state lifetime is about a hundred times longer \cite{Piper1986}. We use a mixture of 5\% NO and 2\% SF$_6$ in Ar, and we also fire the ablation laser, to reproduce the experimental conditions of the AlF measurements. The NO spectral lines have a Gaussian lineshape with a full width at half maximum (FWHM) of 12~MHz. This is an upper limit for the Doppler contribution to the lineshape of the A$^1\Pi \leftarrow \textrm{X}^1\Sigma^+$ band of the heavier AlF molecule. 

The lowest four lines in the Q-branch as well as the isolated R(0) line of the A$^1\Pi, v=0 \leftarrow \textrm{X}^1\Sigma^+, v''=0$ band of AlF are shown in Fig.~\ref{fig:LIF-A-X}. Figure~\ref{fig:LIF-A-X} (a) appears as a low-resolution version of the spectrum shown in Fig.~\ref{fig:High-Resolution-Q2-Branch} (a); the spectrum of the R(0) line shown in Fig.~\ref{fig:LIF-A-X} (b) appears as a low-resolution version of the spectrum of the R$_2$(0) line shown in Fig.~\ref{fig:High-Resolution-Q2-Branch} (b). Based on this resemblance we can assign the $F_1$ and $F$ quantum numbers to the hyperfine components of the $J'=1$ level in the A$^1\Pi, v=0$ state.

To determine both $B_0$ and $q$ in the A$^1\Pi, v=0$ state, we measure the relative frequency of the six lowest rotational lines in the Q-branch and the frequency separation between the closely spaced Q(15), R(0) and Q(16) lines. From the fit to the data we determine the rotational constant to $B_0 = 0.55378 \pm 0.00001$~cm$^{-1}$ and the $\Lambda$-doubling parameter to $q = - 2.94 \pm 0.06$ MHz. The latter value is consistent with the value of $q = -3.0 \pm 0.3$ MHz, found by refitting the original data of Naud\'{e} and Hugo \cite{Naude1953b}.

The spectra presented in Fig.~\ref{fig:LIF-A-X} show the partly resolved hyperfine structure in the A$^1\Pi, v=0$ state. The observed lineshape is well approximated by a Voigt profile. The Gaussian contribution to the Voigt profile is determined from the NO spectra and fixed to 10 MHz (FWHM). We use four parameters to fit to the data: the two hyperfine parameters $a(\textrm{Al})$ and $a(\textrm{F})$, an overall frequency offset and the Lorentzian contribution to the Voigt profile. $B_0$ and $q$ are fixed to the values mentioned above. The simulated spectrum reproduces the data well with $a(\textrm{Al})=113 \pm 5$ MHz and $a(\textrm{F})=181 \pm 5$ MHz. The hyperfine parameters, together with the rotational and $\Lambda$-doubling parameters for the X$^1\Sigma^+, v=0$ state, are presented in Table \ref{tab:Aparam}. The $eq_0Q$ and $eq_2Q$ terms only play a minor role in the energy level structure in the A$^1\Pi$ state and are fixed at 0 MHz and 40 MHz, respectively. The standard deviation of the fit does not change significantly when these parameters are changed by about 10 MHz. The simulated spectra, assuming a rotational temperature of 6 K, are shown as solid blue curves in Fig.~\ref{fig:LIF-A-X} and fit the experimental data well. The Lorentzian contribution to the Voigt profile has a full width at half maximum of 84 $\pm$ 1~MHz, equivalent to a radiative lifetime of the A$^1\Pi, v=0$ state of $1.90 \pm 0.03$~ns. This value is in good agreement with the calculated value of 1.89~ns \cite{Langhoff1988}. 

\begin{table}
\caption{\label{tab:Aparam} Spectroscopic constants obtained from the best fit to the experimental data, together with their standard deviation (SD).}
\begin{ruledtabular}
\begin{tabular}{ld{2.4}d{2.4}}
    Parameter               &   \multicolumn{1}{c}{Value (MHz)}       &   \multicolumn{1}{c}{SD (MHz)} \\
\hline	
	$B_0$                   &   16601.9                 &   0.3     \\
	$q$                     &   -2.94                   &   0.06    \\
	$a(\textrm{Al})$        &   113                     &   5       \\
	$a(\textrm{F})$         &   181                     &   5       \\
\end{tabular}
\end{ruledtabular}
\end{table}

\begin{figure}[htb!]
	\centering
	\includegraphics[width=\linewidth]{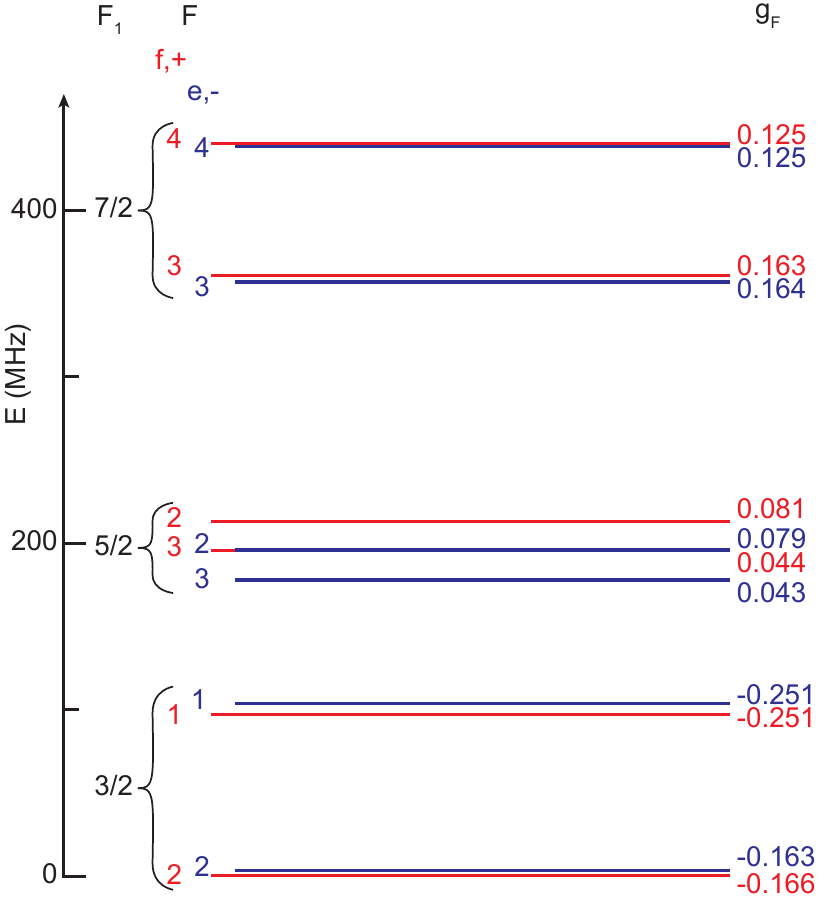}
	\caption{Hyperfine energy level structure for the J=1 rotational level in the A$^1\Pi, v=0$ state together with the calculated $g_F$ factors}
	\label{fig:A_HFS}
\end{figure}

The hyperfine energy level structure for the lowest rotational level in the A$^1\Pi, v=0$ state is shown in Fig.~\ref{fig:A_HFS}. The energies of the hyperfine levels are only known to within 10 MHz.

\section{\label{sec:losses} Optical cycling}
The number of photons an AlF molecule scatters on the Q-lines of the A$^1\Pi,v=0 \leftarrow \textrm{X}^1\Sigma^+,v''=0$ band is limited by branching to states that are not coupled by the excitation laser. In this section we discuss these loss channels in more detail.

\subsection{Rotational branching}
All Q-lines are rotationally closed, due to angular momentum selection rules. However, the magnetic hyperfine interaction of the Al nuclear spin mixes the lowest rotational levels in the A$^1\Pi, v=0$ state, i.e. it mixes states with the same total angular momentum $F$. This mixing causes a small leak from the main optical cycling transition to higher rotational levels in the ground electronic state. The calculated rotational branching ratios for the Q(1) line are listed in table \ref{tab:rotBr}.  
\begin{table}
\caption{\label{tab:rotBr} Calculated branching ratio $r_{JJ''}$ from a given $(F_1,F)$ component of the $J=1$ level in the A$^1\Pi,v=0$ state to the $J''=3$ ($r_{13}$) or to the $J''=1$ ($r_{11}$) level in the ground state.}
\begin{ruledtabular}
\begin{tabular}{cd{2.4}}
	$(F'_1,F')$ &  \multicolumn{1}{c}{$(r_{13}/r_{11})\times 10^6$}\\
	\hline
	$(3/2,2)$   &       0.15   \\
	$(3/2,1)$   &       0.43    \\
	$(5/2,3)$   &       3.0    \\
	$(5/2,2)$   &       0.99    \\
	$(7/2,3)$   &       1.31    \\
	$(7/2,4)$   &       3.42    \\
\end{tabular}
\end{ruledtabular}
\end{table}

\subsection{\label{sec:fc} Vibrational branching}
There are no selection rules that limit the branching to vibrational states in the ground electronic state. To determine the number of vibrational repump lasers that are necessary for laser slowing and cooling, we measure the Einstein coefficient $A_{01}$ of the A$^1\Pi,v=0 \rightarrow \textrm{X}^1\Sigma^+, v''=1$ (0-1) band relative to the Einstein $A_{00}$ coefficient of the 0-0 band. We do this by dispersing the laser-induced fluorescence of the molecules using a compact spectrometer. 
As an alternative method, we measure the ratio of Einstein $B^{ab}_{01}/B^{ab}_{11}$ coefficients by comparing the absorption strengths of rotational lines of the 0-1 band to the absorption strength of the same rotational lines of the 1-1 band. In this case we measure the ratio $A_{01}/A_{11}=\nu^3_{01}B^{ab}_{01}/(\nu^3_{11}B^{ab}_{11})$, with $\nu_{01}$ and $\nu_{11}$ the transition frequencies of the rotational lines of the 0-1 and 1-1 band, respectively. In a molecular beam, the population distribution across vibrational levels in the electronic ground state might change on the timescale of an experiment. Therefore, it is more precise to measure the ratio $A_{01}/A_{11}$ instead of $A_{01}/A_{00}$. The experimental results of the two methods are compared to numerical calculations. 

For both methods, it is useful to increase the molecular flux and the interaction time with the laser beams. Therefore, we use a cryogenic helium buffer gas molecular beam \cite{Maxwell2005, Hutzler2012} instead of the supersonic molecular beam described in Sec. \ref{fig:Experimental-Setup}. The cryogenic beam typically delivers more than two orders of magnitude more molecules per pulse to the detection region with a velocity that is four times lower. The design of the cryogenic beam source is similar to the one presented in the literature \cite{Truppe2018}: a pulsed Nd:YAG laser (Minilite II, 1064 nm, 10-20 mJ, 0.5 mm) ablates a solid aluminum target in the presence of a constant flow of 0.01 sccm room temperature SF$_6$ gas which is mixed with 1 sccm of cryogenic helium gas inside a buffer gas cell. By changing the source conditions we can increase the population in the $\textrm{X}^1\Sigma^+, v''=1$ state from about 4\% to up to 15\% of the overall population. This, however, also increases the mean velocity of the AlF molecules to over 300 m/s. 

\begin{figure}[htb!]
	\centering
	\includegraphics[width=\linewidth]{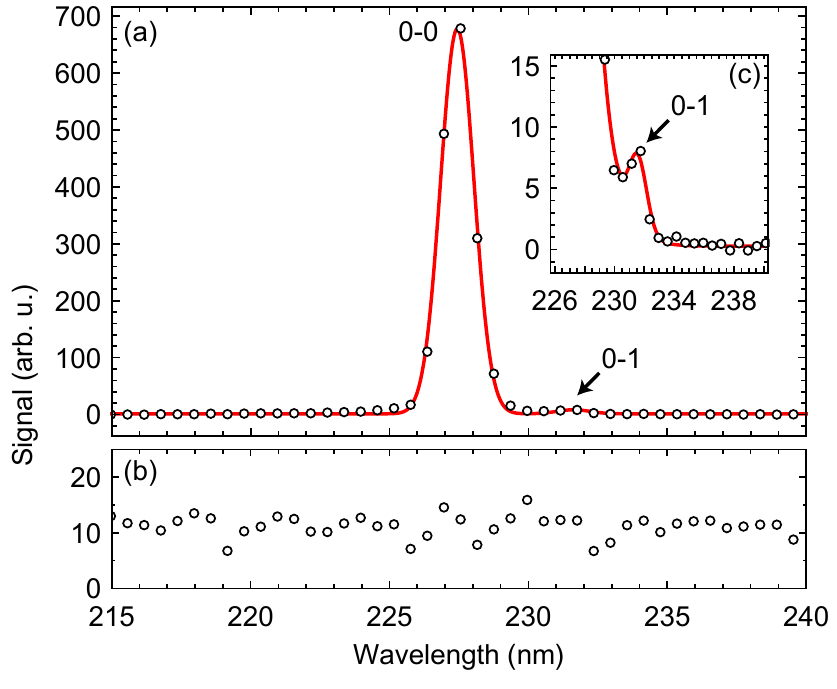}
	\caption{Dispersed fluorescence spectrum after excitation on the Q(1) line of the 0-0 band. The LIF is coupled into an optical fiber and dispersed using a grating spectrometer (a). The red curve shows a fit to the data (circles) using the measured spectral response of the spectrometer. The background spectrum (b) is recorded without the molecular beam but with the excitation laser present. The inset (c) shows the 0-1 emission band on an enlarged scale.}
	\label{fig:spec}
\end{figure}

For the first method, we tune a CW laser to the Q(1) line of the 0-0 band near 227.5 nm (1 mm diameter beam, 90 mW) and cross it with the molecular beam, about 320 mm from the source. The laser-induced fluorescence is collected, coupled into an optical fiber and dispersed using a compact Czerny-Turner spectrometer (Avantes AvaSpec-ULS2048-EVO, 600~\si{\micro\meter} fiber, 25 \si{\micro\meter} entrance slit) with a resolution of 1.5 nm. A typical dispersed fluorescence spectrum is shown in Fig.~\ref{fig:spec} (a). We average 4000 shots and subtract the background, measured without the molecular beam but with 90 mW of CW laser light present (Fig~\ref{fig:spec} (b)). To fit the data, we first determine the line shape of the spectrometer at 227.5 nm and 231.7 nm by coupling narrow-band laser light directly into the spectrometer. We then use the two spectral response functions to fit the molecular spectrum. The only fit parameters are the two emission amplitudes. The fit to the data is shown as the red curve in Fig.~\ref{fig:spec}. Figure \ref{fig:spec} (c) shows the emission of the molecules on the 0-1 band around 231.7 nm more clearly. The fit gives a ratio of emission amplitudes of $\nu_{01}A_{01}/(\nu_{00}A_{00})=\left(7\pm3\right)\times 10^{-3}$. We account for a slightly different detection efficiency of the spectrometer at 231.7 nm  and 227.5 nm. Light at 231.7 nm is detected with a $10\pm3\%$ higher efficiency than light at 227.5nm. To reduce the uncertainty, we repeat this measurement 14 times which gives a final result for the R(2) line of $A_{01}/A_{00}=\left(7.3\pm1\right)\times 10^{-3}$.

\begin{figure}[htb!]
	\centering
	\includegraphics[width=\linewidth]{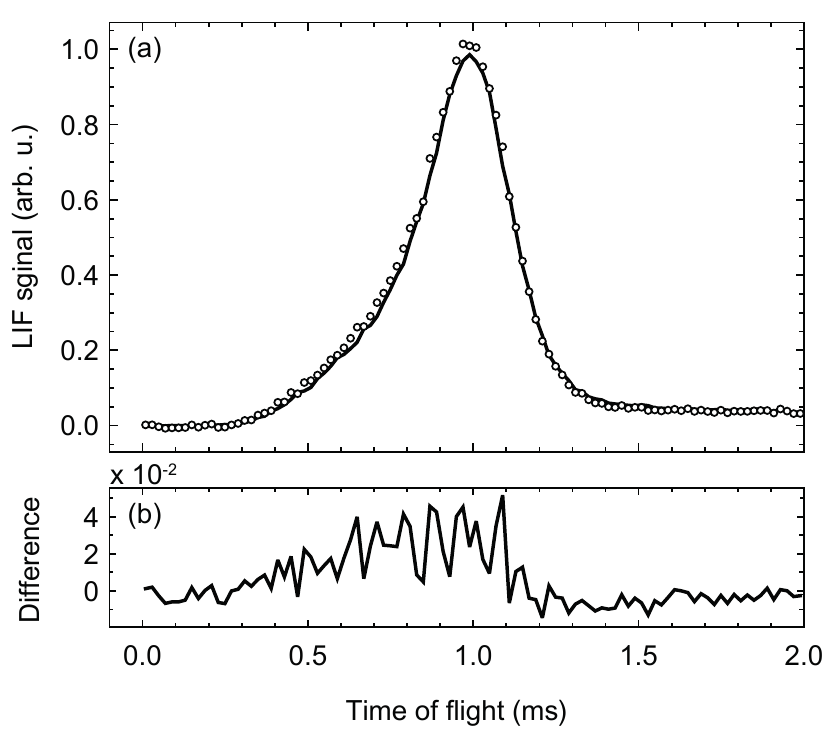}
	\caption{(a) Two ToF profiles. One laser is locked to the R$(2)$ line of the 1-1 band (white circles) and a second laser is locked to the R$(2)$ line of the 0-1 band (solid black curve) of the A$^1\Pi\leftarrow \textrm{X}^1\Sigma^+$ transition. The two laser beams are overlapped and have the same Gaussian intensity profile. The powers are set to 0.045 mW for the 1-1 band and 7.3 mW for the 0-1 band. The difference between the two ToFs is shown in (b)}
	\label{fig:fcLif}
\end{figure}

For the second method, we overlap two lasers and cross them with the molecular beam orthogonally. The frequency of one laser is locked to the center of the R$(2)$ line of the 0-1 band near 231.7 nm and the frequency of the second laser is locked to the R$(2)$ line of the 1-1 band near 227.5 nm. We choose these lines specifically because they appear as isolated single lines. The spatial mode of each laser is cleaned up using pinholes resulting in a Gaussian intensity distribution in the interaction region with a $1/e^2$ diameter of 0.46 mm. The LIF from each laser is imaged on a PMT, recorded on alternate shots to minimize the effect of source fluctuations and averaged over 50 shots. The two ToF profiles are shown in Fig.~\ref{fig:fcLif}. The circles show the ToF profile of the molecules excited on the 0-1 band with a laser power of 7.3 mW. The solid black curve results from excitation on the 1-1 band with a laser power of 0.045 mW. The power in each laser beam is adjusted such that the amplitudes of the two ToF profiles are nearly identical. The weak excitation is necessary to prevent optical pumping into dark rotational states. The difference of the two ToF profiles is shown in Fig.~ \ref{fig:fcLif} (b). The ratio of laser powers directly reflects the ratio of Einstein coefficients $B^{abs}_{01}/B^{abs}_{11}=6.2\times10^{-3}$, because $\nu_{00}A_{00}/(\nu_{11}A_{11})\approx 1$. We repeat this measurement for 16 different pairs of laser powers. The mean and standard error of these measurements gives the final result of $A_{01}/A_{11}=(5.72\pm0.08)\times10^{-3}$. We repeat the measurement using the R(1) rotational lines, giving $A_{01}/A_{11}=(5.59\pm0.02)\times10^{-3}$. The results are summarized in table \ref{tab:ratios} and compared to theoretical calculations. 

First, we derive RKR potentials by using Le Roy's program \cite{LeRoy2017} to fit both Morse and expanded Morse oscillator (EMO) functions to precise spectroscopic data \cite{Barrow1974, Yousefi2018}. These potentials are then used to calculate the Franck-Condon factors (f$_{AX}$) which are listed in Table \ref{tab:fcs} and compared to the Franck-Condon factors derived from \textit{ab initio} calculations \cite{Wells2011}. To predict the transition probability of the weak off-diagonal bands more precisely, we include the variation of the transition dipole moment with internuclear distance \cite{Garland1985}. We use the multi-reference-configuration-interaction (MRCI) method available in MOLPRO 2019 \cite{MOLPRO} to perform {\it ab initio} calculations of the relevant transition dipole moment \footnote{The MRCI calculations are fed with the natural orbitals from a Multi-Configuration Self-Consistent Field (MCSCF) calculation with a complete active space (CAS) consisting of nine orbitals with $A_1$ symmetry, three with $B_1$ symmetry and three with $B_2$ symmetry associated with the point group $C_{2v}$. The calculations employ the AV5Z \cite{BasisSet} basis set for each atom.}. The results of this calculation are listed in Table \ref{tab:ratios}.

\begin{table}
\caption{\label{tab:ratios} Calculated and measured ratios of transition probabilities.}
\begin{ruledtabular}
\begin{tabular}{lcc}
                                        & Measured      &       Calculated       \\
	\hline
	$\left(A_{01}/A_{00}\right)\times 10^{3}$   &    $7.3 \pm 1$  &   $4.7$\\
	$\left(A_{01}/A_{11}\right)\times 10^{3}$   &    $5.59 \pm 0.02$  &   $4.8$\\
	$\left(A_{02}/A_{00}\right)\times 10^{3}$   &         -          &   $0.1$\\
\end{tabular}
\end{ruledtabular}
\end{table}

\begin{table}
\caption{\label{tab:fcs} Calculated Franck-Condon factors.}
\begin{ruledtabular}
\begin{tabular}{lccc}
	$f_{AX}$    & Ab initio \cite{Wells2011}    & EMO       & Morse                \\
	\hline
	$f_{00} $   & 0.99992                       & 0.996              & 0.9948   \\
	$f_{01} $   &$9\times10^{-6}$               & 0.0040             & 0.0005   \\
	$f_{02} $   &$7\times 10^{-5}$              & $2.6\times10^{-5}$ & $2.7\times 10^{-5}$  \\
\end{tabular}
\end{ruledtabular}
\end{table}

\subsection{\label{sec:A-a} Electronic branching - the $A^1\Pi, v=0 \leftarrow a^3\Pi, v'=0$ band}

\begin{figure}
	\centering
	\includegraphics[width=\linewidth]{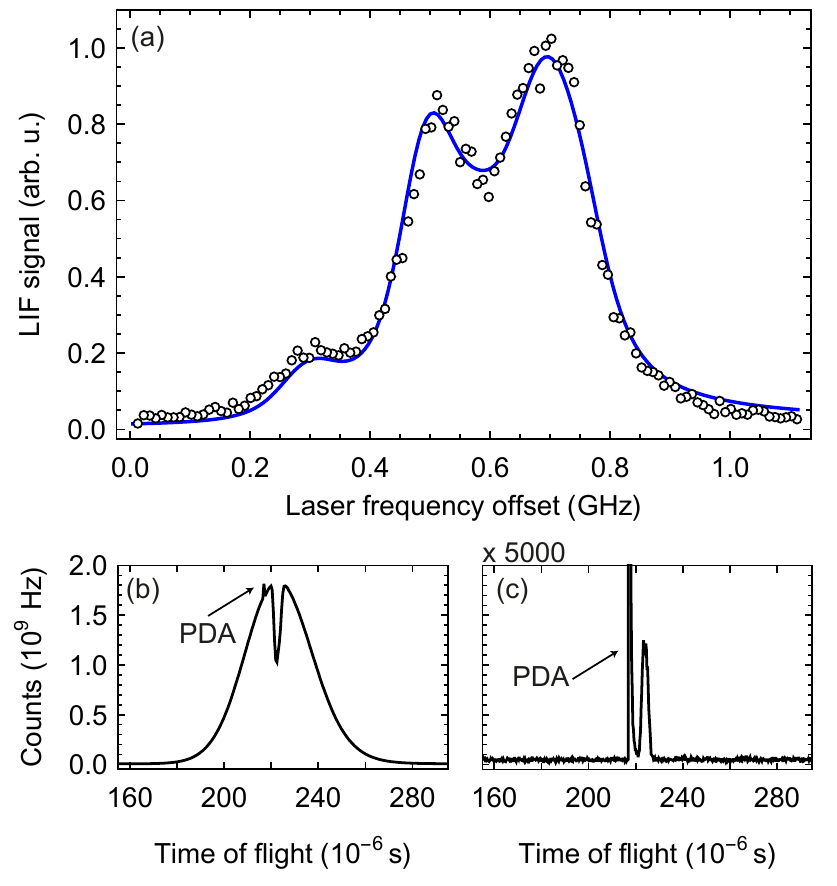}
	\caption{(a) Spectrum of the A$^1\Pi, v=0, J=1, + \leftarrow \textrm{a}^3\Pi_0, v'=0, J'=1, -$ transition near 597 nm. (b) LIF signal as a function of time when the frequency of the 227.5 nm laser is locked to the center of the R(0) line of the A$^1\Pi, v=0 \leftarrow \textrm{X}^1\Sigma^+, v''=0$ band. The laser power is set such that each molecule scatters on average two UV photons. Optical pumping by the PDA on the R$_1(0)$ line of the $\textrm{a}^3\Pi_0, v'=0 \leftarrow \textrm{X}^1\Sigma^+, v''=0$ band creates a dip in the center of the ToF profile. (c) LIF signal as a function of time when the CW dye laser is locked to the highest frequency component of the Q$_1(1)$ line of the A$^1\Pi, v=0 \leftarrow \textrm{a}^3\Pi_0, v'=0$ band. Each excited molecule emits only a single UV photon. The signal is about 2000 times weaker and the stray light from the PDA excitation can be seen as a sharp peak detected 10 \si{\micro\second} before the signal.}
	\label{fig:aAQ}
\end{figure}

Another loss channel from the main optical cycle is branching to the metastable a$^3\Pi$ state. To quantify this loss channel we measure the strength of the spin-forbidden A$^1\Pi,v=0 \leftarrow \textrm{a}^3\Pi,v'=0$ band relative to the A$^1\Pi,v=0 \leftarrow \textrm{X}^1\Sigma^+,v''=0$ band. Figure~\ref{fig:aAQ} (a) shows the spectrum of the A$^1\Pi, v=0, J=1, + \leftarrow \textrm{a}^3\Pi_0, v'=0, J'=1, -$ transition near 597 nm. This is the first time that this very weak band in AlF is observed. To obtain this spectrum we first excite the molecules on the R$_1$(0) line to the $\textrm{a}^3\Pi_0, v'=0, J'=1, -$ level using the PDA. A few mm further downstream, at the center of the LIF detector, the triplet molecules are excited to the A$^1\Pi, v=0, J=1, +$ level by driving the Q$_1$(1) line. We use a CW dye laser with about 300 mW in a 1.6 mm diameter ($1/e^2$) beam to drive this transition. The resulting laser-induced fluorescence on the Q(1) line of the A$^1\Pi, v=0 \rightarrow \textrm{X}^1\Sigma^+, v''=0$ band near 227.5 nm is recorded as a function of the dye laser frequency. The hyperfine structure in both electronic states is known from the measurements presented above, and the simulated spectrum is shown as the solid curve in Fig.~\ref{fig:aAQ} (a). We leave the Doppler width and the population distribution across the hyperfine components in the $\textrm{a}^3\Pi_0, v'=0, J'=1, -$ level as fitting parameters. The best fit is obtained by assuming a Doppler broadening of 82 MHz and an equal population distribution across hyperfine levels. Fig.~\ref{fig:aAQ} (c) shows the ToF profile when the laser frequency is locked to the highest frequency peak of the Q$_1$(1) line (410 mW in a 1.6 mm diameter ($1/e^2$) beam). There is a large peak due to stray light from the PDA, followed by the fluorescence signal of the molecules.
We can now compare the strength of this weak, spin-forbidden transition to the strong R(0) line of the A$^1\Pi, v=0, \leftarrow \textrm{X}^1\Sigma^+, v''=0$ band to quantify this loss channel. We lock the CW UV laser frequency to the center of the R(0) line and set the laser intensity such that each molecule scatters two photons on average (about 2.7 mW in a 2.0 mm ($1/e^2$) diameter beam). For molecules with diagonal Franck-Condon factors, the maximum number of photons emitted per molecule $n$ before being optically pumped into a dark rotational state is determined by the H\"onl-London factor $p$ to $n=1/(1-p)$. For the R(0) line, $p=2/3$ and the maximum number of emitted photons per molecule is $n=3$. The resulting ToF profile is shown in Fig.~\ref{fig:aAQ} (b). The dip in the center of the ToF profile is caused by optical pumping with the PDA on the R$_1(0)$ line of the $\textrm{a}^3\Pi, v'=0 \leftarrow \textrm{X}^1\Sigma^+, v''=0$ band. It measures the fraction of ground-state molecules that are excited to the triplet state. By comparing the amplitude of the dip in Fig.~\ref{fig:aAQ} (b) to the Q$_1(1)$ ToF profile shown in Fig.~\ref{fig:aAQ} (c) we can determine the relative transition strength. The ratio of the two ToF amplitudes is $5 \pm 0.5 \times 10^{-4}$. By accounting for the different laser intensities and wavelengths we measure a decay rate from the A$^1\Pi, v=0, J=1, +/-$ level to the $\textrm{a}^3\Pi_0$ state of $29\pm 3$ s$^{-1}$ or a relative loss from the cycling transition of $(0.56\pm 0.1)\times 10^{-7}$.

As a consistency check, we repeat the experiment for the A$^1\Pi, v=0, J=1, -$ level. Here, we drive the P$_1$(1) transition with the PDA to excite the molecules to the a$^3\Pi_0, v=0, J=0$ level. This is followed by the CW dye laser excitation on the R$_1(0)$ line of the A$^1\Pi, v=0 \leftarrow \textrm{a}^3\Pi, v'=0$ band. The resulting decay rate of the A$^1\Pi, v=0, J=1, +/-$ level to the $\textrm{a}^3\Pi_0$ state is $38\pm4$ s$^{-1}$. In total there are six possible decay channels to six different rotational states in the $\textrm{a}^3\Pi, v'=0$ state. We can neglect decays to higher vibrational levels. By measuring the transition strengths of the Q$_1(1)$ and R$_1(0)$ lines, we can calculate the other four decay channels and give a total decay rate from the A$^1\Pi, v=0, J=1, +/-$ level to the $\textrm{a}^3\Pi$ state of  $53\pm5$ s$^{-1}$, or a relative loss from the cycling transition of $(1.0\pm 0.1)\times 10^{-7}$. 

\subsection{Ionization}
Similar to MOTs of Cd and Mg there is a loss channel due to photoionization. The rate is given by
\begin{equation}
\Gamma_{\textrm{ion}}=\frac{\sigma P(I,\delta) I}{\hbar\omega},
\end{equation}
where $P(I,\delta)$ is the fraction of AlF molecules in the excited state. We estimate the fraction of excited state molecules using a rate model to be about 0.02 for a peak laser intensity of $I=1$ W/cm$^2$. Assuming that the photoionization cross-section from the A$^1\Pi$ state is similar to the one we measure for the a$^3\Pi$ state ($\sigma=3.8\times10^{-17}$ cm$^{-2}$) the expected photo-ionization rate is $\Gamma_{\textrm{ion}}=4$ s$^{-1}$. This is equivalent to a relative loss from the optical cycling transition of $\Gamma_{\textrm{ion}}/\Gamma=2\times 10^{-9}$.

\section{\label{sec:dipole}Electric dipole moment measurements}

\begin{figure}
	\centering
	\includegraphics[width=\linewidth]{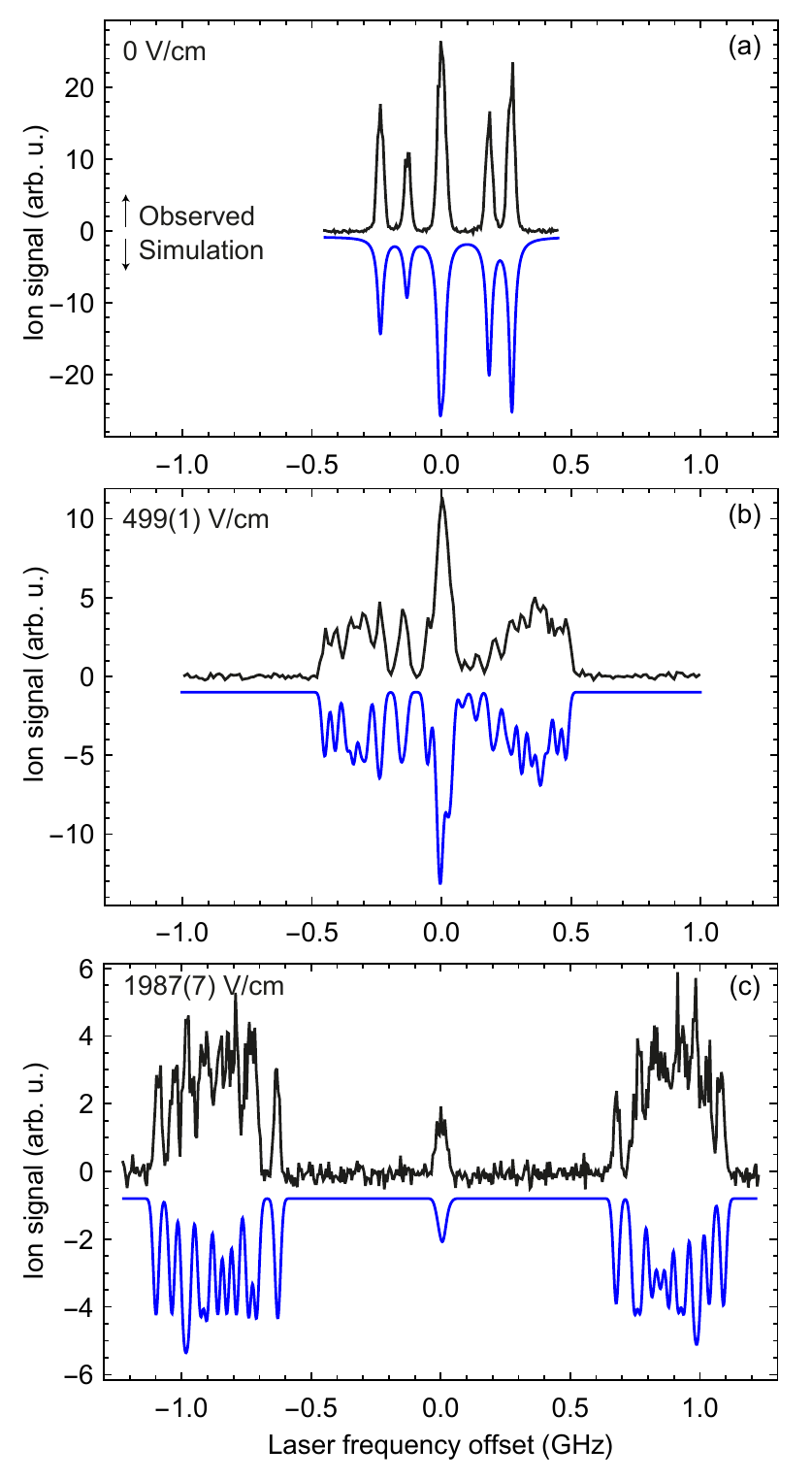}
	\caption{Measurement of the R$_2(0)$ line of the a$^3\Pi, v'=0 \leftarrow \textrm{X}^1\Sigma^+, v''=0$ band under field-free conditions (a), in an electric field of 499~V/cm (b) and in an electric field of 1987~V/cm (c), together with the simulated spectra (blue, pointing downwards), assuming $\mu(\textrm{a})=1.770$ Debye. The overall frequency shift of the spectra due to the negative Stark shift of the $J''=0$ level in the electronic ground state is removed.}
	\label{fig:Stark-Splitting-R2-0}
\end{figure}

We determine the electric dipole moments in the X$^1\Sigma^+, v=0$ ($\mu(\textrm{X})$), the a$^3\Pi, v=0$ ($\mu(\textrm{a})$) and the A$^1\Pi, v=0$ ($\mu(\textrm{A})$) states by measuring how selected rotational lines of optical transitions split and shift in external electric fields. First, we excite the AlF molecules with the CW laser on the R$_2(0)$ line of the a$^3\Pi_1, v'=0 \leftarrow \textrm{X}^1\Sigma^+, v''=0$ band in relatively low electric fields of up to 2~kV/cm. The laser polarization is perpendicular to the direction of the external electric field. In this setup we can only drive transitions with $\Delta M_F=\pm1$. In Fig.~\ref{fig:Stark-Splitting-R2-0}, the measured R$_2(0)$ transition is shown for three different electric field strengths. The observed splitting of the $J=1$ level in the a$^3\Pi, v=0$ state directly determines $\mu(\textrm{a})$. The entire spectrum shifts to higher frequency with increasing electric field due to the negative Stark shift of the $J''=0$ level in the electronic ground state. This overall frequency shift is subtracted in the spectra shown in Fig.~\ref{fig:Stark-Splitting-R2-0}. The frequency axis in the spectra is linearized and scaled by recording the transmission of a temperature stabilized confocal Fabry-P\'{e}rot cavity with a free spectral range of 149.9~MHz. The observed linewidth in the spectra is limited by the residual Doppler broadening to about 30~MHz. The simulated spectra, assuming $\mu(\textrm{a}) = 1.770$ Debye, are shown in blue (pointing downwards) underneath the experimental spectra. The observed structure is reproduced well by the simulations. The $\Delta M_F=\pm 1$ components with positive and negative Stark shift mix through the hyperfine interaction which gives intensity to a small central component. It is not a residual $\Delta M_F= 0$ component due to a small misalignment of the laser polarization and the direction of the electric field. The amplitude of the central peak decreases with increasing Stark splitting. We repeat this measurement for different spacings between the high voltage electrodes. All measurements agree within their experimental uncertainty with a mean value of $\mu(\textrm{a})$ = 1.770 $\pm$ 0.010 Debye.

\begin{figure}
	\centering
	\includegraphics[width=\linewidth]{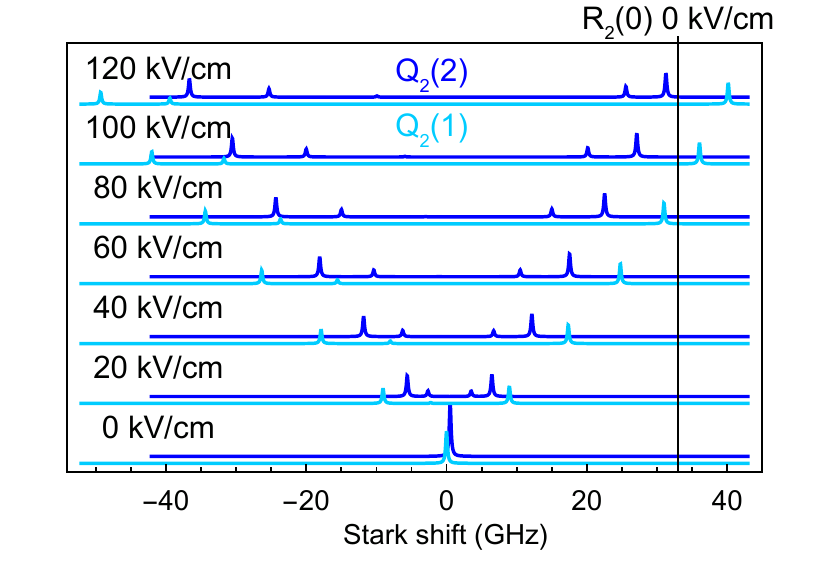}
	\caption{Stark shift of the Q$_2(1)$ and Q$_2(2)$ lines with the laser polarization orthogonal to the electric field. The $\Delta M_F=0$ components of the Q$_2(1)$ and Q$_2(2)$ lines with the largest Stark shift overlap with the R$_2(0)$ line (field-free) at approximately 87 kV/cm and 128 kV/cm, respectively.}
	\label{fig:QsInR}
\end{figure}

Dipole moments can be measured more accurately when the Stark shifts are large, i.e. in high electric fields. However, it is challenging to measure large frequency shifts in optical spectra with high accuracy. In high electric fields, the Stark shifts of AlF are comparable to the spacings between rotational energy levels. These spacings are known to kHz accuracy. Instead of measuring the frequency shift as a function of the electric field strength, we measure the electric field strength at which rotational lines overlap. We can determine the field at which the component of a specific rotational line with positive Stark shift overlaps with the component of a second rotational line with negative Stark shift. Alternatively, we can determine the field at which the Stark shifted components of a specific rotational line overlap with the zero-field center frequency of a second rotational line. For the latter method, we record the ion signal as a function of the excitation frequency in a high electric field and under field-free conditions for each molecular pulse simultaneously. The Stark shift of a rotational line depends linearly on both dipole moments of the two electronic states involved. The slope of this line changes for different rotational lines. Therefore, by comparing the Stark shifts of different rotational lines we can determine the values for both dipole moments. 
First we use this method to determine $\mu(\textrm{a})$ and $\mu(\textrm{X})$, i.e. by driving rotational lines in the a$^3\Pi, v'=0 \leftarrow \textrm{X}^1\Sigma^+, v''=0$ band.  With the laser polarization perpendicular to the electric field, the up-shifted component of the R$_2(0)$ line overlaps with the zero-field center frequency of the R$_2(1)$ line at $58.1 \pm 0.3$ kV/cm. With the laser polarization parallel to the electric field, the $\Delta M_F=0$ component of the Q$_2(1)$ line with the largest Stark shift overlaps with the central hyperfine component of the zero-field R$_2(0)$ line at $86.84 \pm 0.20$~kV/cm. The component of the Q$_2(2)$ line with the largest Stark shift overlaps with the central hyperfine component of the zero-field R$_2$(0) line in an electric field of $128.4 \pm 0.6$~kV/cm. We also determine the electric field strength at which different components of low-$J$ Q$_2$ and R$_2$ lines overlap. This gives additional linear relationships between $\mu(\textrm{a})$ and $\mu(\textrm{X})$. Figure \ref{fig:QsInR} shows simulated Stark spectra of the Q$_2(1)$ and Q$_2(2)$ line with respect to the R$_2(0)$ line.

\begin{figure}
	\centering
	\includegraphics[width=\linewidth]{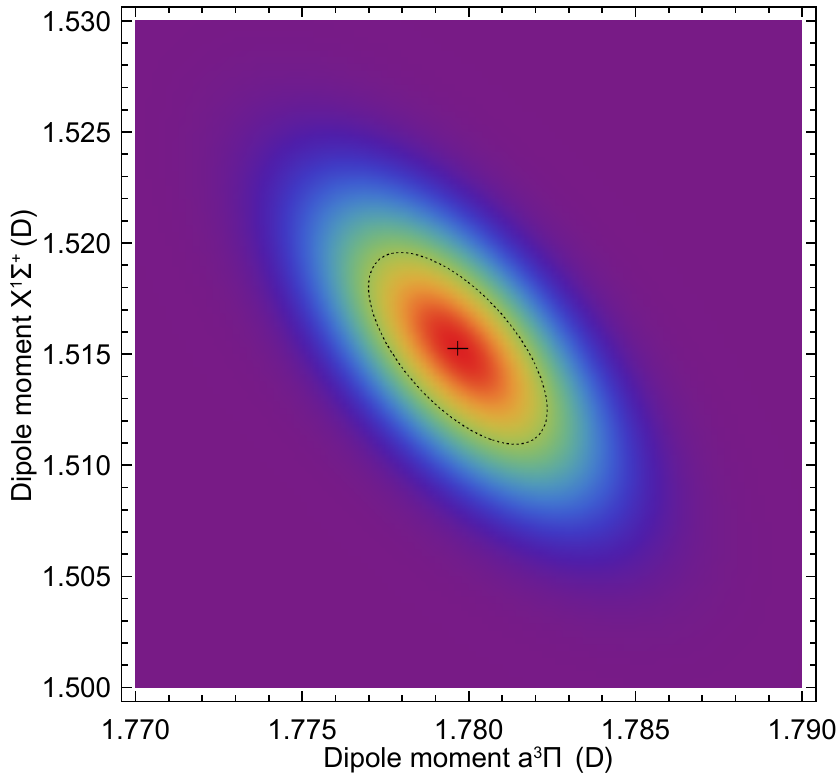}
	\caption{Density plot of $\mu(\textrm{X})$ versus $\mu(\textrm{a})$, depicting the joint probability of the various dipole moment measurements. The cross shows the most probable values for the dipole moments in Debye. The 1$\sigma$ standard deviation is shown as a dashed contour.}
	\label{fig:dipolePlot}
\end{figure}

The resulting density plot of $\mu(\textrm{X})$ versus $\mu(\textrm{a})$ is shown in Fig.~\ref{fig:dipolePlot}. The dashed contour indicates the $1\sigma$ standard deviation. The most probable values for the dipole moments are indicated by the cross as $\mu(\textrm{X})=1.515\pm 0.004$ Debye and $\mu(\textrm{a})=1.780\pm0.003$ Debye. The value of $\mu(\textrm{X})$ reported here is in agreement with the only experimental value reported to date of $1.53 \pm 0.10$~Debye \cite{Lide1965}. Quantum chemistry calculations give a value of $\mu(\textrm{X}) = 1.54$~Debye \cite{Woon2009}.  
Knowing the value of $\mu(\textrm{X})$, a single measurement determines the value for $\mu(\textrm{A})$. We measure the electric field at which a Stark shifted component of the R(0) line of the A$^1\Pi, v=0 \leftarrow \textrm{X}^1\Sigma^+, v''=0$ band overlaps with the field-free R(1) line. We lock the frequency of the 227.5 nm CW laser to the zero-field center frequency of the R(1) line and monitor the LIF signal in the second detection chamber. The CW laser beam interacts with the molecular beam for a second time in the high electric field region in the preparation chamber. At an electric field of $61.85 \pm 0.20$ kV/cm the CW laser is resonant with the Stark shifted component of the R(0) line, thereby depleting the population in the $J''=0$ level in the ground state. We monitor this depletion via two-color ionization in the first detection chamber. This way we determine the dipole moment in the A$^1\Pi, v=0$ state to $\mu(\textrm{A}) = 1.45 \pm 0.02$ Debye.

\begin{figure}
	\centering
	\includegraphics[width=\linewidth]{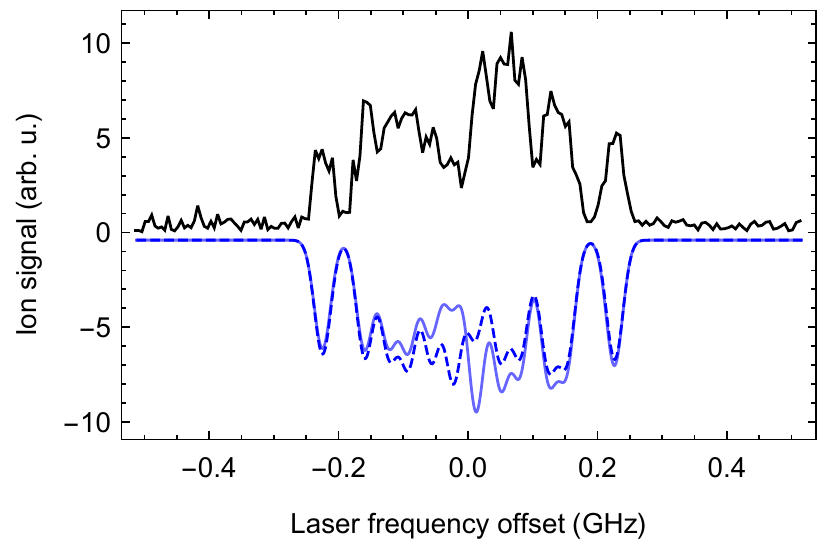}
	\caption{Spectrum of the R$_2(0)$ line of the a$^3\Pi, v'=0 \leftarrow \textrm{X}^1\Sigma^+, v''=0$ band in an electric field of 152.5 kV/cm with the laser polarization parallel to the electric field. The simulated spectrum (pointing downwards) is shown for $\alpha_2= 0$ \si{\cubic\angstrom} (dashed) and for $\alpha_2 = - 1.25$ \si{\cubic\angstrom} (solid).}
	\label{fig:highVoltage}
\end{figure}

So far we have not included effects due to the polarizability of the molecule, which are expected to be small. Figure~\ref{fig:highVoltage} shows a spectrum of a Stark shifted component of the R$_2(0)$ line of the a$^3\Pi, v'=0\leftarrow \textrm{X}^1\Sigma^+, v''=0$ band in an electric field of 152.5 kV/cm, with the laser polarization parallel to the electric field. The simulated spectrum agrees better with the data if a small polarizability anisotropy of $\alpha_2=-1.25$ \si{\cubic\angstrom} is included.  

\section{\label{sec:summary}Summary and Outlook}

In summary, we have characterized the AlF molecule spectroscopically to determine its suitability for laser cooling and trapping.

We have measured the spin-forbidden a$^3\Pi, v'=0 \leftarrow \textrm{X}^1\Sigma^+, v''=0$ band. The spectra show that the band becomes predominantly allowed due to spin-orbit coupling with the A$^1\Pi$ state. The spectra also show that the interaction with $^1 \Sigma$ states is about one order of magnitude weaker. 
We have ionized the molecules from the a$^3\Pi, v=0$ state with ArF excimer light and measured the 193 nm ionization cross-section to be $36\pm 5$ Mbarn. We have used a frequency-doubled pulsed dye laser for state-selective resonance-enhanced multi-photon ionization from the a$^3\Pi, v=0$ state. This enabled us to record laser-radio-frequency-laser triple resonance spectra of AlF molecules in the a$^3\Pi, v=0$ state. We determine the hyperfine energy levels with kHz accuracy and deduce precise hyperfine and rotational constants and the magnetic g-factors.

We have also measured laser-microwave-laser-laser quadruple resonance spectra of the X$^1\Sigma^+, v=0$ state to determine its hyperfine structure and rotational constants. The detailed knowledge of the a$^3\Pi, v=0$ state has enabled us to resolve hyperfine splittings in the X$^1\Sigma^+, v=0$ state that are comparable to the widths of the microwave transitions. The hyperfine splitting in the X$^1\Sigma^+, v=0$ state is smaller than the linewidth of the main laser cooling Q(1) line of the strong A$^1\Pi, v=0 \leftarrow \textrm{X}^1\Sigma^+, v''=0$ band. 

We have measured laser-induced fluorescence excitation spectra of the A$^1\Pi, v=0 \leftarrow \textrm{X}^1\Sigma^+, v''=0$ band to determine the hyperfine interaction, rotational constants and lifetime of the A$^1\Pi, v=0$ state. The spontaneous decay rate from the A$^1\Pi, v=0$ state is measured to be $\Gamma=2\pi\times (83.8\pm 1.3)$ MHz. The hyperfine level structure in the $J=1$ level resembles the structure in the $J=1$ level of the a$^3\Pi_1, v=0$ state. All Q-lines of the A$^1\Pi, v=0 \leftarrow \textrm{X}^1\Sigma^+, v''=0$ band are rotationally closed and can be used for laser cooling. There is a small residual rotational branching due to rotational mixing via the hyperfine interaction in both electronic states. However, the relative loss from the optical cycling transition is calculated to be below $10^{-5}$. We have measured the vibrational branching ratio to X$^1\Sigma^+, v''=1$ to be $(5.6\pm0.02)\times 10^{-3}$ in good agreement with the theoretical prediction. By addressing this vibrational loss channel with a repump laser it is possible to scatter about $10^4$ photons before the molecules are pumped into X$^1\Sigma^+, v''=2$. The weak R$_1$(0) and Q$_1$(1) lines of the spin-forbidden A$^1\Pi, v=0 \leftarrow \textrm{a}^3\Pi, v'=0$ band of AlF have been observed for the first time. From their intensity, the spontaneous decay rate from the A$^1\Pi, v=0, J=1$ level to the a$^3\Pi, v=0$ state is determined as $53\pm 5$ s$^{-1}$. This is equivalent to a relative loss from the optical cycling transition of about $10^{-7}$. 

We have measured laser excitation spectra on the a$^3\Pi, v'=0 \leftarrow \textrm{X}^1\Sigma^+, v''=0$ band and on the A$^1\Pi, v=0 \leftarrow \textrm{X}^1\Sigma^+, v''=0$ band in electric fields to determine the electric dipole moments. In the X$^1\Sigma^+, v=0$ state, in the a$^3\Pi, v=0$ state and in the A$^1\Pi, v=0$ state we determined the electric dipole moments as $\mu(\textrm{X})=1.515\pm 0.004$ Debye, $\mu(\textrm{a})=1.780\pm0.003$ Debye and $\mu(\textrm{A})=1.45\pm 0.02$ Debye, respectively. Knowing the energy level structure and dipole moments in each of the electronic states enabled us to simulate the Stark spectra for fields up to 150 kV/cm. The accurate spectroscopic parameters and electric dipole moments for the lowest singlets and the lowest triplet state of AlF are an excellent benchmark for quantum chemistry calculations. 

We plan to slow AlF molecules from a molecular beam and trap and cool them in a magneto-optical trap to produce a dipolar gas with a high phase space density. The Q-lines on both the A$^1\Pi, v=0 \leftarrow \textrm{X}^1\Sigma^+, v''=0$ band and on the a$^3\Pi, v'=0 \leftarrow \textrm{X}^1\Sigma^+, v''=0$ band are rotationally closed and can be used for laser cooling, slowing and magneto-optical trapping. With one vibrational repump laser it is possible to scatter about 10$^4$ photons on the A$^1\Pi - \textrm{X}^1\Sigma^+$ band and about 4000 photons on the Q$_2(1)$ line of the a$^3\Pi, v'=0 \leftarrow \textrm{X}^1\Sigma^+, v''$ band. Scattering 10$^4$ photons on the A$^1\Pi - \textrm{X}^1\Sigma^+$ band corresponds to a velocity change of 382 m/s, sufficient to slow a cryogenic buffer gas beam or even a supersonic molecular beam to rest. The achievable photon scattering rate is only limited by the currently available laser power in the UV. However, CW lasers with an output power of up to 0.56 W at 229 nm \cite{Kaneda2016} and 1.4 W at 243 nm \cite{Burkley2018} have recently been demonstrated.  
We expect an effective photon scattering rate on the Q(1) line of $\Gamma_{\textrm{eff}}=\Gamma/35=2\pi \times 2.4$~MHz \cite{Tarbutt2013, Norrgard2016}. At this rate the stopping distance is 2 cm for a 150 m/s cryogenic buffer gas beam and 8 cm for a 300 m/s supersonic molecular beam. To compensate for the Doppler shift as the molecules slow down, a Zeeman slower, chirped laser slowing or white light slowing can be used. A static, single frequency type-II MOT can be used to trap and cool a large number of AlF molecules to a temperature of a few mK \cite{Tarbutt2015}. Using a rate model \cite{Tarbutt2013} we estimate the capture velocity to be about 40 m/s, limited by the currently available laser power. Sub-Doppler cooling can be used to lower the temperature to the \si{\micro\kelvin} range. Narrow-line cooling to the recoil limit is feasible on the Q$_2$(1) line of the a$^3\Pi, v'=0 \leftarrow \textrm{X}^1\Sigma^+$ band. However, it is challenging due to the small restoring force resulting from the long radiative lifetime of the $J=1$ level in the a$^3\Pi_1, v=0$ state of approximately 1 ms. 
 
Besides optical manipulation, we can also use electric fields to slow and trap AlF molecules. AlF compares well to metastable CO, which has been used extensively to test and demonstrate various electric field manipulation tools \cite{Bethlem1999, Bethlem2000, Gilijamse2007, Meek2008}. The Stark shift to mass ratio, an important performance metric for Stark manipulation, in the a$^3\Pi_1, v=0, J=1$ state is only $20\%$ smaller than for the corresponding level in CO. A decisive advantage of AlF compared to CO is that it can be controlled with electric fields not only in its triplet but also in its electronic ground state. 

\appendix
\section{Observed and calculated transition frequencies in $a^3\Pi,v=0$}
Table \ref{frequencies} lists the 138 measured radio-frequency and microwave transitions in the a$^3\Pi,v=0$ state of AlF. The measured transition frequency f$_\textrm{o}$ is listed in the first column, followed by the experimental uncertainty $\delta \textrm{f}_\textrm{o}$ (in MHz). The latter includes the statistical uncertainty from the fit and a systematic uncertainty determined by the Zeeman broadening of the line due to the small residual magnetic field in the interaction region. The third column lists $\delta \textrm{f}_\textrm{loop}$, which is used as an experimental consistency check. A combination of three transitions probe an interval that can also be measured directly, in a single transition. Their difference should be zero within the total combined experimental uncertainty. We list this deviation from zero for each line that is part of such a loop as $\delta \textrm{f}_\textrm{loop}$. This deviation from zero is never larger than the combined experimental uncertainty of the four lines that make up a loop. The next twelve columns list the $\Omega$, $J$, parity ($p$) and $F$ labels of the levels involved in the transition, together with their calculated $g_F$-factors. The labels with (without) a prime refer to the upper (lower) level of the transition. However, this labelling scheme is not unique because there can be multiple levels with the same parity and $F$ quantum number within a given $J$ and $\Omega$. To be able to distinguish levels with the same quantum numbers and parity, we introduce an index $n$. The lowest energy level for a given value of $F$ gets the index $n=1$ and is in the a$^3\Pi_0$ manifold. The column with the header $d^2$ (D$^2$) lists the calculated square of the reduced dipole moment $d$ of the transition in Debye$^2$. These transition strengths vary over four orders of magnitude. The difference between the observed and calculated transition frequency, f$_\textrm{o}-\textrm{f}_\textrm{c}$ is given (in MHz) in the second to last column. The last column lists the difference between the calculated $g_F$-factors of the upper and the lower level of the transition. This value correlates with the experimental uncertainty to determine the line center, because transitions with larger $\delta g=g'_F-g_F$ show more Zeeman broadening. \\
We use all 138 transitions in the fit to the eigenvalues of the Hamiltonian to determine the 24 spectroscopic parameters given in Table \ref{tab:a3Param}.

\begin{longtable*}{d{5.3}cccccccd{2.3}cccccd{2.3}d{2.4}d{2.3}d{2.3}}
\caption{Observed and calculated rf and microwave transition frequencies in the a$^3\Pi,v=0$ state of AlF\label{frequencies}, transition dipole moments and magnetic $g_F$ factors.}\\
\hline
\hline
\endfirsthead
\multicolumn{17}{c}{Continuation of Table \ref{frequencies}}\\
\hline
\hline
\multicolumn{1}{c}{f$_\textrm{o}$~(MHz)} &  $\delta $f$_\textrm{o}$ & $\delta $f$_\textrm{loop}$ &
$\Omega'$ & $J'$ & $p'$ & $F'$ & $n'$ &  \multicolumn{1}{c}{$g_F'$} &
$\Omega$ &  $J$ &  $p$ &  $F$ &  $n$ &    \multicolumn{1}{c}{$g_{F}$} &     \multicolumn{1}{c}{d$^2$~(D$^2$)} &    \multicolumn{1}{c}{$\textrm{f}_\textrm{o}$-f$_\textrm{c}$} &  \multicolumn{1}{c}{$g_{F}'$-$g_{F}$} \\
\hline
\endhead
\hline
\hline
\endfoot
\hline
\hline
\endlastfoot
\multicolumn{1}{c}{f$_\textrm{o}$~(MHz)} &  $\delta $f$_\textrm{o}$ & $\delta $f$_\textrm{loop}$ &
$\Omega'$ & $J'$ & $p'$ & $F'$ & $n'$ &  \multicolumn{1}{c}{$g_F'$} &
$\Omega$ &  $J$ &  $p$ &  $F$ &  $n$ &    \multicolumn{1}{c}{$g_{F}$} &    \multicolumn{1}{c}{d$^2$~(D$^2$)} &    \multicolumn{1}{c}{$\textrm{f}_\textrm{o}$-f$_\textrm{c}$} &  \multicolumn{1}{c}{$g_{F}'$-$g_{F}$} \\
\hline
~9815.516&0.003&0.003&0&0&-&~3&~1&-0.028&0&0&+&~2&~1&-0.022&~0.0050&-0.004&-0.006\\
~9818.512&0.003&0.003&0&0&-&~3&~1&-0.028&0&0&+&~3&~1&-0.016&~0.1100&-0.001&-0.012\\
~9822.671&0.003&0.003&0&0&-&~2&~1&-0.039&0&0&+&~3&~1&-0.016&~0.0050&~0.002&-0.023\\
~9819.673&0.003&0.003&0&0&-&~2&~1&-0.039&0&0&+&~2&~1&-0.022&~0.0750&-0.003&-0.017\\
~9686.817&0.005&0.004&0&1&+&~2&~2&-0.024&0&1&-&~3&~3&~0.025&~0.0350&-0.006&-0.049\\
~9692.796&0.004&0.007&0&1&+&~1&~1&-0.040&0&1&-&~2&~3&~0.035&~0.0222&-0.002&-0.075\\
~9807.485&0.003&0.003&0&1&+&~2&~3&~0.015&0&1&-&~3&~3&~0.025&~0.0028&-0.004&-0.010\\
~9808.913&0.003&0.004&0&1&+&~3&~2&~0.011&0&1&-&~3&~3&~0.025&~0.0459&-0.007&-0.014\\
~9812.306&0.002&0.007&0&1&+&~2&~3&~0.015&0&1&-&~2&~3&~0.035&~0.0327&-0.006&-0.020\\
~9821.992&0.004&0.001&0&1&+&~4&~1&~0.006&0&1&-&~3&~3&~0.025&~0.0355&~0.002&-0.019\\
~9822.583&0.004&0.003&0&1&+&~3&~3&~0.008&0&1&-&~3&~3&~0.025&~0.0025&~0.004&-0.017\\
~9827.406&0.003&0.003&0&1&+&~3&~3&~0.008&0&1&-&~2&~3&~0.035&~0.0253&~0.005&-0.027\\
~9841.880&0.003&0.001&0&1&+&~3&~2&~0.011&0&1&-&~4&~1&~0.007&~0.0101&-0.004&~0.004\\
~9845.927&0.004&0.003&0&1&+&~2&~3&~0.015&0&1&-&~3&~2&~0.009&~0.0066&-0.007&~0.006\\
~9854.960&0.005&0.001&0&1&+&~4&~1&~0.006&0&1&-&~4&~1&~0.007&~0.0107&~0.006&-0.001\\
~9861.027&0.005&0.003&0&1&+&~3&~3&~0.008&0&1&-&~3&~2&~0.009&~0.0073&~0.004&-0.001\\
~9894.181&0.001&0.007&0&1&+&~1&~1&-0.040&0&1&-&~1&~1&-0.036&~0.0026&~0.001&-0.004\\
~9899.272&0.002&0.004&0&1&+&~2&~2&-0.024&0&1&-&~2&~2&-0.023&~0.0050&~0.004&-0.001\\
10013.696&0.002&0.007&0&1&+&~2&~3&~0.015&0&1&-&~1&~1&-0.036&~0.0096&~0.003&~0.051\\
10021.372&0.003&0.004&0&1&+&~3&~2&~0.011&0&1&-&~2&~2&-0.023&~0.0155&~0.008&~0.034\\
~9749.561&0.003&0.001&0&2&-&~4&~2&~0.017&0&2&+&~5&~1&~0.016&~0.0018&~0.002&~0.001\\
~9760.598&0.004&~~~~~&0&2&-&~3&~5&~0.022&0&2&+&~4&~3&~0.020&~0.0014&-0.002&~0.002\\
~9840.139&0.004&0.001&0&2&-&~5&~1&~0.016&0&2&+&~5&~1&~0.016&~0.0031&~0.002&~0.000\\
~9851.702&0.003&~~~~~&0&2&-&~4&~3&~0.020&0&2&+&~4&~3&~0.020&~0.0024&~0.003&~0.000\\
~9853.710&0.002&~~~~~&0&2&-&~1&~2&-0.038&0&2&+&~2&~4&~0.007&~0.0018&~0.003&-0.045\\
~9854.234&0.003&0.001&0&2&-&~4&~2&~0.017&0&2&+&~4&~2&~0.019&~0.0010&-0.003&-0.002\\
~9877.260&0.005&~~~~~&0&2&-&~3&~4&~0.015&0&2&+&~3&~4&~0.017&~0.0026&-0.005&-0.002\\
~9881.786&0.004&~~~~~&0&2&-&~2&~5&~0.021&0&2&+&~2&~5&~0.027&~0.0018&-0.005&-0.006\\
~9944.813&0.004&0.001&0&2&-&~5&~1&~0.016&0&2&+&~4&~2&~0.019&~0.0035&-0.002&-0.003\\
~9953.603&0.003&~~~~~&0&2&-&~4&~3&~0.020&0&2&+&~3&~5&~0.025&~0.0027&-0.001&-0.005\\
~9966.078&0.002&~~~~~&0&2&-&~2&~4&~0.006&0&2&+&~1&~2&-0.045&~0.0015&-0.004&~0.051\\	
~~~~1.079&0.002&0.001&1&1&-&~1&~7&-0.227&1&1&+&~1&~7&-0.227&~1.2300&~0.002&~0.000\\
~~~~6.484&0.002&0.007&1&1&-&~3&13&~0.039&1&1&+&~2&12&~0.073&~0.0014&~0.001&-0.034\\
~~~10.902&0.002&0.005&1&1&-&~2&11&-0.151&1&1&+&~2&11&-0.149&~2.6900&~0.002&-0.002\\
~~~13.428&0.003&0.005&1&1&+&~3&14&~0.150&1&1&-&~3&14&~0.148&~7.4000&-0.007&~0.002\\
~~~21.379&0.001&0.007&1&1&-&~2&12&~0.074&1&1&+&~2&12&~0.073&~0.6420&-0.003&~0.001\\
~~~21.744&0.001&0.006&1&1&-&~3&13&~0.039&1&1&+&~3&13&~0.038&~0.4910&-0.003&~0.001\\
~~~21.826&0.003&0.002&1&1&+&~4&13&~0.113&1&1&-&~4&13&~0.113&~9.0800&-0.006&~0.000\\
~~~36.644&0.002&0.005&1&1&-&~2&12&~0.074&1&1&+&~3&13&~0.038&~0.0018&-0.001&~0.036\\
~~~74.353&0.004&0.004&1&1&-&~4&13&~0.113&1&1&+&~3&14&~0.150&~0.0458&~0.006&-0.037\\
~~101.298&0.004&0.002&1&1&+&~1&~7&-0.227&1&1&-&~2&11&-0.151&~0.0128&~0.003&-0.076\\
~~109.610&0.005&0.004&1&1&+&~4&13&~0.113&1&1&-&~3&14&~0.148&~0.1050&-0.004&-0.035\\
~~113.279&0.004&0.005&1&1&-&~1&~7&-0.227&1&1&+&~2&11&-0.149&~0.0373&~0.007&-0.078\\
~~120.377&0.008&0.005&1&1&+&~2&12&~0.073&1&1&-&~1&~7&-0.227&~3.6300&~0.001&~0.300\\
~~142.835&0.007&0.002&1&1&-&~2&12&~0.074&1&1&+&~1&~7&-0.227&~3.6500&~0.001&~0.301\\
~~188.903&0.005&0.007&1&1&+&~3&14&~0.150&1&1&-&~2&12&~0.074&~3.8100&~0.001&~0.076\\
~~196.856&0.006&0.005&1&1&-&~3&14&~0.148&1&1&+&~2&12&~0.073&~3.8300&~0.007&~0.075\\
~~203.805&0.006&0.007&1&1&+&~3&14&~0.150&1&1&-&~3&13&~0.039&~0.1010&~0.005&~0.111\\
~~207.488&0.007&0.006&1&1&+&~3&13&~0.038&1&1&-&~2&11&-0.151&~5.4200&~0.004&~0.189\\
~~212.117&0.006&0.005&1&1&-&~3&14&~0.148&1&1&+&~3&13&~0.038&~0.0245&~0.005&~0.110\\
~~222.754&0.005&0.006&1&1&+&~2&12&~0.073&1&1&-&~2&11&-0.151&~0.0459&~0.007&~0.224\\
~~240.135&0.007&0.005&1&1&-&~3&13&~0.039&1&1&+&~2&11&-0.149&~5.4000&~0.004&~0.188\\
~~255.034&0.006&0.005&1&1&-&~2&12&~0.074&1&1&+&~2&11&-0.149&~0.0321&~0.004&~0.223\\
~~299.902&0.006&0.004&1&1&-&~4&13&~0.113&1&1&+&~3&13&~0.038&~5.4700&~0.009&~0.075\\
~~299.985&0.006&0.004&1&1&+&~4&13&~0.113&1&1&-&~3&13&~0.039&~5.4100&~0.006&~0.074\\
~~430.512&0.010&0.005&1&1&-&~3&14&~0.148&1&1&+&~2&11&-0.149&~0.0160&~0.015&~0.297\\
~~433.038&0.010&0.005&1&1&+&~3&14&~0.150&1&1&-&~2&11&-0.151&~0.0099&~0.007&~0.301\\	
~~~~5.823&0.002&0.002&1&2&+&~4&14&~0.055&1&2&-&~4&14&~0.054&~1.9800&~0.006&~0.001\\
~~~20.560&0.005&~~~~~&1&2&-&~2&13&~0.002&1&2&+&~1&~8&-0.154&~1.0800&~0.000&~0.156\\
~~~23.798&0.003&~~~~~&1&2&-&~4&15&~0.076&1&2&+&~4&15&~0.075&~3.8900&-0.009&~0.001\\
~~~25.431&0.001&~~~~~&1&2&-&~2&14&~0.075&1&2&+&~1&~9&~0.077&~0.9420&-0.005&-0.002\\
~~~28.224&0.001&~~~~~&1&2&+&~3&15&~0.039&1&2&-&~3&15&~0.039&~0.5000&~0.003&~0.000\\
~~~46.617&0.003&~~~~~&1&2&-&~3&15&~0.039&1&2&+&~2&13&~0.001&~1.6200&-0.006&~0.038\\
~~~49.659&0.002&~~~~~&1&2&-&~3&16&~0.076&1&2&+&~2&14&~0.076&~1.0800&-0.003&~0.000\\
~~~61.370&0.003&0.002&1&2&+&~4&15&~0.075&1&2&-&~4&14&~0.054&~0.0343&~0.009&~0.021\\
~~~93.480&0.003&0.002&1&2&-&~5&13&~0.061&1&2&+&~4&15&~0.075&~0.0760&-0.012&-0.014\\
~~115.544&0.003&~~~~~&1&2&+&~5&13&~0.061&1&2&-&~4&14&~0.054&~1.1700&~0.009&~0.007\\
~~116.955&0.003&~~~~~&1&2&+&~1&~9&~0.077&1&2&-&~1&~8&-0.152&~0.0034&-0.004&~0.229\\
~~120.115&0.002&0.002&1&2&+&~4&14&~0.055&1&2&-&~3&15&~0.039&~1.6400&~0.002&~0.016\\
~~133.810&0.004&~~~~~&1&2&-&~3&16&~0.076&1&2&+&~2&13&~0.001&~0.0116&-0.008&~0.075\\
~~149.029&0.003&0.002&1&2&-&~5&13&~0.061&1&2&+&~4&14&~0.055&~1.1100&-0.007&~0.006\\
~~175.662&0.004&0.002&1&2&+&~4&15&~0.075&1&2&-&~3&15&~0.039&~0.0403&~0.005&~0.036\\
~~181.424&0.004&~~~~~&1&2&+&~3&16&~0.076&1&2&-&~2&13&~0.002&~0.0011&-0.005&~0.074\\
~~~26.397&0.002&~~~~~&1&3&-&~4&16&~0.040&1&3&+&~4&16&~0.039&~1.0400&~0.007&~0.001\\
~~~58.145&0.003&~~~~~&1&3&+&~5&15&~0.047&1&3&-&~5&14&~0.040&~0.3110&-0.011&~0.007\\
~~~58.206&0.002&~~~~~&1&3&+&~5&14&~0.039&1&3&-&~4&16&~0.040&~0.6560&-0.011&-0.001\\
~~~89.015&0.002&~~~~~&1&3&-&~4&16&~0.040&1&3&+&~3&17&~0.040&~0.7340&-0.002&~0.000\\
~~110.090&0.003&~~~~~&1&3&+&~5&15&~0.047&1&3&-&~4&16&~0.040&~0.0051&-0.007&~0.007\\
~~~~0.294&0.003&~~~~~&1&7&+&~8&17&~0.016&1&7&-&~8&17&~0.016&~0.9120&~0.000&~0.000\\
~~~~5.226&0.003&~~~~~&1&7&-&~9&13&~0.015&1&7&+&~8&17&~0.016&~0.0178&-0.004&-0.001\\
~~~~5.362&0.002&~~~~~&1&7&-&~7&18&~0.017&1&7&+&~8&16&~0.015&~0.0801&~0.000&~0.002\\
~~~~8.043&0.003&~~~~~&1&7&-&~9&14&~0.015&1&7&+&~8&17&~0.016&~0.0134&~0.001&-0.001\\
~~~11.373&0.003&~~~~~&1&7&+&~8&17&~0.016&1&7&-&~7&19&~0.018&~0.0440&~0.002&-0.002\\
~~~18.589&0.002&~~~~~&1&7&-&~8&16&~0.016&1&7&+&~8&16&~0.015&~0.8280&~0.000&~0.001\\
~~~27.045&0.003&0.002&1&7&+&~9&13&~0.014&1&7&-&10&10&~0.013&~0.0444&~0.000&~0.001\\
~~~33.727&0.003&0.002&1&7&+&~9&13&~0.014&1&7&-&~9&14&~0.015&~0.5340&-0.002&-0.001\\
~~~45.060&0.002&~~~~~&1&7&-&~6&20&~0.020&1&7&+&~7&18&~0.017&~0.0606&~0.000&~0.003\\
~~~46.428&0.003&~~~~~&1&7&+&~9&13&~0.014&1&7&-&~8&16&~0.016&~0.0736&~0.002&-0.002\\
~~~55.504&0.002&0.002&1&7&-&~5&22&~0.025&1&7&+&~6&21&~0.021&~0.0212&-0.001&~0.004\\
~~~62.319&0.002&0.002&1&7&-&~5&23&~0.025&1&7&+&~6&21&~0.021&~0.0097&-0.003&~0.004\\
~~~72.488&0.003&~~~~~&1&7&-&~8&16&~0.016&1&7&+&~7&18&~0.017&~0.0877&~0.002&-0.001\\
~~~75.855&0.002&~~~~~&1&7&-&~5&22&~0.025&1&7&+&~6&20&~0.020&~0.0291&~0.000&~0.005\\
~~~78.724&0.003&~~~~~&1&7&+&~9&14&~0.015&1&7&-&~8&16&~0.016&~0.0006&~0.008&-0.001\\
~~~80.565&0.003&0.002&1&7&-&~4&24&~0.031&1&7&+&~5&23&~0.025&~0.0006&-0.004&~0.006\\
~~~88.363&0.002&0.002&1&7&-&~5&22&~0.025&1&7&+&~5&23&~0.025&~0.3950&~0.000&~0.000\\
~~~95.180&0.003&0.002&1&7&-&~5&23&~0.025&1&7&+&~5&23&~0.025&~0.2150&~0.000&~0.000\\
~~102.205&0.002&~~~~~&1&7&-&~6&20&~0.020&1&7&+&~5&23&~0.025&~0.0154&~0.001&-0.005\\
~~106.043&0.003&0.002&1&7&+&10&10&~0.013&1&7&-&10&10&~0.013&~1.1700&~0.004&~0.000\\
~~109.587&0.003&~~~~~&1&7&-&~6&21&~0.021&1&7&+&~5&23&~0.025&~0.0157&~0.002&-0.004\\
~~112.727&0.003&0.002&1&7&+&10&10&~0.013&1&7&-&~9&14&~0.015&~0.0172&~0.005&-0.002\\
~~131.992&0.003&0.002&1&7&-&~4&24&~0.031&1&7&+&~4&24&~0.031&~0.4810&-0.005&~0.000\\
~~139.792&0.001&0.002&1&7&-&~5&22&~0.025&1&7&+&~4&24&~0.031&~0.0206&~0.001&-0.006\\
~~146.608&0.002&0.001&1&7&-&~5&23&~0.025&1&7&+&~4&24&~0.031&~0.0244&~0.000&-0.006\\
~~~~0.524&0.006&~~~~~&2&4&+&~5&28&~0.166&2&4&-&~5&28&~0.166&~4.5000&~0.001&~0.000\\
~~~~0.569&0.006&~~~~~&2&4&+&~4&30&~0.210&2&4&-&~4&30&~0.210&~3.9700&-0.002&~0.000\\
~~~~1.721&0.005&~~~~~&2&7&-&10&22&~0.041&2&7&+&10&22&~0.041&~4.5700&~0.002&~0.000\\
~~~~2.034&0.006&~~~~~&2&7&-&~9&26&~0.046&2&7&+&~9&26&~0.046&~4.1700&~0.001&~0.000\\
~~~~2.779&0.004&0.001&2&7&-&~9&25&~0.044&2&7&+&~9&25&~0.044&~3.8800&-0.002&~0.000\\
~~~~3.086&0.004&0.004&2&7&-&~8&29&~0.050&2&7&+&~8&29&~0.050&~3.5500&-0.002&~0.000\\
~~~~3.397&0.003&0.001&2&7&-&~8&28&~0.048&2&7&+&~8&28&~0.048&~3.3000&-0.004&~0.000\\
~~~~3.556&0.004&~~~~~&2&7&-&~5&34&~0.075&2&7&+&~5&34&~0.075&~2.1800&-0.006&~0.000\\
~~~~3.666&0.100&0.004&2&7&-&~7&30&~0.054&2&7&+&~7&30&~0.054&~2.8200&-0.006&~0.000\\
~~~~3.666&0.100&0.004&2&7&-&~7&31&~0.056&2&7&+&~7&31&~0.056&~3.0300&-0.005&~0.000\\
~~~~3.684&0.100&~~~~~&2&7&-&~6&32&~0.062&2&7&+&~6&32&~0.062&~2.4400&-0.007&~0.000\\
~~~~3.720&0.005&~~~~~&2&7&-&~4&35&~0.095&2&7&+&~4&35&~0.095&~1.9100&-0.007&~0.000\\
~~~~3.754&0.004&~~~~~&2&7&-&~5&35&~0.078&2&7&+&~5&35&~0.078&~2.3300&-0.004&~0.000\\
~~~~3.861&0.004&~~~~~&2&7&-&~6&33&~0.064&2&7&+&~6&33&~0.064&~2.6200&-0.008&~0.000\\
~~~85.166&0.004&~~~~~&2&7&+&~9&25&~0.044&2&7&-&~8&29&~0.050&~0.0096&-0.001&-0.006\\
~~118.002&0.004&~~~~~&2&7&-&~6&32&~0.062&2&7&+&~5&34&~0.075&~0.2400&-0.002&-0.013\\
~~129.340&0.004&~~~~~&2&7&-&~5&35&~0.078&2&7&+&~4&35&~0.095&~0.0494&~0.000&-0.017\\
~~137.062&0.004&~~~~~&2&7&-&~6&33&~0.064&2&7&+&~5&35&~0.078&~0.1540&~0.000&-0.014\\
~~139.858&0.004&~~~~~&2&7&-&~7&30&~0.054&2&7&+&~6&32&~0.062&~0.3070&-0.002&-0.008\\
~~148.712&0.005&~~~~~&2&7&+&~7&31&~0.056&2&7&-&~6&33&~0.064&~0.2030&~0.011&-0.008\\
~~153.672&0.004&~~~~~&2&7&+&~8&28&~0.048&2&7&-&~7&30&~0.054&~0.3170&~0.008&-0.006\\
~~168.989&0.004&0.004&2&7&+&~8&29&~0.050&2&7&-&~7&31&~0.056&~0.1940&~0.004&-0.006\\
~~174.809&0.004&0.001&2&7&+&~9&25&~0.044&2&7&-&~8&28&~0.048&~0.2700&~0.004&-0.004\\
~~175.737&0.004&0.004&2&7&-&~8&29&~0.050&2&7&+&~7&31&~0.056&~0.1940&-0.007&-0.006\\
~~180.986&0.004&0.001&2&7&-&~9&25&~0.044&2&7&+&~8&28&~0.048&~0.2700&~0.000&-0.004\\
~~195.015&0.004&~~~~~&2&7&-&~9&26&~0.046&2&7&+&~8&29&~0.050&~0.1260&-0.009&-0.004\\
~~200.643&0.005&~~~~~&2&7&-&10&22&~0.041&2&7&+&~9&25&~0.044&~0.1650&-0.003&-0.003\\
64825.398&0.001&~~~~~&0&2&+&~2&~5&~0.027&0&1&-&~2&~3&~0.035&~3.7500&-0.001&-0.008\\
64825.964&0.001&~~~~~&0&2&+&~3&~4&~0.017&0&1&-&~3&~3&~0.025&~5.4100&~0.003&-0.008\\
64830.784&0.002&~~~~~&0&2&+&~3&~4&~0.017&0&1&-&~2&~3&~0.035&~0.1730&~0.001&-0.018\\
65067.937&0.002&~~~~~&0&2&+&~4&~3&~0.020&0&1&-&~3&~2&~0.009&11.4000&-0.004&~0.011\\
65074.328&0.002&~~~~~&0&2&+&~5&~1&~0.016&0&1&-&~4&~1&~0.007&14.3000&-0.001&~0.009\\
66466.114&0.001&~~~~~&1&2&+&~2&14&~0.076&1&1&-&~2&12&~0.074&~2.8500&~0.000&~0.002\\
66471.700&0.001&~~~~~&1&2&+&~3&15&~0.039&1&1&-&~3&13&~0.039&~4.2300&~0.000&~0.000\\
66563.594&0.002&~~~~~&1&2&+&~4&14&~0.055&1&1&-&~3&13&~0.039&~5.5000&-0.001&~0.016\\
\hline
\end{longtable*}
\bibliographystyle{aipnum4-1}

\begin{thebibliography}{112}%
\makeatletter
\providecommand \@ifxundefined [1]{%
 \@ifx{#1\undefined}
}%
\providecommand \@ifnum [1]{%
 \ifnum #1\expandafter \@firstoftwo
 \else \expandafter \@secondoftwo
 \fi
}%
\providecommand \@ifx [1]{%
 \ifx #1\expandafter \@firstoftwo
 \else \expandafter \@secondoftwo
 \fi
}%
\providecommand \natexlab [1]{#1}%
\providecommand \enquote  [1]{``#1''}%
\providecommand \bibnamefont  [1]{#1}%
\providecommand \bibfnamefont [1]{#1}%
\providecommand \citenamefont [1]{#1}%
\providecommand \href@noop [0]{\@secondoftwo}%
\providecommand \href [0]{\begingroup \@sanitize@url \@href}%
\providecommand \@href[1]{\@@startlink{#1}\@@href}%
\providecommand \@@href[1]{\endgroup#1\@@endlink}%
\providecommand \@sanitize@url [0]{\catcode `\\12\catcode `\$12\catcode
  `\&12\catcode `\#12\catcode `\^12\catcode `\_12\catcode `\%12\relax}%
\providecommand \@@startlink[1]{}%
\providecommand \@@endlink[0]{}%
\providecommand \url  [0]{\begingroup\@sanitize@url \@url }%
\providecommand \@url [1]{\endgroup\@href {#1}{\urlprefix }}%
\providecommand \urlprefix  [0]{URL }%
\providecommand \Eprint [0]{\href }%
\providecommand \doibase [0]{http://dx.doi.org/}%
\providecommand \selectlanguage [0]{\@gobble}%
\providecommand \bibinfo  [0]{\@secondoftwo}%
\providecommand \bibfield  [0]{\@secondoftwo}%
\providecommand \translation [1]{[#1]}%
\providecommand \BibitemOpen [0]{}%
\providecommand \bibitemStop [0]{}%
\providecommand \bibitemNoStop [0]{.\EOS\space}%
\providecommand \EOS [0]{\spacefactor3000\relax}%
\providecommand \BibitemShut  [1]{\csname bibitem#1\endcsname}%
\let\auto@bib@innerbib\@empty
\bibitem [{\citenamefont {Carr}\ \emph {et~al.}(2009)\citenamefont {Carr},
  \citenamefont {DeMille}, \citenamefont {Krems},\ and\ \citenamefont
  {Ye}}]{Carr2009}%
  \BibitemOpen
  \bibfield  {author} {\bibinfo {author} {\bibfnamefont {L.~D.}\ \bibnamefont
  {Carr}}, \bibinfo {author} {\bibfnamefont {D.}~\bibnamefont {DeMille}},
  \bibinfo {author} {\bibfnamefont {R.~V.}\ \bibnamefont {Krems}}, \ and\
  \bibinfo {author} {\bibfnamefont {J.}~\bibnamefont {Ye}},\ }\href {\doibase
  10.1088/1367-2630/11/5/055049} {\bibfield  {journal} {\bibinfo  {journal}
  {New Journal of Physics}\ }\textbf {\bibinfo {volume} {11}} (\bibinfo {year}
  {2009}),\ 10.1088/1367-2630/11/5/055049},\ \Eprint
  {http://arxiv.org/abs/0904.3175} {arXiv:0904.3175} \BibitemShut {NoStop}%
\bibitem [{\citenamefont {Bohn}, \citenamefont {Rey},\ and\ \citenamefont
  {Ye}(2017)}]{Bohn2017}%
  \BibitemOpen
  \bibfield  {author} {\bibinfo {author} {\bibfnamefont {J.~L.}\ \bibnamefont
  {Bohn}}, \bibinfo {author} {\bibfnamefont {A.~M.}\ \bibnamefont {Rey}}, \
  and\ \bibinfo {author} {\bibfnamefont {J.}~\bibnamefont {Ye}},\ }\href
  {\doibase 10.1126/science.aam6299} {\bibfield  {journal} {\bibinfo  {journal}
  {Science}\ }\textbf {\bibinfo {volume} {357}},\ \bibinfo {pages} {1002}
  (\bibinfo {year} {2017})},\ \Eprint {http://arxiv.org/abs/1708.02806}
  {arXiv:1708.02806} \BibitemShut {NoStop}%
\bibitem [{\citenamefont {Krems}(2008)}]{Krems2008}%
  \BibitemOpen
  \bibfield  {author} {\bibinfo {author} {\bibfnamefont {R.~V.}\ \bibnamefont
  {Krems}},\ }\href {\doibase 10.1039/b802322k} {\bibfield  {journal} {\bibinfo
   {journal} {Physical Chemistry Chemical Physics}\ }\textbf {\bibinfo {volume}
  {10}},\ \bibinfo {pages} {4079} (\bibinfo {year} {2008})}\BibitemShut
  {NoStop}%
\bibitem [{\citenamefont {Balakrishnan}(2016)}]{Balakrishnan2016}%
  \BibitemOpen
  \bibfield  {author} {\bibinfo {author} {\bibfnamefont {N.}~\bibnamefont
  {Balakrishnan}},\ }\href {\doibase 10.1063/1.4964096} {\bibfield  {journal}
  {\bibinfo  {journal} {The Journal of Chemical Physics}\ }\textbf {\bibinfo
  {volume} {145}},\ \bibinfo {pages} {150901} (\bibinfo {year}
  {2016})}\BibitemShut {NoStop}%
\bibitem [{\citenamefont {DeMille}, \citenamefont {Doyle},\ and\ \citenamefont
  {Sushkov}(2017)}]{DeMille2017}%
  \BibitemOpen
  \bibfield  {author} {\bibinfo {author} {\bibfnamefont {D.}~\bibnamefont
  {DeMille}}, \bibinfo {author} {\bibfnamefont {J.~M.}\ \bibnamefont {Doyle}},
  \ and\ \bibinfo {author} {\bibfnamefont {A.~O.}\ \bibnamefont {Sushkov}},\
  }\href {\doibase 10.1126/science.aal3003} {\bibfield  {journal} {\bibinfo
  {journal} {Science}\ }\textbf {\bibinfo {volume} {357}},\ \bibinfo {pages}
  {990} (\bibinfo {year} {2017})},\ \Eprint {http://arxiv.org/abs/1704.07928}
  {arXiv:1704.07928} \BibitemShut {NoStop}%
\bibitem [{\citenamefont {Safronova}\ \emph {et~al.}(2018)\citenamefont
  {Safronova}, \citenamefont {Budker}, \citenamefont {DeMille}, \citenamefont
  {Kimball}, \citenamefont {Derevianko},\ and\ \citenamefont
  {Clark}}]{Safronova2018}%
  \BibitemOpen
  \bibfield  {author} {\bibinfo {author} {\bibfnamefont {M.~S.}\ \bibnamefont
  {Safronova}}, \bibinfo {author} {\bibfnamefont {D.}~\bibnamefont {Budker}},
  \bibinfo {author} {\bibfnamefont {D.}~\bibnamefont {DeMille}}, \bibinfo
  {author} {\bibfnamefont {D.~F.~J.}\ \bibnamefont {Kimball}}, \bibinfo
  {author} {\bibfnamefont {A.}~\bibnamefont {Derevianko}}, \ and\ \bibinfo
  {author} {\bibfnamefont {C.~W.}\ \bibnamefont {Clark}},\ }\href {\doibase
  10.1103/RevModPhys.90.025008} {\bibfield  {journal} {\bibinfo  {journal}
  {Reviews of Modern Physics}\ }\textbf {\bibinfo {volume} {90}},\ \bibinfo
  {pages} {025008} (\bibinfo {year} {2018})},\ \Eprint
  {http://arxiv.org/abs/1710.01833} {arXiv:1710.01833} \BibitemShut {NoStop}%
\bibitem [{\citenamefont {DeMille}(2002)}]{DeMille2002}%
  \BibitemOpen
  \bibfield  {author} {\bibinfo {author} {\bibfnamefont {D.}~\bibnamefont
  {DeMille}},\ }\href {\doibase 10.1103/PhysRevLett.88.067901} {\bibfield
  {journal} {\bibinfo  {journal} {Physical Review Letters}\ }\textbf {\bibinfo
  {volume} {88}},\ \bibinfo {pages} {067901} (\bibinfo {year} {2002})},\
  \Eprint {http://arxiv.org/abs/0109083} {arXiv:0109083 [quant-ph]}
  \BibitemShut {NoStop}%
\bibitem [{\citenamefont {Moses}\ \emph {et~al.}(2016)\citenamefont {Moses},
  \citenamefont {Covey}, \citenamefont {Miecnikowski}, \citenamefont {Jin},\
  and\ \citenamefont {Ye}}]{Moses2016}%
  \BibitemOpen
  \bibfield  {author} {\bibinfo {author} {\bibfnamefont {S.~A.}\ \bibnamefont
  {Moses}}, \bibinfo {author} {\bibfnamefont {J.~P.}\ \bibnamefont {Covey}},
  \bibinfo {author} {\bibfnamefont {M.~T.}\ \bibnamefont {Miecnikowski}},
  \bibinfo {author} {\bibfnamefont {D.~S.}\ \bibnamefont {Jin}}, \ and\
  \bibinfo {author} {\bibfnamefont {J.}~\bibnamefont {Ye}},\ }\href {\doibase
  10.1038/nphys3985} {\bibfield  {journal} {\bibinfo  {journal} {Nature
  Physics}\ }\textbf {\bibinfo {volume} {13}},\ \bibinfo {pages} {13} (\bibinfo
  {year} {2016})},\ \Eprint {http://arxiv.org/abs/1610.07711}
  {arXiv:1610.07711} \BibitemShut {NoStop}%
\bibitem [{\citenamefont {Blackmore}\ \emph {et~al.}(2018)\citenamefont
  {Blackmore}, \citenamefont {Caldwell}, \citenamefont {Gregory}, \citenamefont
  {Bridge}, \citenamefont {Sawant}, \citenamefont {Aldegunde}, \citenamefont
  {Mur-Petit}, \citenamefont {Jaksch}, \citenamefont {Hutson}, \citenamefont
  {Sauer}, \citenamefont {Tarbutt},\ and\ \citenamefont
  {Cornish}}]{Blackmore2018}%
  \BibitemOpen
  \bibfield  {author} {\bibinfo {author} {\bibfnamefont {J.~A.}\ \bibnamefont
  {Blackmore}}, \bibinfo {author} {\bibfnamefont {L.}~\bibnamefont {Caldwell}},
  \bibinfo {author} {\bibfnamefont {P.~D.}\ \bibnamefont {Gregory}}, \bibinfo
  {author} {\bibfnamefont {E.~M.}\ \bibnamefont {Bridge}}, \bibinfo {author}
  {\bibfnamefont {R.}~\bibnamefont {Sawant}}, \bibinfo {author} {\bibfnamefont
  {J.}~\bibnamefont {Aldegunde}}, \bibinfo {author} {\bibfnamefont
  {J.}~\bibnamefont {Mur-Petit}}, \bibinfo {author} {\bibfnamefont
  {D.}~\bibnamefont {Jaksch}}, \bibinfo {author} {\bibfnamefont {J.~M.}\
  \bibnamefont {Hutson}}, \bibinfo {author} {\bibfnamefont {B.~E.}\
  \bibnamefont {Sauer}}, \bibinfo {author} {\bibfnamefont {M.~R.}\ \bibnamefont
  {Tarbutt}}, \ and\ \bibinfo {author} {\bibfnamefont {S.~L.}\ \bibnamefont
  {Cornish}},\ }\href {\doibase 10.1088/2058-9565/aaee35} {\bibfield  {journal}
  {\bibinfo  {journal} {Quantum Science and Technology}\ }\textbf {\bibinfo
  {volume} {4}},\ \bibinfo {pages} {014010} (\bibinfo {year} {2018})},\ \Eprint
  {http://arxiv.org/abs/1804.02372} {arXiv:1804.02372} \BibitemShut {NoStop}%
\bibitem [{\citenamefont {Ni}, \citenamefont {Rosenband},\ and\ \citenamefont
  {Grimes}(2018)}]{Ni2018}%
  \BibitemOpen
  \bibfield  {author} {\bibinfo {author} {\bibfnamefont {K.-K.}\ \bibnamefont
  {Ni}}, \bibinfo {author} {\bibfnamefont {T.}~\bibnamefont {Rosenband}}, \
  and\ \bibinfo {author} {\bibfnamefont {D.~D.}\ \bibnamefont {Grimes}},\
  }\href {\doibase 10.1039/C8SC02355G} {\bibfield  {journal} {\bibinfo
  {journal} {Chemical Science}\ }\textbf {\bibinfo {volume} {9}},\ \bibinfo
  {pages} {6830} (\bibinfo {year} {2018})}\BibitemShut {NoStop}%
\bibitem [{\citenamefont {Sage}\ \emph {et~al.}(2005)\citenamefont {Sage},
  \citenamefont {Sainis}, \citenamefont {Bergeman},\ and\ \citenamefont
  {DeMille}}]{Sage2005}%
  \BibitemOpen
  \bibfield  {author} {\bibinfo {author} {\bibfnamefont {J.~M.}\ \bibnamefont
  {Sage}}, \bibinfo {author} {\bibfnamefont {S.}~\bibnamefont {Sainis}},
  \bibinfo {author} {\bibfnamefont {T.}~\bibnamefont {Bergeman}}, \ and\
  \bibinfo {author} {\bibfnamefont {D.}~\bibnamefont {DeMille}},\ }\href
  {\doibase 10.1103/PhysRevLett.94.203001} {\bibfield  {journal} {\bibinfo
  {journal} {Physical Review Letters}\ }\textbf {\bibinfo {volume} {94}},\
  \bibinfo {pages} {203001} (\bibinfo {year} {2005})},\ \Eprint
  {http://arxiv.org/abs/0501008} {arXiv:0501008 [physics]} \BibitemShut
  {NoStop}%
\bibitem [{\citenamefont {Ni}\ \emph {et~al.}(2008)\citenamefont {Ni},
  \citenamefont {Ospelkaus}, \citenamefont {de~Miranda}, \citenamefont {Pe'er},
  \citenamefont {Neyenhuis}, \citenamefont {Zirbel}, \citenamefont
  {Kotochigova}, \citenamefont {Julienne}, \citenamefont {Jin},\ and\
  \citenamefont {Ye}}]{Ni2008}%
  \BibitemOpen
  \bibfield  {author} {\bibinfo {author} {\bibfnamefont {K.-K.}\ \bibnamefont
  {Ni}}, \bibinfo {author} {\bibfnamefont {S.}~\bibnamefont {Ospelkaus}},
  \bibinfo {author} {\bibfnamefont {M.~H.~G.}\ \bibnamefont {de~Miranda}},
  \bibinfo {author} {\bibfnamefont {A.}~\bibnamefont {Pe'er}}, \bibinfo
  {author} {\bibfnamefont {B.}~\bibnamefont {Neyenhuis}}, \bibinfo {author}
  {\bibfnamefont {J.~J.}\ \bibnamefont {Zirbel}}, \bibinfo {author}
  {\bibfnamefont {S.}~\bibnamefont {Kotochigova}}, \bibinfo {author}
  {\bibfnamefont {P.~S.}\ \bibnamefont {Julienne}}, \bibinfo {author}
  {\bibfnamefont {D.~S.}\ \bibnamefont {Jin}}, \ and\ \bibinfo {author}
  {\bibfnamefont {J.}~\bibnamefont {Ye}},\ }\href {\doibase
  10.1126/science.1163861} {\bibfield  {journal} {\bibinfo  {journal}
  {Science}\ }\textbf {\bibinfo {volume} {322}},\ \bibinfo {pages} {231}
  (\bibinfo {year} {2008})},\ \Eprint {http://arxiv.org/abs/0808.2963}
  {arXiv:0808.2963} \BibitemShut {NoStop}%
\bibitem [{\citenamefont {Liu}\ \emph {et~al.}(2018)\citenamefont {Liu},
  \citenamefont {Hood}, \citenamefont {Yu}, \citenamefont {Zhang},
  \citenamefont {Hutzler}, \citenamefont {Rosenband},\ and\ \citenamefont
  {Ni}}]{Liu2018}%
  \BibitemOpen
  \bibfield  {author} {\bibinfo {author} {\bibfnamefont {L.~R.}\ \bibnamefont
  {Liu}}, \bibinfo {author} {\bibfnamefont {J.~D.}\ \bibnamefont {Hood}},
  \bibinfo {author} {\bibfnamefont {Y.}~\bibnamefont {Yu}}, \bibinfo {author}
  {\bibfnamefont {J.~T.}\ \bibnamefont {Zhang}}, \bibinfo {author}
  {\bibfnamefont {N.~R.}\ \bibnamefont {Hutzler}}, \bibinfo {author}
  {\bibfnamefont {T.}~\bibnamefont {Rosenband}}, \ and\ \bibinfo {author}
  {\bibfnamefont {K.-K.}\ \bibnamefont {Ni}},\ }\href {\doibase
  10.1126/science.aar7797} {\bibfield  {journal} {\bibinfo  {journal}
  {Science}\ }\textbf {\bibinfo {volume} {360}},\ \bibinfo {pages} {900}
  (\bibinfo {year} {2018})},\ \Eprint {http://arxiv.org/abs/1804.04752}
  {arXiv:1804.04752} \BibitemShut {NoStop}%
\bibitem [{\citenamefont {Weinstein}\ \emph {et~al.}(1998)\citenamefont
  {Weinstein}, \citenamefont {DeCarvalho}, \citenamefont {Guillet},
  \citenamefont {Friedrich},\ and\ \citenamefont {Doyle}}]{Weinstein1998}%
  \BibitemOpen
  \bibfield  {author} {\bibinfo {author} {\bibfnamefont {J.~D.}\ \bibnamefont
  {Weinstein}}, \bibinfo {author} {\bibfnamefont {R.}~\bibnamefont
  {DeCarvalho}}, \bibinfo {author} {\bibfnamefont {T.}~\bibnamefont {Guillet}},
  \bibinfo {author} {\bibfnamefont {B.}~\bibnamefont {Friedrich}}, \ and\
  \bibinfo {author} {\bibfnamefont {J.~M.}\ \bibnamefont {Doyle}},\ }\href
  {\doibase 10.1038/25949} {\bibfield  {journal} {\bibinfo  {journal} {Nature}\
  }\textbf {\bibinfo {volume} {395}},\ \bibinfo {pages} {148} (\bibinfo {year}
  {1998})},\ \Eprint {http://arxiv.org/abs/9810036} {arXiv:9810036 [physics]}
  \BibitemShut {NoStop}%
\bibitem [{\citenamefont {Bethlem}, \citenamefont {Berden},\ and\ \citenamefont
  {Meijer}(1999)}]{Bethlem1999}%
  \BibitemOpen
  \bibfield  {author} {\bibinfo {author} {\bibfnamefont {H.~L.}\ \bibnamefont
  {Bethlem}}, \bibinfo {author} {\bibfnamefont {G.}~\bibnamefont {Berden}}, \
  and\ \bibinfo {author} {\bibfnamefont {G.}~\bibnamefont {Meijer}},\ }\href
  {\doibase 10.1103/PhysRevLett.83.1558} {\bibfield  {journal} {\bibinfo
  {journal} {Physical Review Letters}\ }\textbf {\bibinfo {volume} {83}},\
  \bibinfo {pages} {1558} (\bibinfo {year} {1999})}\BibitemShut {NoStop}%
\bibitem [{\citenamefont {Vanhaecke}\ \emph {et~al.}(2007)\citenamefont
  {Vanhaecke}, \citenamefont {Meier}, \citenamefont {Andrist}, \citenamefont
  {Meier},\ and\ \citenamefont {Merkt}}]{Vanhaecke2007a}%
  \BibitemOpen
  \bibfield  {author} {\bibinfo {author} {\bibfnamefont {N.}~\bibnamefont
  {Vanhaecke}}, \bibinfo {author} {\bibfnamefont {U.}~\bibnamefont {Meier}},
  \bibinfo {author} {\bibfnamefont {M.}~\bibnamefont {Andrist}}, \bibinfo
  {author} {\bibfnamefont {B.~H.}\ \bibnamefont {Meier}}, \ and\ \bibinfo
  {author} {\bibfnamefont {F.}~\bibnamefont {Merkt}},\ }\href {\doibase
  10.1103/PhysRevA.75.031402} {\bibfield  {journal} {\bibinfo  {journal}
  {Physical Review A}\ }\textbf {\bibinfo {volume} {75}},\ \bibinfo {pages}
  {031402} (\bibinfo {year} {2007})}\BibitemShut {NoStop}%
\bibitem [{\citenamefont {Narevicius}\ \emph {et~al.}(2008)\citenamefont
  {Narevicius}, \citenamefont {Libson}, \citenamefont {Parthey}, \citenamefont
  {Chavez}, \citenamefont {Narevicius}, \citenamefont {Even},\ and\
  \citenamefont {Raizen}}]{Narevicius2008}%
  \BibitemOpen
  \bibfield  {author} {\bibinfo {author} {\bibfnamefont {E.}~\bibnamefont
  {Narevicius}}, \bibinfo {author} {\bibfnamefont {A.}~\bibnamefont {Libson}},
  \bibinfo {author} {\bibfnamefont {C.~G.}\ \bibnamefont {Parthey}}, \bibinfo
  {author} {\bibfnamefont {I.}~\bibnamefont {Chavez}}, \bibinfo {author}
  {\bibfnamefont {J.}~\bibnamefont {Narevicius}}, \bibinfo {author}
  {\bibfnamefont {U.}~\bibnamefont {Even}}, \ and\ \bibinfo {author}
  {\bibfnamefont {M.~G.}\ \bibnamefont {Raizen}},\ }\href {\doibase
  10.1103/PhysRevA.77.051401} {\bibfield  {journal} {\bibinfo  {journal}
  {Physical Review A}\ }\textbf {\bibinfo {volume} {77}},\ \bibinfo {pages} {1}
  (\bibinfo {year} {2008})},\ \Eprint {http://arxiv.org/abs/0804.0219}
  {arXiv:0804.0219} \BibitemShut {NoStop}%
\bibitem [{\citenamefont {Zeppenfeld}\ \emph {et~al.}(2012)\citenamefont
  {Zeppenfeld}, \citenamefont {Englert}, \citenamefont {Gl{\"{o}}ckner},
  \citenamefont {Prehn}, \citenamefont {Mielenz}, \citenamefont {Sommer},
  \citenamefont {van Buuren}, \citenamefont {Motsch},\ and\ \citenamefont
  {Rempe}}]{Zeppenfeld2012}%
  \BibitemOpen
  \bibfield  {author} {\bibinfo {author} {\bibfnamefont {M.}~\bibnamefont
  {Zeppenfeld}}, \bibinfo {author} {\bibfnamefont {B.~G.~U.}\ \bibnamefont
  {Englert}}, \bibinfo {author} {\bibfnamefont {R.}~\bibnamefont
  {Gl{\"{o}}ckner}}, \bibinfo {author} {\bibfnamefont {A.}~\bibnamefont
  {Prehn}}, \bibinfo {author} {\bibfnamefont {M.}~\bibnamefont {Mielenz}},
  \bibinfo {author} {\bibfnamefont {C.}~\bibnamefont {Sommer}}, \bibinfo
  {author} {\bibfnamefont {L.~D.}\ \bibnamefont {van Buuren}}, \bibinfo
  {author} {\bibfnamefont {M.}~\bibnamefont {Motsch}}, \ and\ \bibinfo {author}
  {\bibfnamefont {G.}~\bibnamefont {Rempe}},\ }\href {\doibase
  10.1038/nature11595} {\bibfield  {journal} {\bibinfo  {journal} {Nature}\
  }\textbf {\bibinfo {volume} {491}},\ \bibinfo {pages} {570} (\bibinfo {year}
  {2012})},\ \Eprint {http://arxiv.org/abs/1208.0046} {arXiv:1208.0046
  [physics.atom-ph]} \BibitemShut {NoStop}%
\bibitem [{\citenamefont {Shuman}, \citenamefont {Barry},\ and\ \citenamefont
  {DeMille}(2010)}]{Shuman2010}%
  \BibitemOpen
  \bibfield  {author} {\bibinfo {author} {\bibfnamefont {E.~S.}\ \bibnamefont
  {Shuman}}, \bibinfo {author} {\bibfnamefont {J.~F.}\ \bibnamefont {Barry}}, \
  and\ \bibinfo {author} {\bibfnamefont {D.}~\bibnamefont {DeMille}},\ }\href
  {\doibase 10.1038/nature09443} {\bibfield  {journal} {\bibinfo  {journal}
  {Nature}\ }\textbf {\bibinfo {volume} {467}},\ \bibinfo {pages} {820}
  (\bibinfo {year} {2010})},\ \Eprint {http://arxiv.org/abs/1103.6004}
  {arXiv:1103.6004} \BibitemShut {NoStop}%
\bibitem [{\citenamefont {Hummon}\ \emph {et~al.}(2013)\citenamefont {Hummon},
  \citenamefont {Yeo}, \citenamefont {Stuhl}, \citenamefont {Collopy},
  \citenamefont {Xia},\ and\ \citenamefont {Ye}}]{Hummon2013}%
  \BibitemOpen
  \bibfield  {author} {\bibinfo {author} {\bibfnamefont {M.~T.}\ \bibnamefont
  {Hummon}}, \bibinfo {author} {\bibfnamefont {M.}~\bibnamefont {Yeo}},
  \bibinfo {author} {\bibfnamefont {B.~K.}\ \bibnamefont {Stuhl}}, \bibinfo
  {author} {\bibfnamefont {A.~L.}\ \bibnamefont {Collopy}}, \bibinfo {author}
  {\bibfnamefont {Y.}~\bibnamefont {Xia}}, \ and\ \bibinfo {author}
  {\bibfnamefont {J.}~\bibnamefont {Ye}},\ }\href {\doibase
  10.1103/PhysRevLett.110.143001} {\bibfield  {journal} {\bibinfo  {journal}
  {Physical Review Letters}\ }\textbf {\bibinfo {volume} {110}},\ \bibinfo
  {pages} {1} (\bibinfo {year} {2013})},\ \Eprint
  {http://arxiv.org/abs/1209.4069} {arXiv:1209.4069} \BibitemShut {NoStop}%
\bibitem [{\citenamefont {Zhelyazkova}\ \emph {et~al.}(2014)\citenamefont
  {Zhelyazkova}, \citenamefont {Cournol}, \citenamefont {Wall}, \citenamefont
  {Matsushima}, \citenamefont {Hudson}, \citenamefont {Hinds}, \citenamefont
  {Tarbutt},\ and\ \citenamefont {Sauer}}]{Zhelyazkova2014}%
  \BibitemOpen
  \bibfield  {author} {\bibinfo {author} {\bibfnamefont {V.}~\bibnamefont
  {Zhelyazkova}}, \bibinfo {author} {\bibfnamefont {A.}~\bibnamefont
  {Cournol}}, \bibinfo {author} {\bibfnamefont {T.~E.}\ \bibnamefont {Wall}},
  \bibinfo {author} {\bibfnamefont {A.}~\bibnamefont {Matsushima}}, \bibinfo
  {author} {\bibfnamefont {J.~J.}\ \bibnamefont {Hudson}}, \bibinfo {author}
  {\bibfnamefont {E.~A.}\ \bibnamefont {Hinds}}, \bibinfo {author}
  {\bibfnamefont {M.~R.}\ \bibnamefont {Tarbutt}}, \ and\ \bibinfo {author}
  {\bibfnamefont {B.~E.}\ \bibnamefont {Sauer}},\ }\href {\doibase
  10.1103/PhysRevA.89.053416} {\bibfield  {journal} {\bibinfo  {journal}
  {Physical Review A}\ }\textbf {\bibinfo {volume} {89}},\ \bibinfo {pages}
  {053416} (\bibinfo {year} {2014})},\ \Eprint {http://arxiv.org/abs/1308.0421}
  {arXiv:1308.0421} \BibitemShut {NoStop}%
\bibitem [{\citenamefont {Lim}\ \emph {et~al.}(2018)\citenamefont {Lim},
  \citenamefont {Almond}, \citenamefont {Trigatzis}, \citenamefont {Devlin},
  \citenamefont {Fitch}, \citenamefont {Sauer}, \citenamefont {Tarbutt},\ and\
  \citenamefont {Hinds}}]{Lim2018}%
  \BibitemOpen
  \bibfield  {author} {\bibinfo {author} {\bibfnamefont {J.}~\bibnamefont
  {Lim}}, \bibinfo {author} {\bibfnamefont {J.~R.}\ \bibnamefont {Almond}},
  \bibinfo {author} {\bibfnamefont {M.~A.}\ \bibnamefont {Trigatzis}}, \bibinfo
  {author} {\bibfnamefont {J.~A.}\ \bibnamefont {Devlin}}, \bibinfo {author}
  {\bibfnamefont {N.~J.}\ \bibnamefont {Fitch}}, \bibinfo {author}
  {\bibfnamefont {B.~E.}\ \bibnamefont {Sauer}}, \bibinfo {author}
  {\bibfnamefont {M.~R.}\ \bibnamefont {Tarbutt}}, \ and\ \bibinfo {author}
  {\bibfnamefont {E.~A.}\ \bibnamefont {Hinds}},\ }\href {\doibase
  10.1103/PhysRevLett.120.123201} {\bibfield  {journal} {\bibinfo  {journal}
  {Physical Review Letters}\ }\textbf {\bibinfo {volume} {120}},\ \bibinfo
  {pages} {123201} (\bibinfo {year} {2018})},\ \Eprint
  {http://arxiv.org/abs/1712.02868} {arXiv:1712.02868} \BibitemShut {NoStop}%
\bibitem [{\citenamefont {Kozyryev}\ \emph {et~al.}(2017)\citenamefont
  {Kozyryev}, \citenamefont {Baum}, \citenamefont {Matsuda}, \citenamefont
  {Augenbraun}, \citenamefont {Anderegg}, \citenamefont {Sedlack},\ and\
  \citenamefont {Doyle}}]{Kozyryev2017}%
  \BibitemOpen
  \bibfield  {author} {\bibinfo {author} {\bibfnamefont {I.}~\bibnamefont
  {Kozyryev}}, \bibinfo {author} {\bibfnamefont {L.}~\bibnamefont {Baum}},
  \bibinfo {author} {\bibfnamefont {K.}~\bibnamefont {Matsuda}}, \bibinfo
  {author} {\bibfnamefont {B.~L.}\ \bibnamefont {Augenbraun}}, \bibinfo
  {author} {\bibfnamefont {L.}~\bibnamefont {Anderegg}}, \bibinfo {author}
  {\bibfnamefont {A.~P.}\ \bibnamefont {Sedlack}}, \ and\ \bibinfo {author}
  {\bibfnamefont {J.~M.}\ \bibnamefont {Doyle}},\ }\href {\doibase
  10.1103/PhysRevLett.118.173201} {\bibfield  {journal} {\bibinfo  {journal}
  {Physical Review Letters}\ }\textbf {\bibinfo {volume} {118}},\ \bibinfo
  {pages} {173201} (\bibinfo {year} {2017})},\ \Eprint
  {http://arxiv.org/abs/1609.02254} {arXiv:1609.02254} \BibitemShut {NoStop}%
\bibitem [{\citenamefont {Barry}\ \emph {et~al.}(2014)\citenamefont {Barry},
  \citenamefont {McCarron}, \citenamefont {Norrgard}, \citenamefont
  {Steinecker},\ and\ \citenamefont {DeMille}}]{Barry2014}%
  \BibitemOpen
  \bibfield  {author} {\bibinfo {author} {\bibfnamefont {J.~F.}\ \bibnamefont
  {Barry}}, \bibinfo {author} {\bibfnamefont {D.~J.}\ \bibnamefont {McCarron}},
  \bibinfo {author} {\bibfnamefont {E.~B.}\ \bibnamefont {Norrgard}}, \bibinfo
  {author} {\bibfnamefont {M.~H.}\ \bibnamefont {Steinecker}}, \ and\ \bibinfo
  {author} {\bibfnamefont {D.}~\bibnamefont {DeMille}},\ }\href {\doibase
  10.1038/nature13634} {\bibfield  {journal} {\bibinfo  {journal} {Nature}\
  }\textbf {\bibinfo {volume} {512}},\ \bibinfo {pages} {286} (\bibinfo {year}
  {2014})},\ \Eprint {http://arxiv.org/abs/1404.5680} {arXiv:1404.5680}
  \BibitemShut {NoStop}%
\bibitem [{\citenamefont {Williams}\ \emph {et~al.}(2017)\citenamefont
  {Williams}, \citenamefont {Truppe}, \citenamefont {Hambach}, \citenamefont
  {Caldwell}, \citenamefont {Fitch}, \citenamefont {Hinds}, \citenamefont
  {Sauer},\ and\ \citenamefont {Tarbutt}}]{Williams2017}%
  \BibitemOpen
  \bibfield  {author} {\bibinfo {author} {\bibfnamefont {H.~J.}\ \bibnamefont
  {Williams}}, \bibinfo {author} {\bibfnamefont {S.}~\bibnamefont {Truppe}},
  \bibinfo {author} {\bibfnamefont {M.}~\bibnamefont {Hambach}}, \bibinfo
  {author} {\bibfnamefont {L.}~\bibnamefont {Caldwell}}, \bibinfo {author}
  {\bibfnamefont {N.~J.}\ \bibnamefont {Fitch}}, \bibinfo {author}
  {\bibfnamefont {E.~A.}\ \bibnamefont {Hinds}}, \bibinfo {author}
  {\bibfnamefont {B.~E.}\ \bibnamefont {Sauer}}, \ and\ \bibinfo {author}
  {\bibfnamefont {M.~R.}\ \bibnamefont {Tarbutt}},\ }\href {\doibase
  10.1088/1367-2630/aa8e52} {\bibfield  {journal} {\bibinfo  {journal} {New
  Journal of Physics}\ }\textbf {\bibinfo {volume} {19}},\ \bibinfo {pages}
  {113035} (\bibinfo {year} {2017})},\ \Eprint
  {http://arxiv.org/abs/1706.07848} {arXiv:1706.07848} \BibitemShut {NoStop}%
\bibitem [{\citenamefont {Anderegg}\ \emph {et~al.}(2017)\citenamefont
  {Anderegg}, \citenamefont {Augenbraun}, \citenamefont {Chae}, \citenamefont
  {Hemmerling}, \citenamefont {Hutzler}, \citenamefont {Ravi}, \citenamefont
  {Collopy}, \citenamefont {Ye}, \citenamefont {Ketterle},\ and\ \citenamefont
  {Doyle}}]{Anderegg2017}%
  \BibitemOpen
  \bibfield  {author} {\bibinfo {author} {\bibfnamefont {L.}~\bibnamefont
  {Anderegg}}, \bibinfo {author} {\bibfnamefont {B.~L.}\ \bibnamefont
  {Augenbraun}}, \bibinfo {author} {\bibfnamefont {E.}~\bibnamefont {Chae}},
  \bibinfo {author} {\bibfnamefont {B.}~\bibnamefont {Hemmerling}}, \bibinfo
  {author} {\bibfnamefont {N.~R.}\ \bibnamefont {Hutzler}}, \bibinfo {author}
  {\bibfnamefont {A.}~\bibnamefont {Ravi}}, \bibinfo {author} {\bibfnamefont
  {A.}~\bibnamefont {Collopy}}, \bibinfo {author} {\bibfnamefont
  {J.}~\bibnamefont {Ye}}, \bibinfo {author} {\bibfnamefont {W.}~\bibnamefont
  {Ketterle}}, \ and\ \bibinfo {author} {\bibfnamefont {J.~M.}\ \bibnamefont
  {Doyle}},\ }\href {\doibase 10.1103/PhysRevLett.119.103201} {\bibfield
  {journal} {\bibinfo  {journal} {Physical Review Letters}\ }\textbf {\bibinfo
  {volume} {119}},\ \bibinfo {pages} {103201} (\bibinfo {year} {2017})},\
  \Eprint {http://arxiv.org/abs/1705.10288} {arXiv:1705.10288} \BibitemShut
  {NoStop}%
\bibitem [{\citenamefont {Collopy}\ \emph {et~al.}(2018)\citenamefont
  {Collopy}, \citenamefont {Ding}, \citenamefont {Wu}, \citenamefont
  {Finneran}, \citenamefont {Anderegg}, \citenamefont {Augenbraun},
  \citenamefont {Doyle},\ and\ \citenamefont {Ye}}]{Collopy2018}%
  \BibitemOpen
  \bibfield  {author} {\bibinfo {author} {\bibfnamefont {A.~L.}\ \bibnamefont
  {Collopy}}, \bibinfo {author} {\bibfnamefont {S.}~\bibnamefont {Ding}},
  \bibinfo {author} {\bibfnamefont {Y.}~\bibnamefont {Wu}}, \bibinfo {author}
  {\bibfnamefont {I.~A.}\ \bibnamefont {Finneran}}, \bibinfo {author}
  {\bibfnamefont {L.}~\bibnamefont {Anderegg}}, \bibinfo {author}
  {\bibfnamefont {B.~L.}\ \bibnamefont {Augenbraun}}, \bibinfo {author}
  {\bibfnamefont {J.~M.}\ \bibnamefont {Doyle}}, \ and\ \bibinfo {author}
  {\bibfnamefont {J.}~\bibnamefont {Ye}},\ }\href {\doibase
  10.1103/PhysRevLett.121.213201} {\bibfield  {journal} {\bibinfo  {journal}
  {Physical Review Letters}\ }\textbf {\bibinfo {volume} {121}},\ \bibinfo
  {pages} {213201} (\bibinfo {year} {2018})}\BibitemShut {NoStop}%
\bibitem [{\citenamefont {Truppe}\ \emph {et~al.}(2017)\citenamefont {Truppe},
  \citenamefont {Williams}, \citenamefont {Hambach}, \citenamefont {Caldwell},
  \citenamefont {Fitch}, \citenamefont {Hinds}, \citenamefont {Sauer},\ and\
  \citenamefont {Tarbutt}}]{Truppe2017}%
  \BibitemOpen
  \bibfield  {author} {\bibinfo {author} {\bibfnamefont {S.}~\bibnamefont
  {Truppe}}, \bibinfo {author} {\bibfnamefont {H.~J.}\ \bibnamefont
  {Williams}}, \bibinfo {author} {\bibfnamefont {M.}~\bibnamefont {Hambach}},
  \bibinfo {author} {\bibfnamefont {L.}~\bibnamefont {Caldwell}}, \bibinfo
  {author} {\bibfnamefont {N.~J.}\ \bibnamefont {Fitch}}, \bibinfo {author}
  {\bibfnamefont {E.~A.}\ \bibnamefont {Hinds}}, \bibinfo {author}
  {\bibfnamefont {B.~E.}\ \bibnamefont {Sauer}}, \ and\ \bibinfo {author}
  {\bibfnamefont {M.~R.}\ \bibnamefont {Tarbutt}},\ }\href {\doibase
  10.1038/nphys4241} {\bibfield  {journal} {\bibinfo  {journal} {Nature
  Physics}\ }\textbf {\bibinfo {volume} {13}},\ \bibinfo {pages} {1173}
  (\bibinfo {year} {2017})},\ \Eprint {http://arxiv.org/abs/1703.00580}
  {arXiv:1703.00580} \BibitemShut {NoStop}%
\bibitem [{\citenamefont {Chen}, \citenamefont {Bu},\ and\ \citenamefont
  {Yan}(2017)}]{Chen2017}%
  \BibitemOpen
  \bibfield  {author} {\bibinfo {author} {\bibfnamefont {T.}~\bibnamefont
  {Chen}}, \bibinfo {author} {\bibfnamefont {W.}~\bibnamefont {Bu}}, \ and\
  \bibinfo {author} {\bibfnamefont {B.}~\bibnamefont {Yan}},\ }\href {\doibase
  10.1103/PhysRevA.96.053401} {\bibfield  {journal} {\bibinfo  {journal}
  {Physical Review A}\ }\textbf {\bibinfo {volume} {96}},\ \bibinfo {pages}
  {053401} (\bibinfo {year} {2017})},\ \Eprint
  {http://arxiv.org/abs/1709.07602} {arXiv:1709.07602} \BibitemShut {NoStop}%
\bibitem [{\citenamefont {{The NL-eEDM collaboration}}\ \emph
  {et~al.}(2018)\citenamefont {{The NL-eEDM collaboration}}, \citenamefont
  {Aggarwal}, \citenamefont {Bethlem}, \citenamefont {Borschevsky},
  \citenamefont {Denis}, \citenamefont {Esajas}, \citenamefont {Haase},
  \citenamefont {Hao}, \citenamefont {Hoekstra}, \citenamefont {Jungmann},
  \citenamefont {Meijknecht}, \citenamefont {Mooij}, \citenamefont
  {Timmermans}, \citenamefont {Ubachs}, \citenamefont {Willmann},\ and\
  \citenamefont {Zapara}}]{TheNL-eEDMcollaboration2018}%
  \BibitemOpen
  \bibfield  {author} {\bibinfo {author} {\bibnamefont {{The NL-eEDM
  collaboration}}}, \bibinfo {author} {\bibfnamefont {P.}~\bibnamefont
  {Aggarwal}}, \bibinfo {author} {\bibfnamefont {H.~L.}\ \bibnamefont
  {Bethlem}}, \bibinfo {author} {\bibfnamefont {A.}~\bibnamefont
  {Borschevsky}}, \bibinfo {author} {\bibfnamefont {M.}~\bibnamefont {Denis}},
  \bibinfo {author} {\bibfnamefont {K.}~\bibnamefont {Esajas}}, \bibinfo
  {author} {\bibfnamefont {P.~A.~B.}\ \bibnamefont {Haase}}, \bibinfo {author}
  {\bibfnamefont {Y.}~\bibnamefont {Hao}}, \bibinfo {author} {\bibfnamefont
  {S.}~\bibnamefont {Hoekstra}}, \bibinfo {author} {\bibfnamefont
  {K.}~\bibnamefont {Jungmann}}, \bibinfo {author} {\bibfnamefont {T.~B.}\
  \bibnamefont {Meijknecht}}, \bibinfo {author} {\bibfnamefont {M.~C.}\
  \bibnamefont {Mooij}}, \bibinfo {author} {\bibfnamefont {R.~G.~E.}\
  \bibnamefont {Timmermans}}, \bibinfo {author} {\bibfnamefont
  {W.}~\bibnamefont {Ubachs}}, \bibinfo {author} {\bibfnamefont
  {L.}~\bibnamefont {Willmann}}, \ and\ \bibinfo {author} {\bibfnamefont
  {A.}~\bibnamefont {Zapara}},\ }\href {\doibase 10.1140/epjd/e2018-90192-9}
  {\bibfield  {journal} {\bibinfo  {journal} {The European Physical Journal D}\
  }\textbf {\bibinfo {volume} {72}},\ \bibinfo {pages} {197} (\bibinfo {year}
  {2018})}\BibitemShut {NoStop}%
\bibitem [{\citenamefont {Albrecht}\ \emph {et~al.}(2019)\citenamefont
  {Albrecht}, \citenamefont {Scharwaechter}, \citenamefont {Sixt},
  \citenamefont {Hofer},\ and\ \citenamefont {Langen}}]{Albrecht2019}%
  \BibitemOpen
  \bibfield  {author} {\bibinfo {author} {\bibfnamefont {R.}~\bibnamefont
  {Albrecht}}, \bibinfo {author} {\bibfnamefont {M.}~\bibnamefont
  {Scharwaechter}}, \bibinfo {author} {\bibfnamefont {T.}~\bibnamefont {Sixt}},
  \bibinfo {author} {\bibfnamefont {L.}~\bibnamefont {Hofer}}, \ and\ \bibinfo
  {author} {\bibfnamefont {T.}~\bibnamefont {Langen}},\ }\href
  {http://arxiv.org/abs/1906.08798} {\  (\bibinfo {year} {2019})},\ \Eprint
  {http://arxiv.org/abs/1906.08798} {arXiv:1906.08798} \BibitemShut {NoStop}%
\bibitem [{\citenamefont {Iwata}, \citenamefont {McNally},\ and\ \citenamefont
  {Zelevinsky}(2017)}]{Iwata2017}%
  \BibitemOpen
  \bibfield  {author} {\bibinfo {author} {\bibfnamefont {G.~Z.}\ \bibnamefont
  {Iwata}}, \bibinfo {author} {\bibfnamefont {R.~L.}\ \bibnamefont {McNally}},
  \ and\ \bibinfo {author} {\bibfnamefont {T.}~\bibnamefont {Zelevinsky}},\
  }\href {\doibase 10.1103/PhysRevA.96.022509} {\bibfield  {journal} {\bibinfo
  {journal} {Physical Review A}\ }\textbf {\bibinfo {volume} {96}},\ \bibinfo
  {pages} {022509} (\bibinfo {year} {2017})},\ \Eprint
  {http://arxiv.org/abs/1705.00113} {arXiv:1705.00113} \BibitemShut {NoStop}%
\bibitem [{\citenamefont {Norrgard}\ \emph {et~al.}(2017)\citenamefont
  {Norrgard}, \citenamefont {Edwards}, \citenamefont {McCarron}, \citenamefont
  {Steinecker}, \citenamefont {DeMille}, \citenamefont {Alam}, \citenamefont
  {Peck}, \citenamefont {Wadia},\ and\ \citenamefont {Hunter}}]{Norrgard2017}%
  \BibitemOpen
  \bibfield  {author} {\bibinfo {author} {\bibfnamefont {E.~B.}\ \bibnamefont
  {Norrgard}}, \bibinfo {author} {\bibfnamefont {E.~R.}\ \bibnamefont
  {Edwards}}, \bibinfo {author} {\bibfnamefont {D.~J.}\ \bibnamefont
  {McCarron}}, \bibinfo {author} {\bibfnamefont {M.~H.}\ \bibnamefont
  {Steinecker}}, \bibinfo {author} {\bibfnamefont {D.}~\bibnamefont {DeMille}},
  \bibinfo {author} {\bibfnamefont {S.~S.}\ \bibnamefont {Alam}}, \bibinfo
  {author} {\bibfnamefont {S.~K.}\ \bibnamefont {Peck}}, \bibinfo {author}
  {\bibfnamefont {N.~S.}\ \bibnamefont {Wadia}}, \ and\ \bibinfo {author}
  {\bibfnamefont {L.~R.}\ \bibnamefont {Hunter}},\ }\href {\doibase
  10.1103/PhysRevA.95.062506} {\bibfield  {journal} {\bibinfo  {journal}
  {Physical Review A}\ }\textbf {\bibinfo {volume} {95}},\ \bibinfo {pages}
  {062506} (\bibinfo {year} {2017})},\ \Eprint
  {http://arxiv.org/abs/1702.02548} {arXiv:1702.02548} \BibitemShut {NoStop}%
\bibitem [{\citenamefont {Xu}\ \emph {et~al.}(2016)\citenamefont {Xu},
  \citenamefont {Yin}, \citenamefont {Wei}, \citenamefont {Xia},\ and\
  \citenamefont {Yin}}]{Xu2016}%
  \BibitemOpen
  \bibfield  {author} {\bibinfo {author} {\bibfnamefont {L.}~\bibnamefont
  {Xu}}, \bibinfo {author} {\bibfnamefont {Y.}~\bibnamefont {Yin}}, \bibinfo
  {author} {\bibfnamefont {B.}~\bibnamefont {Wei}}, \bibinfo {author}
  {\bibfnamefont {Y.}~\bibnamefont {Xia}}, \ and\ \bibinfo {author}
  {\bibfnamefont {J.}~\bibnamefont {Yin}},\ }\href {\doibase
  10.1103/PhysRevA.93.013408} {\bibfield  {journal} {\bibinfo  {journal}
  {Physical Review A}\ }\textbf {\bibinfo {volume} {93}},\ \bibinfo {pages}
  {013408} (\bibinfo {year} {2016})}\BibitemShut {NoStop}%
\bibitem [{\citenamefont {Kozyryev}\ \emph {et~al.}(2019)\citenamefont
  {Kozyryev}, \citenamefont {Steimle}, \citenamefont {Yu}, \citenamefont
  {Nguyen},\ and\ \citenamefont {Doyle}}]{Kozyryev2019}%
  \BibitemOpen
  \bibfield  {author} {\bibinfo {author} {\bibfnamefont {I.}~\bibnamefont
  {Kozyryev}}, \bibinfo {author} {\bibfnamefont {T.~C.}\ \bibnamefont
  {Steimle}}, \bibinfo {author} {\bibfnamefont {P.}~\bibnamefont {Yu}},
  \bibinfo {author} {\bibfnamefont {D.-T.}\ \bibnamefont {Nguyen}}, \ and\
  \bibinfo {author} {\bibfnamefont {J.~M.}\ \bibnamefont {Doyle}},\ }\href
  {\doibase 10.1088/1367-2630/ab19d7} {\bibfield  {journal} {\bibinfo
  {journal} {New Journal of Physics}\ }\textbf {\bibinfo {volume} {21}},\
  \bibinfo {pages} {052002} (\bibinfo {year} {2019})}\BibitemShut {NoStop}%
\bibitem [{\citenamefont {Nakhate}\ \emph {et~al.}(2019)\citenamefont
  {Nakhate}, \citenamefont {Steimle}, \citenamefont {Pilgram},\ and\
  \citenamefont {Hutzler}}]{Nakhate2019}%
  \BibitemOpen
  \bibfield  {author} {\bibinfo {author} {\bibfnamefont {S.}~\bibnamefont
  {Nakhate}}, \bibinfo {author} {\bibfnamefont {T.~C.}\ \bibnamefont
  {Steimle}}, \bibinfo {author} {\bibfnamefont {N.~H.}\ \bibnamefont
  {Pilgram}}, \ and\ \bibinfo {author} {\bibfnamefont {N.~R.}\ \bibnamefont
  {Hutzler}},\ }\href {\doibase 10.1016/j.cplett.2018.11.030} {\bibfield
  {journal} {\bibinfo  {journal} {Chemical Physics Letters}\ }\textbf {\bibinfo
  {volume} {715}},\ \bibinfo {pages} {105} (\bibinfo {year}
  {2019})}\BibitemShut {NoStop}%
\bibitem [{\citenamefont {Williams}\ \emph {et~al.}(2018)\citenamefont
  {Williams}, \citenamefont {Caldwell}, \citenamefont {Fitch}, \citenamefont
  {Truppe}, \citenamefont {Rodewald}, \citenamefont {Hinds}, \citenamefont
  {Sauer},\ and\ \citenamefont {Tarbutt}}]{Williams2018}%
  \BibitemOpen
  \bibfield  {author} {\bibinfo {author} {\bibfnamefont {H.~J.}\ \bibnamefont
  {Williams}}, \bibinfo {author} {\bibfnamefont {L.}~\bibnamefont {Caldwell}},
  \bibinfo {author} {\bibfnamefont {N.~J.}\ \bibnamefont {Fitch}}, \bibinfo
  {author} {\bibfnamefont {S.}~\bibnamefont {Truppe}}, \bibinfo {author}
  {\bibfnamefont {J.}~\bibnamefont {Rodewald}}, \bibinfo {author}
  {\bibfnamefont {E.~A.}\ \bibnamefont {Hinds}}, \bibinfo {author}
  {\bibfnamefont {B.~E.}\ \bibnamefont {Sauer}}, \ and\ \bibinfo {author}
  {\bibfnamefont {M.~R.}\ \bibnamefont {Tarbutt}},\ }\href {\doibase
  10.1103/PhysRevLett.120.163201} {\bibfield  {journal} {\bibinfo  {journal}
  {Physical Review Letters}\ }\textbf {\bibinfo {volume} {120}},\ \bibinfo
  {pages} {163201} (\bibinfo {year} {2018})},\ \Eprint
  {http://arxiv.org/abs/1711.07355} {arXiv:1711.07355} \BibitemShut {NoStop}%
\bibitem [{\citenamefont {McCarron}\ \emph {et~al.}(2018)\citenamefont
  {McCarron}, \citenamefont {Steinecker}, \citenamefont {Zhu},\ and\
  \citenamefont {DeMille}}]{McCarron2018}%
  \BibitemOpen
  \bibfield  {author} {\bibinfo {author} {\bibfnamefont {D.~J.}\ \bibnamefont
  {McCarron}}, \bibinfo {author} {\bibfnamefont {M.~H.}\ \bibnamefont
  {Steinecker}}, \bibinfo {author} {\bibfnamefont {Y.}~\bibnamefont {Zhu}}, \
  and\ \bibinfo {author} {\bibfnamefont {D.}~\bibnamefont {DeMille}},\ }\href
  {\doibase 10.1103/PhysRevLett.121.013202} {\bibfield  {journal} {\bibinfo
  {journal} {Physical Review Letters}\ }\textbf {\bibinfo {volume} {121}},\
  \bibinfo {pages} {013202} (\bibinfo {year} {2018})},\ \Eprint
  {http://arxiv.org/abs/1712.01462} {arXiv:1712.01462} \BibitemShut {NoStop}%
\bibitem [{\citenamefont {Anderegg}\ \emph {et~al.}(2018)\citenamefont
  {Anderegg}, \citenamefont {Augenbraun}, \citenamefont {Bao}, \citenamefont
  {Burchesky}, \citenamefont {Cheuk}, \citenamefont {Ketterle},\ and\
  \citenamefont {Doyle}}]{Anderegg2018}%
  \BibitemOpen
  \bibfield  {author} {\bibinfo {author} {\bibfnamefont {L.}~\bibnamefont
  {Anderegg}}, \bibinfo {author} {\bibfnamefont {B.~L.}\ \bibnamefont
  {Augenbraun}}, \bibinfo {author} {\bibfnamefont {Y.}~\bibnamefont {Bao}},
  \bibinfo {author} {\bibfnamefont {S.}~\bibnamefont {Burchesky}}, \bibinfo
  {author} {\bibfnamefont {L.~W.}\ \bibnamefont {Cheuk}}, \bibinfo {author}
  {\bibfnamefont {W.}~\bibnamefont {Ketterle}}, \ and\ \bibinfo {author}
  {\bibfnamefont {J.~M.}\ \bibnamefont {Doyle}},\ }\href {\doibase
  10.1038/s41567-018-0191-z} {\bibfield  {journal} {\bibinfo  {journal} {Nature
  Physics}\ }\textbf {\bibinfo {volume} {14}},\ \bibinfo {pages} {890}
  (\bibinfo {year} {2018})},\ \Eprint {http://arxiv.org/abs/1803.04571}
  {arXiv:1803.04571} \BibitemShut {NoStop}%
\bibitem [{\citenamefont {Anderegg}\ \emph {et~al.}(2019)\citenamefont
  {Anderegg}, \citenamefont {Cheuk}, \citenamefont {Bao}, \citenamefont
  {Burchesky}, \citenamefont {Ketterle}, \citenamefont {Ni},\ and\
  \citenamefont {Doyle}}]{Anderegg2019}%
  \BibitemOpen
  \bibfield  {author} {\bibinfo {author} {\bibfnamefont {L.}~\bibnamefont
  {Anderegg}}, \bibinfo {author} {\bibfnamefont {L.~W.}\ \bibnamefont {Cheuk}},
  \bibinfo {author} {\bibfnamefont {Y.}~\bibnamefont {Bao}}, \bibinfo {author}
  {\bibfnamefont {S.}~\bibnamefont {Burchesky}}, \bibinfo {author}
  {\bibfnamefont {W.}~\bibnamefont {Ketterle}}, \bibinfo {author}
  {\bibfnamefont {K.-K.}\ \bibnamefont {Ni}}, \ and\ \bibinfo {author}
  {\bibfnamefont {J.~M.}\ \bibnamefont {Doyle}},\ }\href
  {http://arxiv.org/abs/1902.00497} {\ ,\ \bibinfo {pages} {1} (\bibinfo {year}
  {2019})},\ \Eprint {http://arxiv.org/abs/1902.00497} {arXiv:1902.00497}
  \BibitemShut {NoStop}%
\bibitem [{\citenamefont {Lim}\ \emph {et~al.}(2015)\citenamefont {Lim},
  \citenamefont {Frye}, \citenamefont {Hutson},\ and\ \citenamefont
  {Tarbutt}}]{Lim2015}%
  \BibitemOpen
  \bibfield  {author} {\bibinfo {author} {\bibfnamefont {J.}~\bibnamefont
  {Lim}}, \bibinfo {author} {\bibfnamefont {M.~D.}\ \bibnamefont {Frye}},
  \bibinfo {author} {\bibfnamefont {J.~M.}\ \bibnamefont {Hutson}}, \ and\
  \bibinfo {author} {\bibfnamefont {M.~R.}\ \bibnamefont {Tarbutt}},\ }\href
  {\doibase 10.1103/PhysRevA.92.053419} {\bibfield  {journal} {\bibinfo
  {journal} {Physical Review A}\ }\textbf {\bibinfo {volume} {92}},\ \bibinfo
  {pages} {053419} (\bibinfo {year} {2015})},\ \Eprint
  {http://arxiv.org/abs/1508.03987} {arXiv:1508.03987} \BibitemShut {NoStop}%
\bibitem [{\citenamefont {Son}\ \emph {et~al.}(2019)\citenamefont {Son},
  \citenamefont {Park}, \citenamefont {Ketterle},\ and\ \citenamefont
  {Jamison}}]{Son2019}%
  \BibitemOpen
  \bibfield  {author} {\bibinfo {author} {\bibfnamefont {H.}~\bibnamefont
  {Son}}, \bibinfo {author} {\bibfnamefont {J.~J.}\ \bibnamefont {Park}},
  \bibinfo {author} {\bibfnamefont {W.}~\bibnamefont {Ketterle}}, \ and\
  \bibinfo {author} {\bibfnamefont {A.~O.}\ \bibnamefont {Jamison}},\ }\href
  {http://arxiv.org/abs/1907.09630} {\ ,\ \bibinfo {pages} {1} (\bibinfo {year}
  {2019})},\ \Eprint {http://arxiv.org/abs/1907.09630} {arXiv:1907.09630}
  \BibitemShut {NoStop}%
\bibitem [{\citenamefont {Stuhl}\ \emph {et~al.}(2012)\citenamefont {Stuhl},
  \citenamefont {Hummon}, \citenamefont {Yeo}, \citenamefont
  {Qu{\'{e}}m{\'{e}}ner}, \citenamefont {Bohn},\ and\ \citenamefont
  {Ye}}]{Stuhl2012}%
  \BibitemOpen
  \bibfield  {author} {\bibinfo {author} {\bibfnamefont {B.~K.}\ \bibnamefont
  {Stuhl}}, \bibinfo {author} {\bibfnamefont {M.~T.}\ \bibnamefont {Hummon}},
  \bibinfo {author} {\bibfnamefont {M.}~\bibnamefont {Yeo}}, \bibinfo {author}
  {\bibfnamefont {G.}~\bibnamefont {Qu{\'{e}}m{\'{e}}ner}}, \bibinfo {author}
  {\bibfnamefont {J.~L.}\ \bibnamefont {Bohn}}, \ and\ \bibinfo {author}
  {\bibfnamefont {J.}~\bibnamefont {Ye}},\ }\href {\doibase
  10.1038/nature11718} {\bibfield  {journal} {\bibinfo  {journal} {Nature}\
  }\textbf {\bibinfo {volume} {492}},\ \bibinfo {pages} {396} (\bibinfo {year}
  {2012})},\ \Eprint {http://arxiv.org/abs/1209.6343} {arXiv:1209.6343}
  \BibitemShut {NoStop}%
\bibitem [{\citenamefont {{De Marco}}\ \emph {et~al.}(2019)\citenamefont {{De
  Marco}}, \citenamefont {Valtolina}, \citenamefont {Matsuda}, \citenamefont
  {Tobias}, \citenamefont {Covey},\ and\ \citenamefont {Ye}}]{DeMarco2019}%
  \BibitemOpen
  \bibfield  {author} {\bibinfo {author} {\bibfnamefont {L.}~\bibnamefont {{De
  Marco}}}, \bibinfo {author} {\bibfnamefont {G.}~\bibnamefont {Valtolina}},
  \bibinfo {author} {\bibfnamefont {K.}~\bibnamefont {Matsuda}}, \bibinfo
  {author} {\bibfnamefont {W.~G.}\ \bibnamefont {Tobias}}, \bibinfo {author}
  {\bibfnamefont {J.~P.}\ \bibnamefont {Covey}}, \ and\ \bibinfo {author}
  {\bibfnamefont {J.}~\bibnamefont {Ye}},\ }\href {\doibase
  10.1126/science.aau7230} {\bibfield  {journal} {\bibinfo  {journal}
  {Science}\ }\textbf {\bibinfo {volume} {7230}} (\bibinfo {year} {2019}),\
  10.1126/science.aau7230}\BibitemShut {NoStop}%
\bibitem [{\citenamefont {Fitch}\ and\ \citenamefont
  {Tarbutt}(2016)}]{Fitch2016}%
  \BibitemOpen
  \bibfield  {author} {\bibinfo {author} {\bibfnamefont {N.~J.}\ \bibnamefont
  {Fitch}}\ and\ \bibinfo {author} {\bibfnamefont {M.~R.}\ \bibnamefont
  {Tarbutt}},\ }\href {\doibase 10.1002/cphc.201600656} {\bibfield  {journal}
  {\bibinfo  {journal} {ChemPhysChem}\ }\textbf {\bibinfo {volume} {17}},\
  \bibinfo {pages} {3609} (\bibinfo {year} {2016})},\ \Eprint
  {http://arxiv.org/abs/1609.05823} {arXiv:1609.05823} \BibitemShut {NoStop}%
\bibitem [{\citenamefont {Jayich}, \citenamefont {Long},\ and\ \citenamefont
  {Campbell}(2016)}]{Jayich2016}%
  \BibitemOpen
  \bibfield  {author} {\bibinfo {author} {\bibfnamefont {A.~M.}\ \bibnamefont
  {Jayich}}, \bibinfo {author} {\bibfnamefont {X.}~\bibnamefont {Long}}, \ and\
  \bibinfo {author} {\bibfnamefont {W.~C.}\ \bibnamefont {Campbell}},\ }\href
  {\doibase 10.1103/PhysRevX.6.041004} {\bibfield  {journal} {\bibinfo
  {journal} {Physical Review X}\ }\textbf {\bibinfo {volume} {6}},\ \bibinfo
  {pages} {041004} (\bibinfo {year} {2016})},\ \Eprint
  {http://arxiv.org/abs/1603.08053} {arXiv:1603.08053} \BibitemShut {NoStop}%
\bibitem [{\citenamefont {Wu}\ \emph {et~al.}(2017)\citenamefont {Wu},
  \citenamefont {Gantner}, \citenamefont {Koller}, \citenamefont {Zeppenfeld},
  \citenamefont {Chervenkov},\ and\ \citenamefont {Rempe}}]{Wu2017}%
  \BibitemOpen
  \bibfield  {author} {\bibinfo {author} {\bibfnamefont {X.}~\bibnamefont
  {Wu}}, \bibinfo {author} {\bibfnamefont {T.}~\bibnamefont {Gantner}},
  \bibinfo {author} {\bibfnamefont {M.}~\bibnamefont {Koller}}, \bibinfo
  {author} {\bibfnamefont {M.}~\bibnamefont {Zeppenfeld}}, \bibinfo {author}
  {\bibfnamefont {S.}~\bibnamefont {Chervenkov}}, \ and\ \bibinfo {author}
  {\bibfnamefont {G.}~\bibnamefont {Rempe}},\ }\href {\doibase
  10.1126/science.aan3029} {\bibfield  {journal} {\bibinfo  {journal}
  {Science}\ }\textbf {\bibinfo {volume} {358}},\ \bibinfo {pages} {645}
  (\bibinfo {year} {2017})},\ \Eprint {http://arxiv.org/abs/1710.05988}
  {arXiv:1710.05988} \BibitemShut {NoStop}%
\bibitem [{\citenamefont {Galica}\ \emph {et~al.}(2018)\citenamefont {Galica},
  \citenamefont {Aldridge}, \citenamefont {McCarron}, \citenamefont {Eyler},\
  and\ \citenamefont {Gould}}]{Galica2018}%
  \BibitemOpen
  \bibfield  {author} {\bibinfo {author} {\bibfnamefont {S.~E.}\ \bibnamefont
  {Galica}}, \bibinfo {author} {\bibfnamefont {L.}~\bibnamefont {Aldridge}},
  \bibinfo {author} {\bibfnamefont {D.~J.}\ \bibnamefont {McCarron}}, \bibinfo
  {author} {\bibfnamefont {E.~E.}\ \bibnamefont {Eyler}}, \ and\ \bibinfo
  {author} {\bibfnamefont {P.~L.}\ \bibnamefont {Gould}},\ }\href {\doibase
  10.1103/PhysRevA.98.023408} {\bibfield  {journal} {\bibinfo  {journal}
  {Physical Review A}\ }\textbf {\bibinfo {volume} {98}},\ \bibinfo {pages}
  {023408} (\bibinfo {year} {2018})},\ \Eprint
  {http://arxiv.org/abs/1804.04184} {arXiv:1804.04184} \BibitemShut {NoStop}%
\bibitem [{\citenamefont {Petzold}\ \emph {et~al.}(2018)\citenamefont
  {Petzold}, \citenamefont {Kaebert}, \citenamefont {Gersema}, \citenamefont
  {Siercke},\ and\ \citenamefont {Ospelkaus}}]{Petzold2018}%
  \BibitemOpen
  \bibfield  {author} {\bibinfo {author} {\bibfnamefont {M.}~\bibnamefont
  {Petzold}}, \bibinfo {author} {\bibfnamefont {P.}~\bibnamefont {Kaebert}},
  \bibinfo {author} {\bibfnamefont {P.}~\bibnamefont {Gersema}}, \bibinfo
  {author} {\bibfnamefont {M.}~\bibnamefont {Siercke}}, \ and\ \bibinfo
  {author} {\bibfnamefont {S.}~\bibnamefont {Ospelkaus}},\ }\href {\doibase
  10.1088/1367-2630/aab9f5} {\bibfield  {journal} {\bibinfo  {journal} {New
  Journal of Physics}\ }\textbf {\bibinfo {volume} {20}} (\bibinfo {year}
  {2018}),\ 10.1088/1367-2630/aab9f5},\ \Eprint
  {http://arxiv.org/abs/1712.05157} {arXiv:1712.05157} \BibitemShut {NoStop}%
\bibitem [{\citenamefont {McCarron}(2018)}]{McCarron2018a}%
  \BibitemOpen
  \bibfield  {author} {\bibinfo {author} {\bibfnamefont {D.~J.}\ \bibnamefont
  {McCarron}},\ }\href {http://aapt.scitation.org/doi/10.1119/1.1578063}
  {\bibfield  {journal} {\bibinfo  {journal} {Journal of Physics B: Atomic,
  Molecular and Optical Physics}\ }\textbf {\bibinfo {volume} {51}},\ \bibinfo
  {pages} {212001} (\bibinfo {year} {2018})}\BibitemShut {NoStop}%
\bibitem [{\citenamefont {Stuhl}\ \emph {et~al.}(2008)\citenamefont {Stuhl},
  \citenamefont {Sawyer}, \citenamefont {Wang},\ and\ \citenamefont
  {Ye}}]{Stuhl2008}%
  \BibitemOpen
  \bibfield  {author} {\bibinfo {author} {\bibfnamefont {B.~K.}\ \bibnamefont
  {Stuhl}}, \bibinfo {author} {\bibfnamefont {B.~C.}\ \bibnamefont {Sawyer}},
  \bibinfo {author} {\bibfnamefont {D.}~\bibnamefont {Wang}}, \ and\ \bibinfo
  {author} {\bibfnamefont {J.}~\bibnamefont {Ye}},\ }\href {\doibase
  10.1103/PhysRevLett.101.243002} {\bibfield  {journal} {\bibinfo  {journal}
  {Physical Review Letters}\ }\textbf {\bibinfo {volume} {101}},\ \bibinfo
  {pages} {243002} (\bibinfo {year} {2008})},\ \Eprint
  {http://arxiv.org/abs/0808.2171} {arXiv:0808.2171} \BibitemShut {NoStop}%
\bibitem [{\citenamefont {{Di Rosa}}(2004)}]{DiRosa2004}%
  \BibitemOpen
  \bibfield  {author} {\bibinfo {author} {\bibfnamefont {M.~D.}\ \bibnamefont
  {{Di Rosa}}},\ }\href {\doibase 10.1140/epjd/e2004-00167-2} {\bibfield
  {journal} {\bibinfo  {journal} {The European Physical Journal D}\ }\textbf
  {\bibinfo {volume} {31}},\ \bibinfo {pages} {395} (\bibinfo {year}
  {2004})}\BibitemShut {NoStop}%
\bibitem [{\citenamefont {Hendricks}\ \emph {et~al.}(2014)\citenamefont
  {Hendricks}, \citenamefont {Holland}, \citenamefont {Truppe}, \citenamefont
  {Sauer},\ and\ \citenamefont {Tarbutt}}]{Hendricks2014}%
  \BibitemOpen
  \bibfield  {author} {\bibinfo {author} {\bibfnamefont {R.~J.}\ \bibnamefont
  {Hendricks}}, \bibinfo {author} {\bibfnamefont {D.~A.}\ \bibnamefont
  {Holland}}, \bibinfo {author} {\bibfnamefont {S.}~\bibnamefont {Truppe}},
  \bibinfo {author} {\bibfnamefont {B.~E.}\ \bibnamefont {Sauer}}, \ and\
  \bibinfo {author} {\bibfnamefont {M.~R.}\ \bibnamefont {Tarbutt}},\ }\href
  {\doibase 10.3389/fphy.2014.00051} {\bibfield  {journal} {\bibinfo  {journal}
  {Frontiers in Physics}\ }\textbf {\bibinfo {volume} {2}},\ \bibinfo {pages}
  {1} (\bibinfo {year} {2014})},\ \Eprint {http://arxiv.org/abs/1404.6174v1}
  {arXiv:1404.6174v1} \BibitemShut {NoStop}%
\bibitem [{\citenamefont {Tarbutt}(2018)}]{Tarbutt2018}%
  \BibitemOpen
  \bibfield  {author} {\bibinfo {author} {\bibfnamefont {M.~R.}\ \bibnamefont
  {Tarbutt}},\ }\href {\doibase 10.1080/00107514.2018.1576338} {\bibfield
  {journal} {\bibinfo  {journal} {Contemporary Physics}\ }\textbf {\bibinfo
  {volume} {59}},\ \bibinfo {pages} {356} (\bibinfo {year} {2018})}\BibitemShut
  {NoStop}%
\bibitem [{\citenamefont {Wells}\ and\ \citenamefont {Lane}(2011)}]{Wells2011}%
  \BibitemOpen
  \bibfield  {author} {\bibinfo {author} {\bibfnamefont {N.}~\bibnamefont
  {Wells}}\ and\ \bibinfo {author} {\bibfnamefont {I.~C.}\ \bibnamefont
  {Lane}},\ }\href {\doibase 10.1039/c1cp21313j} {\bibfield  {journal}
  {\bibinfo  {journal} {Physical Chemistry Chemical Physics}\ }\textbf
  {\bibinfo {volume} {13}},\ \bibinfo {pages} {19018} (\bibinfo {year}
  {2011})}\BibitemShut {NoStop}%
\bibitem [{\citenamefont {Langhoff}, \citenamefont {Bauschlicher},\ and\
  \citenamefont {Taylor}(1988)}]{Langhoff1988}%
  \BibitemOpen
  \bibfield  {author} {\bibinfo {author} {\bibfnamefont {S.~R.}\ \bibnamefont
  {Langhoff}}, \bibinfo {author} {\bibfnamefont {C.~W.}\ \bibnamefont
  {Bauschlicher}}, \ and\ \bibinfo {author} {\bibfnamefont {P.~R.}\
  \bibnamefont {Taylor}},\ }\href {\doibase 10.1063/1.454531} {\bibfield
  {journal} {\bibinfo  {journal} {The Journal of Chemical Physics}\ }\textbf
  {\bibinfo {volume} {88}},\ \bibinfo {pages} {5715} (\bibinfo {year}
  {1988})}\BibitemShut {NoStop}%
\bibitem [{\citenamefont {Ko}\ \emph {et~al.}(1965)\citenamefont {Ko},
  \citenamefont {Greenbaum}, \citenamefont {Blauer},\ and\ \citenamefont
  {Farber}}]{Ko1965}%
  \BibitemOpen
  \bibfield  {author} {\bibinfo {author} {\bibfnamefont {H.~C.}\ \bibnamefont
  {Ko}}, \bibinfo {author} {\bibfnamefont {M.~A.}\ \bibnamefont {Greenbaum}},
  \bibinfo {author} {\bibfnamefont {J.~A.}\ \bibnamefont {Blauer}}, \ and\
  \bibinfo {author} {\bibfnamefont {M.}~\bibnamefont {Farber}},\ }\href
  {\doibase 10.1021/j100891a030} {\bibfield  {journal} {\bibinfo  {journal}
  {The Journal of Physical Chemistry}\ }\textbf {\bibinfo {volume} {69}},\
  \bibinfo {pages} {2311} (\bibinfo {year} {1965})}\BibitemShut {NoStop}%
\bibitem [{\citenamefont {Barrow}, \citenamefont {Kopp},\ and\ \citenamefont
  {Malmberg}(1974)}]{Barrow1974}%
  \BibitemOpen
  \bibfield  {author} {\bibinfo {author} {\bibfnamefont {R.~F.}\ \bibnamefont
  {Barrow}}, \bibinfo {author} {\bibfnamefont {I.}~\bibnamefont {Kopp}}, \ and\
  \bibinfo {author} {\bibfnamefont {C.}~\bibnamefont {Malmberg}},\ }\href
  {\doibase 10.1088/0031-8949/10/1-2/008} {\bibfield  {journal} {\bibinfo
  {journal} {Physica Scripta}\ }\textbf {\bibinfo {volume} {10}},\ \bibinfo
  {pages} {86} (\bibinfo {year} {1974})}\BibitemShut {NoStop}%
\bibitem [{\citenamefont {Brown}\ \emph {et~al.}(1978)\citenamefont {Brown},
  \citenamefont {Kopp}, \citenamefont {Malmberg},\ and\ \citenamefont
  {Rydh}}]{Brown1978}%
  \BibitemOpen
  \bibfield  {author} {\bibinfo {author} {\bibfnamefont {J.~M.}\ \bibnamefont
  {Brown}}, \bibinfo {author} {\bibfnamefont {I.}~\bibnamefont {Kopp}},
  \bibinfo {author} {\bibfnamefont {C.}~\bibnamefont {Malmberg}}, \ and\
  \bibinfo {author} {\bibfnamefont {B.}~\bibnamefont {Rydh}},\ }\href {\doibase
  10.1088/0031-8949/17/2/003} {\bibfield  {journal} {\bibinfo  {journal}
  {Physica Scripta}\ }\textbf {\bibinfo {volume} {17}},\ \bibinfo {pages} {55}
  (\bibinfo {year} {1978})}\BibitemShut {NoStop}%
\bibitem [{\citenamefont {Lide}(1963)}]{Lide1963}%
  \BibitemOpen
  \bibfield  {author} {\bibinfo {author} {\bibfnamefont {D.~R.}\ \bibnamefont
  {Lide}},\ }\href {\doibase 10.1063/1.1733914} {\bibfield  {journal} {\bibinfo
   {journal} {The Journal of Chemical Physics}\ }\textbf {\bibinfo {volume}
  {38}},\ \bibinfo {pages} {2027} (\bibinfo {year} {1963})}\BibitemShut
  {NoStop}%
\bibitem [{\citenamefont {Lide}(1965)}]{Lide1965}%
  \BibitemOpen
  \bibfield  {author} {\bibinfo {author} {\bibfnamefont {D.~R.}\ \bibnamefont
  {Lide}},\ }\href {\doibase 10.1063/1.1696035} {\bibfield  {journal} {\bibinfo
   {journal} {The Journal of Chemical Physics}\ }\textbf {\bibinfo {volume}
  {42}},\ \bibinfo {pages} {1013} (\bibinfo {year} {1965})}\BibitemShut
  {NoStop}%
\bibitem [{\citenamefont {Wyse}, \citenamefont {Gordy},\ and\ \citenamefont
  {Pearson}(1970)}]{Wyse1970}%
  \BibitemOpen
  \bibfield  {author} {\bibinfo {author} {\bibfnamefont {F.~C.}\ \bibnamefont
  {Wyse}}, \bibinfo {author} {\bibfnamefont {W.}~\bibnamefont {Gordy}}, \ and\
  \bibinfo {author} {\bibfnamefont {E.~F.}\ \bibnamefont {Pearson}},\ }\href
  {\doibase 10.1063/1.1673587} {\bibfield  {journal} {\bibinfo  {journal} {The
  Journal of Chemical Physics}\ }\textbf {\bibinfo {volume} {52}},\ \bibinfo
  {pages} {3887} (\bibinfo {year} {1970})}\BibitemShut {NoStop}%
\bibitem [{\citenamefont {Hoeft}\ \emph {et~al.}(1970)\citenamefont {Hoeft},
  \citenamefont {Lovas}, \citenamefont {Tiemann},\ and\ \citenamefont
  {T{\"{o}}rring}}]{Hoeft1970}%
  \BibitemOpen
  \bibfield  {author} {\bibinfo {author} {\bibfnamefont {J.}~\bibnamefont
  {Hoeft}}, \bibinfo {author} {\bibfnamefont {F.~J.}\ \bibnamefont {Lovas}},
  \bibinfo {author} {\bibfnamefont {E.}~\bibnamefont {Tiemann}}, \ and\
  \bibinfo {author} {\bibfnamefont {T.}~\bibnamefont {T{\"{o}}rring}},\ }\href
  {\doibase 10.1515/zna-1970-0706} {\bibfield  {journal} {\bibinfo  {journal}
  {Zeitschrift f{\"{u}}r Naturforschung A}\ }\textbf {\bibinfo {volume} {25}},\
  \bibinfo {pages} {1029} (\bibinfo {year} {1970})}\BibitemShut {NoStop}%
\bibitem [{\citenamefont {Honerj{\"{a}}ger}\ and\ \citenamefont
  {Tischer}(1974)}]{Honerjager1974}%
  \BibitemOpen
  \bibfield  {author} {\bibinfo {author} {\bibfnamefont {R.}~\bibnamefont
  {Honerj{\"{a}}ger}}\ and\ \bibinfo {author} {\bibfnamefont {R.}~\bibnamefont
  {Tischer}},\ }\href {\doibase 10.1515/zna-1974-0224} {\bibfield  {journal}
  {\bibinfo  {journal} {Zeitschrift f{\"{u}}r Naturforschung A}\ }\textbf
  {\bibinfo {volume} {29}},\ \bibinfo {pages} {342} (\bibinfo {year}
  {1974})}\BibitemShut {NoStop}%
\bibitem [{\citenamefont {Rochester}(1939)}]{Rochester1939}%
  \BibitemOpen
  \bibfield  {author} {\bibinfo {author} {\bibfnamefont {G.~D.}\ \bibnamefont
  {Rochester}},\ }\href {\doibase 10.1103/PhysRev.56.305} {\bibfield  {journal}
  {\bibinfo  {journal} {Physical Review}\ }\textbf {\bibinfo {volume} {56}},\
  \bibinfo {pages} {305} (\bibinfo {year} {1939})}\BibitemShut {NoStop}%
\bibitem [{\citenamefont {Rowlinson}\ and\ \citenamefont
  {Barrow}(1953)}]{Rowlinson1953}%
  \BibitemOpen
  \bibfield  {author} {\bibinfo {author} {\bibfnamefont {H.~C.}\ \bibnamefont
  {Rowlinson}}\ and\ \bibinfo {author} {\bibfnamefont {R.~F.}\ \bibnamefont
  {Barrow}},\ }\href {\doibase 10.1088/0370-1298/66/5/303} {\bibfield
  {journal} {\bibinfo  {journal} {Proceedings of the Physical Society. Section
  A}\ }\textbf {\bibinfo {volume} {66}},\ \bibinfo {pages} {437} (\bibinfo
  {year} {1953})}\BibitemShut {NoStop}%
\bibitem [{\citenamefont {Naud{\'{e}}}\ and\ \citenamefont
  {Hugo}(1953{\natexlab{a}})}]{Naude1953a}%
  \BibitemOpen
  \bibfield  {author} {\bibinfo {author} {\bibfnamefont {S.~M.}\ \bibnamefont
  {Naud{\'{e}}}}\ and\ \bibinfo {author} {\bibfnamefont {T.~J.}\ \bibnamefont
  {Hugo}},\ }\href {\doibase 10.1103/PhysRev.90.318} {\bibfield  {journal}
  {\bibinfo  {journal} {Physical Review}\ }\textbf {\bibinfo {volume} {90}},\
  \bibinfo {pages} {318} (\bibinfo {year} {1953}{\natexlab{a}})}\BibitemShut
  {NoStop}%
\bibitem [{\citenamefont {Naud{\'{e}}}\ and\ \citenamefont
  {Hugo}(1953{\natexlab{b}})}]{Naude1953b}%
  \BibitemOpen
  \bibfield  {author} {\bibinfo {author} {\bibfnamefont {S.~M.}\ \bibnamefont
  {Naud{\'{e}}}}\ and\ \bibinfo {author} {\bibfnamefont {T.~J.}\ \bibnamefont
  {Hugo}},\ }\href {\doibase 10.1139/p53-095} {\bibfield  {journal} {\bibinfo
  {journal} {Canadian Journal of Physics}\ }\textbf {\bibinfo {volume} {31}},\
  \bibinfo {pages} {1106} (\bibinfo {year} {1953}{\natexlab{b}})}\BibitemShut
  {NoStop}%
\bibitem [{\citenamefont {Naud{\'{e}}}\ and\ \citenamefont
  {Hugo}(1954)}]{Naude1954}%
  \BibitemOpen
  \bibfield  {author} {\bibinfo {author} {\bibfnamefont {S.~M.}\ \bibnamefont
  {Naud{\'{e}}}}\ and\ \bibinfo {author} {\bibfnamefont {T.~J.}\ \bibnamefont
  {Hugo}},\ }\href {\doibase 10.1139/p54-023} {\bibfield  {journal} {\bibinfo
  {journal} {Canadian Journal of Physics}\ }\textbf {\bibinfo {volume} {32}},\
  \bibinfo {pages} {246} (\bibinfo {year} {1954})}\BibitemShut {NoStop}%
\bibitem [{\citenamefont {Rosenwaks}, \citenamefont {Steele},\ and\
  \citenamefont {Broida}(1976)}]{Rosenwaks1976}%
  \BibitemOpen
  \bibfield  {author} {\bibinfo {author} {\bibfnamefont {S.}~\bibnamefont
  {Rosenwaks}}, \bibinfo {author} {\bibfnamefont {R.}~\bibnamefont {Steele}}, \
  and\ \bibinfo {author} {\bibfnamefont {H.}~\bibnamefont {Broida}},\ }\href
  {\doibase 10.1016/0009-2614(76)80270-9} {\bibfield  {journal} {\bibinfo
  {journal} {Chemical Physics Letters}\ }\textbf {\bibinfo {volume} {38}},\
  \bibinfo {pages} {121} (\bibinfo {year} {1976})}\BibitemShut {NoStop}%
\bibitem [{\citenamefont {Kopp}, \citenamefont {Lindgren},\ and\ \citenamefont
  {Malmberg}(1976)}]{Kopp1976}%
  \BibitemOpen
  \bibfield  {author} {\bibinfo {author} {\bibfnamefont {I.}~\bibnamefont
  {Kopp}}, \bibinfo {author} {\bibfnamefont {B.}~\bibnamefont {Lindgren}}, \
  and\ \bibinfo {author} {\bibfnamefont {C.}~\bibnamefont {Malmberg}},\ }\href
  {\doibase 10.1088/0031-8949/14/4/008} {\bibfield  {journal} {\bibinfo
  {journal} {Physica Scripta}\ }\textbf {\bibinfo {volume} {14}},\ \bibinfo
  {pages} {170} (\bibinfo {year} {1976})}\BibitemShut {NoStop}%
\bibitem [{\citenamefont {Gross}\ \emph {et~al.}(1948)\citenamefont {Gross},
  \citenamefont {Campbell}, \citenamefont {Kent},\ and\ \citenamefont
  {Levi}}]{Gross1948}%
  \BibitemOpen
  \bibfield  {author} {\bibinfo {author} {\bibfnamefont {P.}~\bibnamefont
  {Gross}}, \bibinfo {author} {\bibfnamefont {C.~S.}\ \bibnamefont {Campbell}},
  \bibinfo {author} {\bibfnamefont {P.~J.~C.}\ \bibnamefont {Kent}}, \ and\
  \bibinfo {author} {\bibfnamefont {D.~L.}\ \bibnamefont {Levi}},\ }\href
  {\doibase 10.1039/df9480400206} {\bibfield  {journal} {\bibinfo  {journal}
  {Discussions of the Faraday Society}\ }\textbf {\bibinfo {volume} {4}},\
  \bibinfo {pages} {206} (\bibinfo {year} {1948})}\BibitemShut {NoStop}%
\bibitem [{\citenamefont {Gross}, \citenamefont {Hayman},\ and\ \citenamefont
  {Levi}(1954)}]{Gross1954}%
  \BibitemOpen
  \bibfield  {author} {\bibinfo {author} {\bibfnamefont {P.}~\bibnamefont
  {Gross}}, \bibinfo {author} {\bibfnamefont {C.}~\bibnamefont {Hayman}}, \
  and\ \bibinfo {author} {\bibfnamefont {D.~L.}\ \bibnamefont {Levi}},\ }\href
  {\doibase 10.1039/tf9545000477} {\bibfield  {journal} {\bibinfo  {journal}
  {Transactions of the Faraday Society}\ }\textbf {\bibinfo {volume} {50}},\
  \bibinfo {pages} {477} (\bibinfo {year} {1954})}\BibitemShut {NoStop}%
\bibitem [{\citenamefont {Barrow}, \citenamefont {Johns},\ and\ \citenamefont
  {Smith}(1956)}]{Barrow1956}%
  \BibitemOpen
  \bibfield  {author} {\bibinfo {author} {\bibfnamefont {R.~F.}\ \bibnamefont
  {Barrow}}, \bibinfo {author} {\bibfnamefont {J.~W.~C.}\ \bibnamefont
  {Johns}}, \ and\ \bibinfo {author} {\bibfnamefont {F.~J.}\ \bibnamefont
  {Smith}},\ }\href {\doibase 10.1039/tf9565200913} {\bibfield  {journal}
  {\bibinfo  {journal} {Transactions of the Faraday Society}\ }\textbf
  {\bibinfo {volume} {52}},\ \bibinfo {pages} {913} (\bibinfo {year}
  {1956})}\BibitemShut {NoStop}%
\bibitem [{\citenamefont {Zhang}(1997)}]{Zhang1997}%
  \BibitemOpen
  \bibfield  {author} {\bibinfo {author} {\bibfnamefont {K.}~\bibnamefont
  {Zhang}},\ }\emph {\bibinfo {title} {{Infrared Spectroscopy of Molecules in
  the Gas Phase}}},\ \href
  {https://www.collectionscanada.gc.ca/obj/s4/f2/dsk3/ftp04/nq21402.pdf} {Ph.D.
  thesis},\ \bibinfo  {school} {Waterloo} (\bibinfo {year} {1997})\BibitemShut
  {NoStop}%
\bibitem [{\citenamefont {Freund}\ and\ \citenamefont
  {Klemperer}(1965)}]{Freund1965}%
  \BibitemOpen
  \bibfield  {author} {\bibinfo {author} {\bibfnamefont {R.~S.}\ \bibnamefont
  {Freund}}\ and\ \bibinfo {author} {\bibfnamefont {W.}~\bibnamefont
  {Klemperer}},\ }\href {\doibase 10.1063/1.1697141} {\bibfield  {journal}
  {\bibinfo  {journal} {The Journal of Chemical Physics}\ }\textbf {\bibinfo
  {volume} {43}},\ \bibinfo {pages} {2422} (\bibinfo {year}
  {1965})}\BibitemShut {NoStop}%
\bibitem [{\citenamefont {Frosch}\ and\ \citenamefont
  {Foley}(1952)}]{Frosch1952}%
  \BibitemOpen
  \bibfield  {author} {\bibinfo {author} {\bibfnamefont {R.~A.}\ \bibnamefont
  {Frosch}}\ and\ \bibinfo {author} {\bibfnamefont {H.~M.}\ \bibnamefont
  {Foley}},\ }\href {\doibase 10.1103/PhysRev.88.1337} {\bibfield  {journal}
  {\bibinfo  {journal} {Physical Review}\ }\textbf {\bibinfo {volume} {88}},\
  \bibinfo {pages} {1337} (\bibinfo {year} {1952})}\BibitemShut {NoStop}%
\bibitem [{\citenamefont {Zare}\ \emph {et~al.}(1973)\citenamefont {Zare},
  \citenamefont {Schmeltekopf}, \citenamefont {Harrop},\ and\ \citenamefont
  {Albritton}}]{Zare1973}%
  \BibitemOpen
  \bibfield  {author} {\bibinfo {author} {\bibfnamefont {R.}~\bibnamefont
  {Zare}}, \bibinfo {author} {\bibfnamefont {A.}~\bibnamefont {Schmeltekopf}},
  \bibinfo {author} {\bibfnamefont {W.}~\bibnamefont {Harrop}}, \ and\ \bibinfo
  {author} {\bibfnamefont {D.}~\bibnamefont {Albritton}},\ }\href {\doibase
  10.1016/0022-2852(73)90025-8} {\bibfield  {journal} {\bibinfo  {journal}
  {Journal of Molecular Spectroscopy}\ }\textbf {\bibinfo {volume} {46}},\
  \bibinfo {pages} {37} (\bibinfo {year} {1973})}\BibitemShut {NoStop}%
\bibitem [{\citenamefont {Brown}\ \emph {et~al.}(1979)\citenamefont {Brown},
  \citenamefont {Colbourn}, \citenamefont {Watson},\ and\ \citenamefont
  {Wayne}}]{Brown1979a}%
  \BibitemOpen
  \bibfield  {author} {\bibinfo {author} {\bibfnamefont {J.}~\bibnamefont
  {Brown}}, \bibinfo {author} {\bibfnamefont {E.}~\bibnamefont {Colbourn}},
  \bibinfo {author} {\bibfnamefont {J.}~\bibnamefont {Watson}}, \ and\ \bibinfo
  {author} {\bibfnamefont {F.}~\bibnamefont {Wayne}},\ }\href {\doibase
  10.1016/0022-2852(79)90059-6} {\bibfield  {journal} {\bibinfo  {journal}
  {Journal of Molecular Spectroscopy}\ }\textbf {\bibinfo {volume} {74}},\
  \bibinfo {pages} {294} (\bibinfo {year} {1979})}\BibitemShut {NoStop}%
\bibitem [{\citenamefont {Brown}\ and\ \citenamefont
  {Merer}(1979)}]{Brown1979b}%
  \BibitemOpen
  \bibfield  {author} {\bibinfo {author} {\bibfnamefont {J.}~\bibnamefont
  {Brown}}\ and\ \bibinfo {author} {\bibfnamefont {A.}~\bibnamefont {Merer}},\
  }\href {\doibase 10.1016/0022-2852(79)90172-3} {\bibfield  {journal}
  {\bibinfo  {journal} {Journal of Molecular Spectroscopy}\ }\textbf {\bibinfo
  {volume} {74}},\ \bibinfo {pages} {488} (\bibinfo {year} {1979})}\BibitemShut
  {NoStop}%
\bibitem [{\citenamefont {Brazier}, \citenamefont {Ram},\ and\ \citenamefont
  {Bernath}(1986)}]{Brazier1986}%
  \BibitemOpen
  \bibfield  {author} {\bibinfo {author} {\bibfnamefont {C.}~\bibnamefont
  {Brazier}}, \bibinfo {author} {\bibfnamefont {R.}~\bibnamefont {Ram}}, \ and\
  \bibinfo {author} {\bibfnamefont {P.}~\bibnamefont {Bernath}},\ }\href
  {\doibase 10.1016/0022-2852(86)90012-3} {\bibfield  {journal} {\bibinfo
  {journal} {Journal of Molecular Spectroscopy}\ }\textbf {\bibinfo {volume}
  {120}},\ \bibinfo {pages} {381} (\bibinfo {year} {1986})}\BibitemShut
  {NoStop}%
\bibitem [{\citenamefont {Gordy}\ and\ \citenamefont {Cook}(1984)}]{Gordy1984}%
  \BibitemOpen
  \bibfield  {author} {\bibinfo {author} {\bibfnamefont {W.}~\bibnamefont
  {Gordy}}\ and\ \bibinfo {author} {\bibfnamefont {R.~L.}\ \bibnamefont
  {Cook}},\ }\href@noop {} {\emph {\bibinfo {title} {{Microwave molecular
  spectra}}}}\ (\bibinfo  {publisher} {Wiley},\ \bibinfo {address} {New York},\
  \bibinfo {year} {1984})\BibitemShut {NoStop}%
\bibitem [{\citenamefont {Carrington}\ and\ \citenamefont
  {Brown}(2003)}]{Carrington2003}%
  \BibitemOpen
  \bibfield  {author} {\bibinfo {author} {\bibfnamefont {A.}~\bibnamefont
  {Carrington}}\ and\ \bibinfo {author} {\bibfnamefont {J.~A.}\ \bibnamefont
  {Brown}},\ }\href@noop {} {\emph {\bibinfo {title} {{Rotational spectroscopy
  of diatomic molecules}}}}\ (\bibinfo  {publisher} {Cambridge University
  Press},\ \bibinfo {year} {2003})\BibitemShut {NoStop}%
\bibitem [{\citenamefont {Townes}\ and\ \citenamefont
  {Schawlow}(1975)}]{Townes1975}%
  \BibitemOpen
  \bibfield  {author} {\bibinfo {author} {\bibfnamefont {C.~H.}\ \bibnamefont
  {Townes}}\ and\ \bibinfo {author} {\bibfnamefont {A.~L.}\ \bibnamefont
  {Schawlow}},\ }\href@noop {} {\emph {\bibinfo {title} {{Microwave
  Spectroscopy}}}}\ (\bibinfo  {publisher} {Dover publications, Inc.},\
  \bibinfo {address} {New York},\ \bibinfo {year} {1975})\BibitemShut {NoStop}%
\bibitem [{\citenamefont {Klaus}, \citenamefont {Takano},\ and\ \citenamefont
  {Winnewisser}(1997)}]{Klaus1997}%
  \BibitemOpen
  \bibfield  {author} {\bibinfo {author} {\bibfnamefont {T.}~\bibnamefont
  {Klaus}}, \bibinfo {author} {\bibfnamefont {S.}~\bibnamefont {Takano}}, \
  and\ \bibinfo {author} {\bibfnamefont {G.}~\bibnamefont {Winnewisser}},\
  }\href@noop {} {\bibfield  {journal} {\bibinfo  {journal} {Astronomy and
  Astrophysics}\ }\textbf {\bibinfo {volume} {322}},\ \bibinfo {pages} {L1}
  (\bibinfo {year} {1997})}\BibitemShut {NoStop}%
\bibitem [{\citenamefont {Champion}, \citenamefont {Loete},\ and\ \citenamefont
  {Pierre}(1992)}]{Champion1992}%
  \BibitemOpen
  \bibfield  {author} {\bibinfo {author} {\bibfnamefont {J.~P.}\ \bibnamefont
  {Champion}}, \bibinfo {author} {\bibfnamefont {M.}~\bibnamefont {Loete}}, \
  and\ \bibinfo {author} {\bibfnamefont {G.}~\bibnamefont {Pierre}},\
  }\href@noop {} {\emph {\bibinfo {title} {{Spectroscopy of the Earth's
  atmosphere and interstellar medium}}}},\ edited by\ \bibinfo {editor}
  {\bibfnamefont {K.~N.}\ \bibnamefont {Rao}}\ and\ \bibinfo {editor}
  {\bibfnamefont {A.}~\bibnamefont {Weber}}\ (\bibinfo  {publisher} {Academic
  Press},\ \bibinfo {address} {Columbus, OH},\ \bibinfo {year} {1992})\ pp.\
  \bibinfo {pages} {339--422}\BibitemShut {NoStop}%
\bibitem [{\citenamefont {Veldhoven}\ \emph {et~al.}(2004)\citenamefont
  {Veldhoven}, \citenamefont {K{\"{u}}pper}, \citenamefont {Bethlem},
  \citenamefont {Sartakov}, \citenamefont {Roij},\ and\ \citenamefont
  {Meijer}}]{Veldhoven2004}%
  \BibitemOpen
  \bibfield  {author} {\bibinfo {author} {\bibfnamefont {J.}~\bibnamefont
  {Veldhoven}}, \bibinfo {author} {\bibfnamefont {J.}~\bibnamefont
  {K{\"{u}}pper}}, \bibinfo {author} {\bibfnamefont {H.~L.}\ \bibnamefont
  {Bethlem}}, \bibinfo {author} {\bibfnamefont {B.}~\bibnamefont {Sartakov}},
  \bibinfo {author} {\bibfnamefont {A.~J.~A.}\ \bibnamefont {Roij}}, \ and\
  \bibinfo {author} {\bibfnamefont {G.}~\bibnamefont {Meijer}},\ }\href
  {\doibase 10.1140/epjd/e2004-00160-9} {\bibfield  {journal} {\bibinfo
  {journal} {The European Physical Journal D}\ }\textbf {\bibinfo {volume}
  {31}},\ \bibinfo {pages} {337} (\bibinfo {year} {2004})}\BibitemShut
  {NoStop}%
\bibitem [{\citenamefont {Dearden}, \citenamefont {Johnson},\ and\
  \citenamefont {Hudgens}(1991)}]{Dearden1991}%
  \BibitemOpen
  \bibfield  {author} {\bibinfo {author} {\bibfnamefont {D.~V.}\ \bibnamefont
  {Dearden}}, \bibinfo {author} {\bibfnamefont {R.~D.}\ \bibnamefont
  {Johnson}}, \ and\ \bibinfo {author} {\bibfnamefont {J.~W.}\ \bibnamefont
  {Hudgens}},\ }\href {\doibase 10.1021/j100164a022} {\bibfield  {journal}
  {\bibinfo  {journal} {The Journal of Physical Chemistry}\ }\textbf {\bibinfo
  {volume} {95}},\ \bibinfo {pages} {4291} (\bibinfo {year}
  {1991})}\BibitemShut {NoStop}%
\bibitem [{\citenamefont {Kohl}\ and\ \citenamefont
  {Parkinson}(1973)}]{Kohl1973}%
  \BibitemOpen
  \bibfield  {author} {\bibinfo {author} {\bibfnamefont {J.~L.}\ \bibnamefont
  {Kohl}}\ and\ \bibinfo {author} {\bibfnamefont {W.~H.}\ \bibnamefont
  {Parkinson}},\ }\href {\doibase 10.1086/152356} {\bibfield  {journal}
  {\bibinfo  {journal} {The Astrophysical Journal}\ }\textbf {\bibinfo {volume}
  {184}},\ \bibinfo {pages} {641} (\bibinfo {year} {1973})}\BibitemShut
  {NoStop}%
\bibitem [{\citenamefont {Gilijamse}\ \emph {et~al.}(2007)\citenamefont
  {Gilijamse}, \citenamefont {Hoekstra}, \citenamefont {Meek}, \citenamefont
  {Mets{\"{a}}l{\"{a}}}, \citenamefont {van~de Meerakker}, \citenamefont
  {Meijer},\ and\ \citenamefont {Groenenboom}}]{Gilijamse2007}%
  \BibitemOpen
  \bibfield  {author} {\bibinfo {author} {\bibfnamefont {J.~J.}\ \bibnamefont
  {Gilijamse}}, \bibinfo {author} {\bibfnamefont {S.}~\bibnamefont {Hoekstra}},
  \bibinfo {author} {\bibfnamefont {S.~A.}\ \bibnamefont {Meek}}, \bibinfo
  {author} {\bibfnamefont {M.}~\bibnamefont {Mets{\"{a}}l{\"{a}}}}, \bibinfo
  {author} {\bibfnamefont {S.~Y.~T.}\ \bibnamefont {van~de Meerakker}},
  \bibinfo {author} {\bibfnamefont {G.}~\bibnamefont {Meijer}}, \ and\ \bibinfo
  {author} {\bibfnamefont {G.~C.}\ \bibnamefont {Groenenboom}},\ }\href
  {\doibase 10.1063/1.2813888} {\bibfield  {journal} {\bibinfo  {journal} {The
  Journal of Chemical Physics}\ }\textbf {\bibinfo {volume} {127}},\ \bibinfo
  {pages} {221102} (\bibinfo {year} {2007})},\ \Eprint
  {http://arxiv.org/abs/0710.2240} {arXiv:0710.2240} \BibitemShut {NoStop}%
\bibitem [{\citenamefont {James}(1971)}]{James1971}%
  \BibitemOpen
  \bibfield  {author} {\bibinfo {author} {\bibfnamefont {T.~C.}\ \bibnamefont
  {James}},\ }\href {\doibase 10.1063/1.1676710} {\bibfield  {journal}
  {\bibinfo  {journal} {The Journal of Chemical Physics}\ }\textbf {\bibinfo
  {volume} {55}},\ \bibinfo {pages} {4118} (\bibinfo {year}
  {1971})}\BibitemShut {NoStop}%
\bibitem [{\citenamefont {Brown}\ \emph {et~al.}(1975)\citenamefont {Brown},
  \citenamefont {Hougen}, \citenamefont {Huber}, \citenamefont {Johns},
  \citenamefont {Kopp}, \citenamefont {Lefebvre-Brion}, \citenamefont {Merer},
  \citenamefont {Ramsay}, \citenamefont {Rostas},\ and\ \citenamefont
  {Zare}}]{Brown1975}%
  \BibitemOpen
  \bibfield  {author} {\bibinfo {author} {\bibfnamefont {J.}~\bibnamefont
  {Brown}}, \bibinfo {author} {\bibfnamefont {J.}~\bibnamefont {Hougen}},
  \bibinfo {author} {\bibfnamefont {K.-P.}\ \bibnamefont {Huber}}, \bibinfo
  {author} {\bibfnamefont {J.}~\bibnamefont {Johns}}, \bibinfo {author}
  {\bibfnamefont {I.}~\bibnamefont {Kopp}}, \bibinfo {author} {\bibfnamefont
  {H.}~\bibnamefont {Lefebvre-Brion}}, \bibinfo {author} {\bibfnamefont
  {A.}~\bibnamefont {Merer}}, \bibinfo {author} {\bibfnamefont
  {D.}~\bibnamefont {Ramsay}}, \bibinfo {author} {\bibfnamefont
  {J.}~\bibnamefont {Rostas}}, \ and\ \bibinfo {author} {\bibfnamefont
  {R.}~\bibnamefont {Zare}},\ }\href {\doibase 10.1016/0022-2852(75)90291-X}
  {\bibfield  {journal} {\bibinfo  {journal} {Journal of Molecular
  Spectroscopy}\ }\textbf {\bibinfo {volume} {55}},\ \bibinfo {pages} {500}
  (\bibinfo {year} {1975})}\BibitemShut {NoStop}%
\bibitem [{\citenamefont {Watson}(1977)}]{Watson1977}%
  \BibitemOpen
  \bibfield  {author} {\bibinfo {author} {\bibfnamefont {J.~K.}\ \bibnamefont
  {Watson}},\ }\href {\doibase 10.1016/0022-2852(77)90308-3} {\bibfield
  {journal} {\bibinfo  {journal} {Journal of Molecular Spectroscopy}\ }\textbf
  {\bibinfo {volume} {66}},\ \bibinfo {pages} {500} (\bibinfo {year}
  {1977})}\BibitemShut {NoStop}%
\bibitem [{\citenamefont {Veseth}(1970)}]{Veseth1970}%
  \BibitemOpen
  \bibfield  {author} {\bibinfo {author} {\bibfnamefont {L.}~\bibnamefont
  {Veseth}},\ }\href {\doibase 10.1088/0022-3700/3/12/012} {\bibfield
  {journal} {\bibinfo  {journal} {Journal of Physics B: Atomic and Molecular
  Physics}\ }\textbf {\bibinfo {volume} {3}},\ \bibinfo {pages} {1677}
  (\bibinfo {year} {1970})}\BibitemShut {NoStop}%
\bibitem [{\citenamefont {Piper}\ and\ \citenamefont
  {Cowles}(1986)}]{Piper1986}%
  \BibitemOpen
  \bibfield  {author} {\bibinfo {author} {\bibfnamefont {L.~G.}\ \bibnamefont
  {Piper}}\ and\ \bibinfo {author} {\bibfnamefont {L.~M.}\ \bibnamefont
  {Cowles}},\ }\href {\doibase 10.1063/1.451098} {\bibfield  {journal}
  {\bibinfo  {journal} {The Journal of Chemical Physics}\ }\textbf {\bibinfo
  {volume} {85}},\ \bibinfo {pages} {2419} (\bibinfo {year}
  {1986})}\BibitemShut {NoStop}%
\bibitem [{\citenamefont {Maxwell}\ \emph {et~al.}(2005)\citenamefont
  {Maxwell}, \citenamefont {Brahms}, \citenamefont {DeCarvalho}, \citenamefont
  {Glenn}, \citenamefont {Helton}, \citenamefont {Nguyen}, \citenamefont
  {Patterson}, \citenamefont {Petricka}, \citenamefont {DeMille},\ and\
  \citenamefont {Doyle}}]{Maxwell2005}%
  \BibitemOpen
  \bibfield  {author} {\bibinfo {author} {\bibfnamefont {S.~E.}\ \bibnamefont
  {Maxwell}}, \bibinfo {author} {\bibfnamefont {N.}~\bibnamefont {Brahms}},
  \bibinfo {author} {\bibfnamefont {R.}~\bibnamefont {DeCarvalho}}, \bibinfo
  {author} {\bibfnamefont {D.~R.}\ \bibnamefont {Glenn}}, \bibinfo {author}
  {\bibfnamefont {J.~S.}\ \bibnamefont {Helton}}, \bibinfo {author}
  {\bibfnamefont {S.~V.}\ \bibnamefont {Nguyen}}, \bibinfo {author}
  {\bibfnamefont {D.}~\bibnamefont {Patterson}}, \bibinfo {author}
  {\bibfnamefont {J.}~\bibnamefont {Petricka}}, \bibinfo {author}
  {\bibfnamefont {D.}~\bibnamefont {DeMille}}, \ and\ \bibinfo {author}
  {\bibfnamefont {J.~M.}\ \bibnamefont {Doyle}},\ }\href {\doibase
  10.1103/PhysRevLett.95.173201} {\bibfield  {journal} {\bibinfo  {journal}
  {Physical Review Letters}\ }\textbf {\bibinfo {volume} {95}},\ \bibinfo
  {pages} {173201} (\bibinfo {year} {2005})}\BibitemShut {NoStop}%
\bibitem [{\citenamefont {Hutzler}, \citenamefont {Lu},\ and\ \citenamefont
  {Doyle}(2012)}]{Hutzler2012}%
  \BibitemOpen
  \bibfield  {author} {\bibinfo {author} {\bibfnamefont {N.~R.}\ \bibnamefont
  {Hutzler}}, \bibinfo {author} {\bibfnamefont {H.-I.}\ \bibnamefont {Lu}}, \
  and\ \bibinfo {author} {\bibfnamefont {J.~M.}\ \bibnamefont {Doyle}},\ }\href
  {\doibase 10.1021/cr200362u} {\bibfield  {journal} {\bibinfo  {journal}
  {Chemical Reviews}\ }\textbf {\bibinfo {volume} {112}},\ \bibinfo {pages}
  {4803} (\bibinfo {year} {2012})},\ \Eprint {http://arxiv.org/abs/1111.2841}
  {arXiv:1111.2841} \BibitemShut {NoStop}%
\bibitem [{\citenamefont {Truppe}\ \emph {et~al.}(2018)\citenamefont {Truppe},
  \citenamefont {Hambach}, \citenamefont {Skoff}, \citenamefont {Bulleid},
  \citenamefont {Bumby}, \citenamefont {Hendricks}, \citenamefont {Hinds},
  \citenamefont {Sauer},\ and\ \citenamefont {Tarbutt}}]{Truppe2018}%
  \BibitemOpen
  \bibfield  {author} {\bibinfo {author} {\bibfnamefont {S.}~\bibnamefont
  {Truppe}}, \bibinfo {author} {\bibfnamefont {M.}~\bibnamefont {Hambach}},
  \bibinfo {author} {\bibfnamefont {S.~M.}\ \bibnamefont {Skoff}}, \bibinfo
  {author} {\bibfnamefont {N.~E.}\ \bibnamefont {Bulleid}}, \bibinfo {author}
  {\bibfnamefont {J.~S.}\ \bibnamefont {Bumby}}, \bibinfo {author}
  {\bibfnamefont {R.~J.}\ \bibnamefont {Hendricks}}, \bibinfo {author}
  {\bibfnamefont {E.~A.}\ \bibnamefont {Hinds}}, \bibinfo {author}
  {\bibfnamefont {B.~E.}\ \bibnamefont {Sauer}}, \ and\ \bibinfo {author}
  {\bibfnamefont {M.~R.}\ \bibnamefont {Tarbutt}},\ }\href {\doibase
  10.1080/09500340.2017.1384516} {\bibfield  {journal} {\bibinfo  {journal}
  {Journal of Modern Optics}\ }\textbf {\bibinfo {volume} {65}},\ \bibinfo
  {pages} {648} (\bibinfo {year} {2018})}\BibitemShut {NoStop}%
\bibitem [{\citenamefont {{Le Roy}}(2017)}]{LeRoy2017}%
  \BibitemOpen
  \bibfield  {author} {\bibinfo {author} {\bibfnamefont {R.~J.}\ \bibnamefont
  {{Le Roy}}},\ }\href {\doibase 10.1016/j.jqsrt.2016.05.028} {\bibfield
  {journal} {\bibinfo  {journal} {Journal of Quantitative Spectroscopy and
  Radiative Transfer}\ }\textbf {\bibinfo {volume} {186}},\ \bibinfo {pages}
  {167} (\bibinfo {year} {2017})}\BibitemShut {NoStop}%
\bibitem [{\citenamefont {Yousefi}\ and\ \citenamefont
  {Bernath}(2018)}]{Yousefi2018}%
  \BibitemOpen
  \bibfield  {author} {\bibinfo {author} {\bibfnamefont {M.}~\bibnamefont
  {Yousefi}}\ and\ \bibinfo {author} {\bibfnamefont {P.~F.}\ \bibnamefont
  {Bernath}},\ }\href {\doibase 10.3847/1538-4365/aacc6a} {\bibfield  {journal}
  {\bibinfo  {journal} {The Astrophysical Journal Supplement Series}\ }\textbf
  {\bibinfo {volume} {237}},\ \bibinfo {pages} {7} (\bibinfo {year}
  {2018})}\BibitemShut {NoStop}%
\bibitem [{\citenamefont {Garland}\ and\ \citenamefont
  {Crosley}(1985)}]{Garland1985}%
  \BibitemOpen
  \bibfield  {author} {\bibinfo {author} {\bibfnamefont {N.~L.}\ \bibnamefont
  {Garland}}\ and\ \bibinfo {author} {\bibfnamefont {D.~R.}\ \bibnamefont
  {Crosley}},\ }\href {\doibase 10.1016/0022-4073(85)90026-3} {\bibfield
  {journal} {\bibinfo  {journal} {Journal of Quantitative Spectroscopy and
  Radiative Transfer}\ }\textbf {\bibinfo {volume} {33}},\ \bibinfo {pages}
  {591} (\bibinfo {year} {1985})}\BibitemShut {NoStop}%
\bibitem [{\citenamefont {Werner}\ \emph {et~al.}(2019)\citenamefont {Werner},
  \citenamefont {Knowles}, \citenamefont {Knizia}, \citenamefont {Manby},
  \citenamefont {Sch{\"{u}}tz}, \citenamefont {Celani}, \citenamefont
  {Gy{\"{o}}rffy}, \citenamefont {Kats}, \citenamefont {Korona}, \citenamefont
  {Lindh}, \citenamefont {Mitrushenkov}, \citenamefont {Rauhut}, \citenamefont
  {Shamasundar}, \citenamefont {Adler}, \citenamefont {Amos}, \citenamefont
  {Bennie}, \citenamefont {Bernhardsson}, \citenamefont {Berning},
  \citenamefont {Cooper}, \citenamefont {Deegan}, \citenamefont {Dobbyn},
  \citenamefont {Eckert}, \citenamefont {Goll}, \citenamefont {Hampel},
  \citenamefont {Hesselmann}, \citenamefont {Hetzer}, \citenamefont {Hrenar},
  \citenamefont {Jansen}, \citenamefont {K{\"{o}}ppl}, \citenamefont {Lee},
  \citenamefont {Liu}, \citenamefont {Lloyd}, \citenamefont {Ma}, \citenamefont
  {Mata}, \citenamefont {May}, \citenamefont {McNicholas}, \citenamefont
  {Meyer}, \citenamefont {{Miller III}}, \citenamefont {Mura}, \citenamefont
  {Nicklass}, \citenamefont {O'Neill}, \citenamefont {Palmieri}, \citenamefont
  {Peng}, \citenamefont {Pfl{\"{u}}ger}, \citenamefont {Pitzer}, \citenamefont
  {Reiher}, \citenamefont {Shiozaki}, \citenamefont {Stoll}, \citenamefont
  {Stone}, \citenamefont {Tarroni}, \citenamefont {Thorsteinsson},
  \citenamefont {Wang},\ and\ \citenamefont {Welborn}}]{MOLPRO}%
  \BibitemOpen
  \bibfield  {author} {\bibinfo {author} {\bibfnamefont {H.-J.}\ \bibnamefont
  {Werner}}, \bibinfo {author} {\bibfnamefont {P.~J.}\ \bibnamefont {Knowles}},
  \bibinfo {author} {\bibfnamefont {G.}~\bibnamefont {Knizia}}, \bibinfo
  {author} {\bibfnamefont {F.~R.}\ \bibnamefont {Manby}}, \bibinfo {author}
  {\bibfnamefont {M.}~\bibnamefont {Sch{\"{u}}tz}}, \bibinfo {author}
  {\bibfnamefont {P.}~\bibnamefont {Celani}}, \bibinfo {author} {\bibfnamefont
  {W.}~\bibnamefont {Gy{\"{o}}rffy}}, \bibinfo {author} {\bibfnamefont
  {D.}~\bibnamefont {Kats}}, \bibinfo {author} {\bibfnamefont {T.}~\bibnamefont
  {Korona}}, \bibinfo {author} {\bibfnamefont {R.}~\bibnamefont {Lindh}},
  \bibinfo {author} {\bibfnamefont {A.}~\bibnamefont {Mitrushenkov}}, \bibinfo
  {author} {\bibfnamefont {G.}~\bibnamefont {Rauhut}}, \bibinfo {author}
  {\bibfnamefont {K.~R.}\ \bibnamefont {Shamasundar}}, \bibinfo {author}
  {\bibfnamefont {T.~B.}\ \bibnamefont {Adler}}, \bibinfo {author}
  {\bibfnamefont {R.~D.}\ \bibnamefont {Amos}}, \bibinfo {author}
  {\bibfnamefont {S.~J.}\ \bibnamefont {Bennie}}, \bibinfo {author}
  {\bibfnamefont {A.}~\bibnamefont {Bernhardsson}}, \bibinfo {author}
  {\bibfnamefont {A.}~\bibnamefont {Berning}}, \bibinfo {author} {\bibfnamefont
  {D.~L.}\ \bibnamefont {Cooper}}, \bibinfo {author} {\bibfnamefont {M.~J.~O.}\
  \bibnamefont {Deegan}}, \bibinfo {author} {\bibfnamefont {A.~J.}\
  \bibnamefont {Dobbyn}}, \bibinfo {author} {\bibfnamefont {F.}~\bibnamefont
  {Eckert}}, \bibinfo {author} {\bibfnamefont {E.}~\bibnamefont {Goll}},
  \bibinfo {author} {\bibfnamefont {C.}~\bibnamefont {Hampel}}, \bibinfo
  {author} {\bibfnamefont {A.}~\bibnamefont {Hesselmann}}, \bibinfo {author}
  {\bibfnamefont {G.}~\bibnamefont {Hetzer}}, \bibinfo {author} {\bibfnamefont
  {T.}~\bibnamefont {Hrenar}}, \bibinfo {author} {\bibfnamefont
  {G.}~\bibnamefont {Jansen}}, \bibinfo {author} {\bibfnamefont
  {C.}~\bibnamefont {K{\"{o}}ppl}}, \bibinfo {author} {\bibfnamefont
  {S.~J.~R.}\ \bibnamefont {Lee}}, \bibinfo {author} {\bibfnamefont
  {Y.}~\bibnamefont {Liu}}, \bibinfo {author} {\bibfnamefont {A.~W.}\
  \bibnamefont {Lloyd}}, \bibinfo {author} {\bibfnamefont {Q.}~\bibnamefont
  {Ma}}, \bibinfo {author} {\bibfnamefont {R.~A.}\ \bibnamefont {Mata}},
  \bibinfo {author} {\bibfnamefont {A.~J.}\ \bibnamefont {May}}, \bibinfo
  {author} {\bibfnamefont {S.~J.}\ \bibnamefont {McNicholas}}, \bibinfo
  {author} {\bibfnamefont {W.}~\bibnamefont {Meyer}}, \bibinfo {author}
  {\bibfnamefont {T.~F.}\ \bibnamefont {{Miller III}}}, \bibinfo {author}
  {\bibfnamefont {M.~E.}\ \bibnamefont {Mura}}, \bibinfo {author}
  {\bibfnamefont {A.}~\bibnamefont {Nicklass}}, \bibinfo {author}
  {\bibfnamefont {D.~P.}\ \bibnamefont {O'Neill}}, \bibinfo {author}
  {\bibfnamefont {P.}~\bibnamefont {Palmieri}}, \bibinfo {author}
  {\bibfnamefont {D.}~\bibnamefont {Peng}}, \bibinfo {author} {\bibfnamefont
  {K.}~\bibnamefont {Pfl{\"{u}}ger}}, \bibinfo {author} {\bibfnamefont
  {R.}~\bibnamefont {Pitzer}}, \bibinfo {author} {\bibfnamefont
  {M.}~\bibnamefont {Reiher}}, \bibinfo {author} {\bibfnamefont
  {T.}~\bibnamefont {Shiozaki}}, \bibinfo {author} {\bibfnamefont
  {H.}~\bibnamefont {Stoll}}, \bibinfo {author} {\bibfnamefont {A.~J.}\
  \bibnamefont {Stone}}, \bibinfo {author} {\bibfnamefont {R.}~\bibnamefont
  {Tarroni}}, \bibinfo {author} {\bibfnamefont {T.}~\bibnamefont
  {Thorsteinsson}}, \bibinfo {author} {\bibfnamefont {M.}~\bibnamefont {Wang}},
  \ and\ \bibinfo {author} {\bibfnamefont {M.}~\bibnamefont {Welborn}},\
  }\href@noop {} {\enquote {\bibinfo {title} {{MOLPRO, version 2019.2, a
  package of ab initio programs}},}\ } (\bibinfo {year} {2019})\BibitemShut
  {NoStop}%
\bibitem [{Note1()}]{Note1}%
  \BibitemOpen
  \bibinfo {note} {The MRCI calculations are fed with the natural orbitals from
  a Multi-Configuration Self-Consistent Field (MCSCF) calculation with a
  complete active space (CAS) consisting of nine orbitals with $A_1$ symmetry,
  three with $B_1$ symmetry and three with $B_2$ symmetry associated with the
  point group $C_{2v}$. The calculations employ the AV5Z \cite {BasisSet} basis
  set for each atom.}\BibitemShut {Stop}%
\bibitem [{\citenamefont {Woon}\ and\ \citenamefont {Herbst}(2009)}]{Woon2009}%
  \BibitemOpen
  \bibfield  {author} {\bibinfo {author} {\bibfnamefont {D.~E.}\ \bibnamefont
  {Woon}}\ and\ \bibinfo {author} {\bibfnamefont {E.}~\bibnamefont {Herbst}},\
  }\href {\doibase 10.1088/0067-0049/185/2/273} {\bibfield  {journal} {\bibinfo
   {journal} {Astrophysical Journal, Supplement Series}\ }\textbf {\bibinfo
  {volume} {185}},\ \bibinfo {pages} {273} (\bibinfo {year}
  {2009})}\BibitemShut {NoStop}%
\bibitem [{\citenamefont {Kaneda}\ \emph {et~al.}(2016)\citenamefont {Kaneda},
  \citenamefont {Yarborough}, \citenamefont {Merzlyak}, \citenamefont
  {Yamaguchi}, \citenamefont {Hayashida}, \citenamefont {Ohmae},\ and\
  \citenamefont {Katori}}]{Kaneda2016}%
  \BibitemOpen
  \bibfield  {author} {\bibinfo {author} {\bibfnamefont {Y.}~\bibnamefont
  {Kaneda}}, \bibinfo {author} {\bibfnamefont {J.~M.}\ \bibnamefont
  {Yarborough}}, \bibinfo {author} {\bibfnamefont {Y.}~\bibnamefont
  {Merzlyak}}, \bibinfo {author} {\bibfnamefont {A.}~\bibnamefont {Yamaguchi}},
  \bibinfo {author} {\bibfnamefont {K.}~\bibnamefont {Hayashida}}, \bibinfo
  {author} {\bibfnamefont {N.}~\bibnamefont {Ohmae}}, \ and\ \bibinfo {author}
  {\bibfnamefont {H.}~\bibnamefont {Katori}},\ }\href {\doibase
  10.1364/OL.41.000705} {\bibfield  {journal} {\bibinfo  {journal} {Optics
  Letters}\ }\textbf {\bibinfo {volume} {41}},\ \bibinfo {pages} {705}
  (\bibinfo {year} {2016})}\BibitemShut {NoStop}%
\bibitem [{\citenamefont {Burkley}\ \emph {et~al.}(2018)\citenamefont
  {Burkley}, \citenamefont {Brandt}, \citenamefont {Rasor}, \citenamefont
  {Cooper},\ and\ \citenamefont {Yost}}]{Burkley2018}%
  \BibitemOpen
  \bibfield  {author} {\bibinfo {author} {\bibfnamefont {Z.}~\bibnamefont
  {Burkley}}, \bibinfo {author} {\bibfnamefont {A.~D.}\ \bibnamefont {Brandt}},
  \bibinfo {author} {\bibfnamefont {C.}~\bibnamefont {Rasor}}, \bibinfo
  {author} {\bibfnamefont {S.~F.}\ \bibnamefont {Cooper}}, \ and\ \bibinfo
  {author} {\bibfnamefont {D.~C.}\ \bibnamefont {Yost}},\ }\href
  {http://arxiv.org/abs/1811.09874} {\ \textbf {\bibinfo {volume} {80523}},\
  \bibinfo {pages} {1} (\bibinfo {year} {2018})},\ \Eprint
  {http://arxiv.org/abs/1811.09874} {arXiv:1811.09874} \BibitemShut {NoStop}%
\bibitem [{\citenamefont {Tarbutt}\ \emph {et~al.}(2013)\citenamefont
  {Tarbutt}, \citenamefont {Sauer}, \citenamefont {Hudson},\ and\ \citenamefont
  {Hinds}}]{Tarbutt2013}%
  \BibitemOpen
  \bibfield  {author} {\bibinfo {author} {\bibfnamefont {M.~R.}\ \bibnamefont
  {Tarbutt}}, \bibinfo {author} {\bibfnamefont {B.~E.}\ \bibnamefont {Sauer}},
  \bibinfo {author} {\bibfnamefont {J.~J.}\ \bibnamefont {Hudson}}, \ and\
  \bibinfo {author} {\bibfnamefont {E.~A.}\ \bibnamefont {Hinds}},\ }\href
  {\doibase 10.1088/1367-2630/15/5/053034} {\bibfield  {journal} {\bibinfo
  {journal} {New Journal of Physics}\ }\textbf {\bibinfo {volume} {15}},\
  \bibinfo {pages} {053034} (\bibinfo {year} {2013})},\ \Eprint
  {http://arxiv.org/abs/1302.2870} {arXiv:1302.2870} \BibitemShut {NoStop}%
\bibitem [{\citenamefont {Norrgard}\ \emph {et~al.}(2016)\citenamefont
  {Norrgard}, \citenamefont {McCarron}, \citenamefont {Steinecker},
  \citenamefont {Tarbutt},\ and\ \citenamefont {DeMille}}]{Norrgard2016}%
  \BibitemOpen
  \bibfield  {author} {\bibinfo {author} {\bibfnamefont {E.~B.}\ \bibnamefont
  {Norrgard}}, \bibinfo {author} {\bibfnamefont {D.~J.}\ \bibnamefont
  {McCarron}}, \bibinfo {author} {\bibfnamefont {M.~H.}\ \bibnamefont
  {Steinecker}}, \bibinfo {author} {\bibfnamefont {M.~R.}\ \bibnamefont
  {Tarbutt}}, \ and\ \bibinfo {author} {\bibfnamefont {D.}~\bibnamefont
  {DeMille}},\ }\href {\doibase 10.1103/PhysRevLett.116.063004} {\bibfield
  {journal} {\bibinfo  {journal} {Physical Review Letters}\ }\textbf {\bibinfo
  {volume} {116}},\ \bibinfo {pages} {063004} (\bibinfo {year}
  {2016})}\BibitemShut {NoStop}%
\bibitem [{\citenamefont {Tarbutt}(2015)}]{Tarbutt2015}%
  \BibitemOpen
  \bibfield  {author} {\bibinfo {author} {\bibfnamefont {M.~R.}\ \bibnamefont
  {Tarbutt}},\ }\href {\doibase 10.1088/1367-2630/17/1/015007} {\bibfield
  {journal} {\bibinfo  {journal} {New Journal of Physics}\ }\textbf {\bibinfo
  {volume} {17}},\ \bibinfo {pages} {015007} (\bibinfo {year} {2015})},\
  \Eprint {http://arxiv.org/abs/1409.0244} {arXiv:1409.0244} \BibitemShut
  {NoStop}%
\bibitem [{\citenamefont {Bethlem}\ \emph {et~al.}(2000)\citenamefont
  {Bethlem}, \citenamefont {Berden}, \citenamefont {van Roij}, \citenamefont
  {Crompvoets},\ and\ \citenamefont {Meijer}}]{Bethlem2000}%
  \BibitemOpen
  \bibfield  {author} {\bibinfo {author} {\bibfnamefont {H.~L.}\ \bibnamefont
  {Bethlem}}, \bibinfo {author} {\bibfnamefont {G.}~\bibnamefont {Berden}},
  \bibinfo {author} {\bibfnamefont {A.~J.~A.}\ \bibnamefont {van Roij}},
  \bibinfo {author} {\bibfnamefont {F.~M.~H.}\ \bibnamefont {Crompvoets}}, \
  and\ \bibinfo {author} {\bibfnamefont {G.}~\bibnamefont {Meijer}},\ }\href
  {\doibase 10.1103/PhysRevLett.84.5744} {\bibfield  {journal} {\bibinfo
  {journal} {Physical Review Letters}\ }\textbf {\bibinfo {volume} {84}},\
  \bibinfo {pages} {5744} (\bibinfo {year} {2000})},\ \Eprint
  {http://arxiv.org/abs/0801.2943} {arXiv:0801.2943} \BibitemShut {NoStop}%
\bibitem [{\citenamefont {Meek}\ \emph {et~al.}(2008)\citenamefont {Meek},
  \citenamefont {Bethlem}, \citenamefont {Conrad},\ and\ \citenamefont
  {Meijer}}]{Meek2008}%
  \BibitemOpen
  \bibfield  {author} {\bibinfo {author} {\bibfnamefont {S.~A.}\ \bibnamefont
  {Meek}}, \bibinfo {author} {\bibfnamefont {H.~L.}\ \bibnamefont {Bethlem}},
  \bibinfo {author} {\bibfnamefont {H.}~\bibnamefont {Conrad}}, \ and\ \bibinfo
  {author} {\bibfnamefont {G.}~\bibnamefont {Meijer}},\ }\href {\doibase
  10.1103/PhysRevLett.100.153003} {\bibfield  {journal} {\bibinfo  {journal}
  {Physical Review Letters}\ }\textbf {\bibinfo {volume} {100}},\ \bibinfo
  {pages} {153003} (\bibinfo {year} {2008})},\ \Eprint
  {http://arxiv.org/abs/0801.2943} {arXiv:0801.2943} \BibitemShut {NoStop}%
\bibitem [{\citenamefont {Schuchardt}\ \emph {et~al.}(2007)\citenamefont
  {Schuchardt}, \citenamefont {Didier}, \citenamefont {Elsethagen},
  \citenamefont {Sun}, \citenamefont {Gurumoorthi}, \citenamefont {Chase},
  \citenamefont {Li},\ and\ \citenamefont {Windus}}]{BasisSet}%
  \BibitemOpen
  \bibfield  {author} {\bibinfo {author} {\bibfnamefont {K.~L.}\ \bibnamefont
  {Schuchardt}}, \bibinfo {author} {\bibfnamefont {B.~T.}\ \bibnamefont
  {Didier}}, \bibinfo {author} {\bibfnamefont {T.}~\bibnamefont {Elsethagen}},
  \bibinfo {author} {\bibfnamefont {L.}~\bibnamefont {Sun}}, \bibinfo {author}
  {\bibfnamefont {V.}~\bibnamefont {Gurumoorthi}}, \bibinfo {author}
  {\bibfnamefont {J.}~\bibnamefont {Chase}}, \bibinfo {author} {\bibfnamefont
  {J.}~\bibnamefont {Li}}, \ and\ \bibinfo {author} {\bibfnamefont {T.~L.}\
  \bibnamefont {Windus}},\ }\href {\doibase 10.1021/ci600510j} {\bibfield
  {journal} {\bibinfo  {journal} {Journal of Chemical Information and
  Modeling}\ }\textbf {\bibinfo {volume} {47}},\ \bibinfo {pages} {1045}
  (\bibinfo {year} {2007})}\BibitemShut {NoStop}%
\end{thebibliography}

\end{document}